\documentclass[pdflscape,lscape,iop,apj,numberedappendix,appendixfloats,stfloats,url,microtype]{emulateapj}

\usepackage{paralist,amsmath}
\usepackage{float}
\usepackage{morefloats}

\usepackage{color,soul}

\graphicspath{{/Users/craiganderson/Dropbox/LaTeX/images/}}

\shorttitle{Broadband Polarimetry of 563 Radio Sources}
\shortauthors{Anderson et al.}

\begin{document}


\title{Broadband Radio Polarimetry and Faraday Rotation of 563 Extragalactic Radio Sources}


\author{
C. S. Anderson\altaffilmark{1,2,}$^{\dagger}$,
B. M. Gaensler\altaffilmark{1,3},
I. J. Feain\altaffilmark{1,4},
T. M. O. Franzen\altaffilmark{2,5}
}

\altaffiltext{$\dagger$}{{\bf craiga@physics.usyd.edu.au}}
\altaffiltext{1}{Sydney Institute for Astronomy (SIfA), School of Physics, University of Sydney, NSW 2006, Australia}
\altaffiltext{2}{CSIRO Astronomy \& Space Science, Epping, NSW 1710, Australia}
\altaffiltext{3}{Dunlap Institute for Astronomy \& Astrophysics, University of Toronto, Ontario, Canada}
\altaffiltext{4}{Central Clinical School, School of Medicine, University of Sydney, NSW 2006, Australia}
\altaffiltext{5}{International Centre for Radio Astronomy Research (ICRAR), Curtin University, Bentley, WA 6102, Australia}


\begin{abstract}
We present a broadband spectropolarimetric survey of 563 discrete, mostly unresolved radio sources between 1.3 \& 2.0 GHz using data taken with the Australia Telescope Compact Array (ATCA). We have used rotation measure synthesis to identify Faraday complex polarized sources --- i.e. objects whose frequency-dependent polarization behaviour indicates the presence of material possessing complicated magnetoionic structure along the line of sight (LOS). For sources classified as Faraday complex, we have analyzed a number of their radio and multiwavelength properties to determine whether they differ from Faraday simple polarized sources (i.e. sources for which LOS magnetoionic structures are comparatively simple) in these properties. We use this information to constrain the physical nature of the magnetoionic structures responsible for generating the observed complexity. We detect Faraday complexity in 12\% of polarized sources at $\sim1'$ resolution, but demonstrate that underlying signal-to-noise limitations mean the true percentage is likely to be significantly higher in the polarized radio source population. We find that the properties of Faraday complex objects are diverse, but that complexity is most often associated with depolarization of extended radio sources possessing a relatively steep total intensity spectrum. We find an association between Faraday complexity and LOS structure in the Galactic interstellar medium (ISM), and claim that a significant proportion of the Faraday complexity we observe may be generated at interfaces of the ISM associated with ionization fronts near neutral hydrogen structures. Galaxy clusters environments and internally generated Faraday complexity provide possible alternative explanations in some cases.

\end{abstract}






\section{Introduction}\label{sec-intro}

Magnetized plasma is central to our understanding of important astrophysical processes in diverse cosmic environments. One of the key observational tools used to study these plasmas is Faraday rotation. When linearly polarized radiation passes through a magnetized thermal plasma (a \emph{Faraday screen}), its plane of polarization is rotated according to

 \begin{equation}
\chi_{\text{obs}} = \chi_0 + \text{RM} \lambda_{\text{obs}}^2
\label{eq:FaradayRotation}
 \end{equation}
 
where $\chi_0$ and $\chi_{\text{obs}}$ are the emitted and observed polarization angles, and $\lambda_{\text{obs}}$ is the observing wavelength. The magnitude of the effect is parameterized by the rotation measure (RM). Defined as the gradient of $\chi_{\text{obs}}$ vs. $\lambda_{\text{obs}}$, the RM of a source can be related to the electron density $n_e$ [cm$^{-3}$], magnetic field $\boldsymbol{B}$ [$\mu$G] and displacement $\boldsymbol{s}$ [parsecs] between it and the observer as:
 
  \begin{equation}
\text{RM} = 0.812 \int_{\text{source}}^{\text{telescope}} n_e\boldsymbol{B}.\text{d}\boldsymbol{s}~\text{rad m}^{-2}
\label{eq:RotationMeasure}
 \end{equation}
 
RMs are routinely used to study Faraday screens in the radio jets of AGN (e.g. Hovatta et al. 2012), the immediate vicinity of AGN (e.g. G\'{o}mez et al. 2011), the environment surrounding radio lobes (e.g. Rudnick \& Blundell 2003, Feain et al. 2009), intercluster material (e.g. Feretti et al. 2012), intervening galaxies (e.g. Kronberg et al. 1992, Bernet et al. 2008, Farnes et al. 2014b) or the Galactic interstellar medium (ISM) (e.g. Taylor, Stil \& Sunstrum 2009, Harvey-Smith et al. 2011, Haverkorn 2014 and refs therein).
 
However, RMs alone can sometimes give an incomplete (e.g. Farnsworth et al. 2011) or even erroneous (e.g. O'Sullivan et al. 2012) picture of magnetoionic structure along the LOS if a) a source is observed through a Faraday screen with non-uniformities in either $n_e$ or $\boldsymbol{B}$, b) there exists multiple synchrotron emitting regions along the line of sight, each possessing different RMs, or c) there is mixing of the synchrotron emitting and Faraday rotating plasma a the source. To handle these situations, Burn (1966) derived a more generally applicable mathematical framework as follows. First, a quantity known as Faraday depth is defined by:

 \begin{equation}
\text{$\phi$($L$)} = 0.812 \int_{L}^{\text{telescope}} n_e\boldsymbol{B}.\text{d}\boldsymbol{s}~\text{rad m}^{-2}
\label{eq:FaradayDepth}
 \end{equation}

where $L$ is the position along a line of sight (LOS). Next, a complex polarization vector $\boldsymbol{P}$ is defined, related to the Stokes parameters $Q$ \& $U$, the polarization angle $\chi$, the fractional polarization $p$ and the total intensity $I$ as:

 \begin{equation}
\boldsymbol{P} = Q + iU = pIe^{2i\chi}
\label{eq:ComplexPolVec}
 \end{equation}

 The Stokes parameters are related to the polarization angle by:

 \begin{equation}
\chi = \frac{1}{2} \text{tan}^{-1} \left(\frac{U}{Q}\right)
\label{eq:StokesPolAng}
 \end{equation}
 
The observable linear polarization along a LOS is then given by:

 \begin{equation}
\boldsymbol{P}(\lambda^2) = \int_{-\infty}^{\infty} F(\phi) e^{2i\phi\lambda^2} d\phi
\label{eq:SumPol}
 \end{equation}
 
 where $F(\phi)$ is the complex polarized surface brightness per unit Faraday depth along the LOS, possessing units of Jy m$^2$ rad$^{-3}$. 
 
 The range of possible behaviours that $\boldsymbol{P}(\lambda^2)$ can show can be divided into two broad categories. The first occurs when $F(\phi)$ is non-zero at a single value of $\phi$ only, which corresponds to polarized emission undergoing pure Faraday rotation by a uniform Faraday screen in the foreground. Eqns. \ref{eq:ComplexPolVec}, \ref{eq:StokesPolAng} \& \ref{eq:SumPol} show that under these circumstances $\chi(\lambda^2)$ is linear (modulo $\pi$ radians), Stokes $Q(\lambda^2)/I(\lambda^2)$ \& $U(\lambda^2)/I(\lambda^2)$ vary sinusoidally, and $p(\lambda^2)$ is constant. This is an idealization since a perfectly uniform Faraday screen will never be fully realized in Nature. However, finite observational sensitivity can render deviations from these behaviours undetectable, even for polarized emission components with S/N ratios $>100$ (O'Sullivan et al. 2012). For the purposes of our study, we refer to sources as being \emph{Faraday simple} when deviations from the behaviours described above are undetectable. Note that this is an observational rather than physical classification, in the same way that objects can be classified as point sources in aperture synthesis images despite having non-zero physical extent in reality.
 
 
 In contrast, when $F(\phi)$ is non-zero at multiple values of $\phi$, Stokes $Q(\lambda^2)/I(\lambda^2)$ \& $U(\lambda^2)/I(\lambda^2)$, $\chi(\lambda^2)$ and $p(\lambda^2)$ must (or in the case of $\chi(\lambda^2)$, can) deviate from the idealized Faraday simple behaviours described above. When we detect these deviations in sources, we refer to them as being \emph{Faraday complex}. While observations of Faraday complexity are not new (e.g. Slysh 1965, Conway et al. 1974,  Goldstein \& Reid 1984), the capability to detect Faraday complexity and characterize $\boldsymbol{P}(\lambda^2)$ has historically been limited by the narrow bandwidths of previous generation radio telescopes, which typically only allowed narrowband or sparsely sampled studies of $p$($\lambda^2$) or $\chi(\lambda^2)$. In recent years however, these limitations have been substantially lifted by the advent of GHz-bandwidth spectropolarimetry. Broadband studies now suggest that Faraday complexity is commonplace among polarized radio sources in the frequency range 300 MHz -- 3 GHz (Law et al. 2011, Farnsworth et al. 2011 and O'Sullivan et al. 2012), while Farnes et al. (2014a) have built on the results of Conway et al. (1974) to conclusively show that \emph{most} radio sources appear Faraday complex when examined over ultra wide ($\sim10$ GHz) bands. Furthermore, broadband spectropolarimetric modelling has allowed Law et al. (2011), O'Sullivan et al. (2012) \& Farnes et al. (2014a) to argue that in some cases, complexity is generated internally to active galactic nuclei (AGN). If this turns out to be common, broadband data from future surveys such as POSSUM (the Polarization Sky Survey of the Universe's Magnetism; Gaensler et al. 2009) will allow aspects of the magnetized structure of vast numbers of AGN to be resolved spectrally rather than spatially, opening up entirely new avenues for studying magnetic processes in these objects.
  
 At the present time however, broadband analysis of Faraday complexity remains in its infancy. The amount of Faraday complexity observable in the radio source population remains uncertain, in terms of both degree and prevalence, especially among fainter, sub-Jy sources. Little is known about whether different types of Faraday complexity exist, and if so, whether these differences are associated with different types of sources. While the aforementioned studies argue that complexity can be internally generated in radio sources, relatively few examples have been studied to date. It is therefore not clear what proportion of sources this might apply to, nor is it clear where complexity is generated along the LOS for sources where it does not apply. In this work, we begin to address these outstanding issues by performing a blind survey for Faraday complexity with dense spectral sampling from 1.3--2.0 GHz for a large sample of galaxies ($563$ sources). Based on the observed polarization data, we classify each polarized source as either Faraday complex or simple (i.e. no complexity is detected), then focus on answering the following questions: How readily can we detect Faraday complexity in sources using broadband spectropolarimetric analysis techniques such as RM synthesis? In what proportion of the general source population do we detect complexity? Do the polarimetric and non-polarimetric properties of Faraday complex sources differ from those of Faraday simple sources? Is Faraday complexity limited to particular types of sources? What is the physical origin of the Faraday rotation responsible for complexity in these objects? 

Our paper is set out as follows. We describe our observations and their calibration in Section \ref{sec-observations}, our source-finding procedure in Section \ref{sec-srcfind}, and our spectropolarimetric analysis in Section \ref{sec-specpol}. In Section \ref{sec-additional} we describe ancillary, non-spectropolarimetric characterization of our sample. We present our results in Section \ref{sec-results}. In Section \ref{sec-discussion} we consider the prevalence of complex sources and the physical nature of the complexity-inducing Faraday screens. We summarize our work and present our conclusions in Section \ref{sec-conclusion}, and elaborate on technical aspects of our work in an Appendix.

\section{Observations and calibration}\label{sec-observations}

 We obtained mosaiced observations of a 30 deg$^2$ region of sky using the CABB correlator on the Australia Telescope Compact Array (ATCA; Wilson et al. 2011). Our observations were performed using the \lq CFB 1M' mode, which generates all polarization products from 1.1 to 3.1 GHz with 1 MHz channel widths. The mosaic consisted of 342 pointings laid out in a hexagonal grid. This grid spanned $7.5^{\circ}$ in RA and $5.5^{\circ}$ in Dec, and was centered on RA $= 03^{h}29^{m}40^{s}$ \& Dec $= -36^{d}16^{m}30^{s}$ (J2000) in Fornax. The angular separation of the mosaic pointings was 0.323 degrees, and therefore spatially Nyquist-sampled at $1.4$ GHz. To obtain adequate $uv$ coverage, we broke the full mosaic up into 7 sub-mosaics and observed each sub-mosaic on consecutive days. We completed the full 7-day observing run twice --- once in each of the 1.5B \& 750B array configurations, from 2011 May 5th--11th \& 2011 June 10th--16th, respectively.  Each pointing in the mosaic received an average integration time of 30 minutes divided between $\sim20$ $uv$ cuts, resulting in a theoretical sensitivity of $\sim22~\mu$Jy beam$^{-1}$ (over the full 2 GHz bandwidth, assuming six antennas and natural weighting employed; the actual sensitivities achieved on individual image \& data products are listed in the relevant sections).
 
 We calibrated our data with {\sc miriad} (Sault et al. 1995) following standard procedures for CABB data. Prior to the calibration, we flagged the data for antenna shadowing and poor sensitivity at the band edges (100 channels at each end). Radio-frequency interference (RFI) was flagged iteratively during the calibration process using the {\sc sumthreshold} algorithm (Offringa et al. 2010). The bandpass response and absolute flux scale were calibrated using daily observations of PKS B1934$-$638, while PKS B0402$-$362 was observed at 60 minute intervals to calibrate the time-dependent complex antenna gains and on-axis polarization leakage. Independent calibration solutions were derived at 128 MHz intervals across the full band due to the frequency dependence of the gain and leakage solutions (Schnitzeler et al. 2011). On the basis of analysis that we present in Appendix \ref{sec-discussion-soundness-leak}, we estimate that post-calibration, on-axis leakage is limited to $<0.1$\% of Stokes $I$ when averaged over the band. 

After applying the calibration to the target data, we derived and cross-applied RFI flags for Stokes $I$, $Q$, $U$ \& $V$. The sub-bands 1.1--1.35 GHz \& 1.48--1.64 GHz were heavily RFI-afflicted so we flagged them completely, leaving a total usable bandwidth of 1.59 GHz in the 1.1--3.1 GHz band. Outside the flagged bands, $\sim35$\% of the data were flagged.

\section{Source finding}\label{sec-srcfind}

We created a high sensitivity ($\sim 60 ~\mu$Jy beam$^{-1}$), 15" resolution mosaic of the entire field to locate sources for subsequent spectropolarimetric analysis. Data from individual pointings between 1.35 and 2.1 GHz were imaged using multifrequency synthesis then {\sc clean}ed, restored, primary beam-corrected and linearly mosaiced. We refer to this mosaic henceforth as `the source finding image'. The lower sensitivity of the source finding image (compared to the theoretical sensitivity quoted in the preceding section) is due to the robust$=0.5$ weighting scheme and the limited bandwidth employed in its creation.
 
 We used {\sc selavy} (Whiting \& Humphreys, 2012) to detect and catalogue sources with Stokes $I$ flux densities $>3$ mJy beam$^{-1}$ in the source finding image --- a limit imposed because robust detections of polarized emission in fainter sources were unlikely. We estimate that the positional uncertainty for all catalogued sources caused by calibration errors and thermal noise is substantially smaller than an arcsecond. We excluded sources lying $<1^\circ$ from the core of the bright radio galaxy Fornax A from the catalogue due to sidelobe confusion and dynamic range errors generated by its radio lobes. Names for each polarized source (sources are classified as polarized or unpolarized in Section \ref{sec-compprev}) and their RA \& Dec (J2000) positions are recorded in columns 1, 2 \& 3 of Table \ref{tab:allsourcedata} respectively. We have also recorded the angular separation and position angle of each source from the position of its nearest mosaic pointing centre in columns 4 \& 5 of Table \ref{tab:allsourcedata}, which are required for our characterization of instrumental polarization (discussed in Section \ref{sec-impol}).

\section{Spectropolarimetric  analysis}\label{sec-specpol}

\subsection{Spectropolarimetric imaging}\label{sec-imaging}
 
To obtain the frequency-dependent source flux densities required for our spectropolarimetric analysis, we generated a second set of mosaic images of the field in Stokes $I$, $Q$, $U$ \& $V$ at 8 MHz frequency intervals between 1.35 \& 3.1 GHz. We refer to these as our `spectropolarimetric images'. We chose an 8 MHz imaging sub-band to achieve acceptable imaging times while preventing bandwidth depolarization for Faraday depths of $\phi\leq4500$ rad m$^2$. For each 8 MHz sub-band, we imaged each of the 342 mosaic pointings as follows: 

We generated the Stokes $I$, $Q$, $U$ \& $V$ dirty maps out to 3 times the primary beam width (beyond which the primary beam response is negligible) with 7 resolution elements across the synthesized beam. We employed a Briggs (1995) robust weighting scheme with robustness $= -2$. We discarded data for baselines incorporating antenna 6 (thereby reducing the resolution of the spectropolarimetric images by a factor of $\sim$4 from that of the source finding image --- see below for final resolution) to decrease the amplitude of sidelobe structure and to keep the image processing and data storage requirements achievable. The heavily RFI-affected visibilities in the $uv$ distance range 0 to 1 k$\lambda$ were also discarded.

We deconvolved the dirty maps using {\sc clean}. For Stokes $Q$ \& $U$ images, we imposed a flux density cutoff threshold of 4 times the RMS noise of the Stokes $V$ map. For the Stokes $I$ images, our dynamic range limit of $\sim 100$ (limited by $uv$ coverage) meant that {\sc clean} sometimes diverged when deconvolving the brightest sources in our field. To handle this, we set the {\sc clean} cutoff threshold for Stokes $I$ images to whichever was greater out of: a) 8 times the Stokes $V$ noise or b) 1\% of the flux density of the brightest source in the field. This resulted in sidelobes being {\sc clean}ed to a level of $\sim$ twice the noise level measured in Stokes $V$ in almost all pointings --- below the typical measured noise of $\sim 3$--$5$ mJy beam$^{-1}$ in the Stokes $I$ image at each frequency. Compared to theoretical and source finding image sensitivities quoted in Sections \ref{sec-observations} \& \ref{sec-srcfind} respectively, the sensitivity of our spectropolarimetric images is reduced mainly due to the effect of the robust $= -2$ weighting scheme.

We created restored maps then smoothed them to a common resolution of $90"\times 45"$ --- slightly lower than the resolution of our images at 1.35 GHz produced under the weighting scheme described above.
 
After imaging each of the 342 mosaic pointings in this way, we primary beam-corrected and linearly combined the maps to form the final mosaic image of the field at each frequency. 
 
 \subsection{Stokes parameter extraction}\label{sec-extract}
 
We extracted the frequency-dependent Stokes $I$, $Q$, $U$ \& $V$ flux densities for each source catalogued in Table \ref{tab:allsourcedata} (Section \ref{sec-srcfind}) directly from the spectropolarimetric images. We determined whether sources were resolved by assessing the goodness of fit of a $90" \times 45"$ Gaussian (i.e. the restoring beam of the spectropolarimetric images) to the source. The result is recorded in column 6 of Table \ref{tab:allsourcedata}. For unresolved sources, flux densities were extracted by measuring the value of the image pixel centred on the source positions obtained in Section \ref{sec-srcfind}. For resolved sources, we generated an image mask for each source by smoothing the source finding image to $90" \times 45"$ and masking pixels for which the signal was $< 5\sigma$. We then applied this mask to the Stokes $I$, $Q$, $U$ \& $V$ spectropolarimetric mosaics at each frequency, summing the spectral brightness in the unmasked regions to obtain the integrated flux density for each Stokes parameter in units of Jy. 
   
 For both resolved and unresolved sources, the uncertainties on the data were estimated via direct measurement of the RMS noise adjacent to the source in the  Stokes $I$, $Q$ \& $U$ {\sc clean} residual images and the Stokes $V$ dirty map. These values were appropriately adjusted for resolved sources to reflect the number of statistically independent spatial regions sampled in the unmasked region.
 
\subsection{Off-axis instrumental polarization}\label{sec-impol}

A robust protocol for calibrating off-axis polarization leakage is not currently available for ATCA data. Instead, we used a statistical analysis of the target data (described in Section \ref{sec-discussion-soundness-leak}) to estimate upper limits on the leakage as a function of both frequency and position in the primary beam. On this basis, we excluded from subsequent analysis data at frequencies $> 2$ GHz and sources lying $> 0.2^\circ$ from the beam centre (the latter condition affected sources at the mosaic edges only). While some frequency-dependent leakage persists in the remaining data, its impact on our work is diminished due to incoherent summation under RM synthesis (Section \ref{sec-rmsynth}). An analysis of each of the sources in the field (also presented in Section \ref{sec-discussion-soundness-leak}) shows that the maximum polarization leakage that appears in RM synthesis spectra is $\lesssim0.3\%$ of Stokes $I$ for sources $\lesssim0.16^\circ$ from the beam centre (which applies to $>80$\% of our sample). Due to the limited number of sufficiently bright sources $\gtrsim0.16^\circ$ from the beam centre, we can only derive a weak leakage limit of $<1\%$ of Stokes $I$ for sources occupying these beam positions.

\subsection{RM synthesis and {\sc rmclean}}\label{sec-rmsynth}

RM synthesis (Brentjens \& DeBruyn 2005, Burn 1966) exploits the following Fourier relationship to directly calculate the intensity of polarized emission from a source over a vector of $\phi$ values:\\
 
 \begin{equation}
\boldsymbol{F}_\phi = \sum\limits_{j=1}^n w_j \boldsymbol{p}_je^{-2i\phi (\lambda_j^2-\lambda_0^2)} \Biggm/ \sum\limits_{j=1}^n w_j 
\label{eq:RMs}
 \end{equation}
 
where $F_\phi$ is the complex value of the FDS at Faraday depth $\phi$, n is the number of channels, $\boldsymbol{p}_j$ is the complex value of the fractional polarization vector in channel $j$ possessing a mean wavelength $\lambda_j$, $w_j$ is a weighting term acting on $\boldsymbol{p}_j$, and $\lambda_0$ is the weighted mean of $\lambda^2$ over all channels:\\

 \begin{equation}
\lambda_0^2 = \sum\limits_{j=1}^n w_j \lambda_j^2 \Biggm/ \sum\limits_{j=1}^n w_j
\label{eq:lSQ_0}
 \end{equation}
 
 We use $\lambda_0$ as a convenient reference wavelength (recorded in column 7 of Table \ref{tab:allsourcedata}) at which we evaluate various models \& quantities. We denote such quantities using a $\lambda_0$ subscript (e.g. $I_{\lambda_{0}}$ --- the interpolated value of Stokes $I$ at $\lambda_0$).

The ability of RM synthesis to reconstruct polarized emission is characterized by three main quantities: The maximum detectable Faraday depth, the maximum detectable scale of emission structures in $\phi$ space, and the resolution in $\phi$ space (see Brentjens \& DeBruyn 2005). For our data, the values are (respectively) $\phi_{max}\approx4100$ rad m$^{-2}$,  $\phi_{max-scale}\approx140$ rad m$^{-2}$ and $\delta \phi\approx120$ rad m$^{-2}$.
 
 We applied RM synthesis to each source as follows: 
 
 \begin{enumerate}
\item Fit a polynomial of order 1 to $\partial\text{log}(I) vs. \partial\text{log}(\nu)$ to get I$_{model}(\lambda)$ --- a power law model of the total intensity spectrum. 

\item Divide out first-order spectral index effects by dividing Stokes $Q$ \& $U$ by I$_{model}$ to obtain Stokes $q$ \& $u$.

\item Apply RM synthesis to the Stokes $q$ \& $u$ spectra, using either constant weighting (i.e. $w_j=1$) or weighting by the inverse of its variance (our results vary depending on the weighting scheme adopted --- see Section \ref{sec-weighteffect}). 

\item Deconvolve the FDS output by RM synthesis using {\sc rmclean} (Heald et al. 2009). We ran our analysis using three different {\sc rmclean} cutoff levels (the FDS amplitude below which the deconvolution procedure ceases to be applied); the values adopted for these levels and the reasons for their use are discussed in the following section.
\end{enumerate}

The FDS and associated {\sc rmclean} component model can then be used to assess the Faraday complexity of a polarized radio source. 

\subsection{Method for detecting Faraday complexity}\label{sec-compcat}

The question of how to best detect Faraday complexity is a subtle and complicated one, and an optimal approach has not yet been found. Previous studies have identified complexity using non-linearity in $\chi$ vs. $\lambda^2$ (e.g. Goldstein \& Reed 1984), frequency-dependant change in $p$ (e.g. Farnes et al. 2014a), non-sinusoidal variation in Stokes $q$ \& $u$ (e.g. O'Sullivan et al. 2012), or detection of multiple components in the FDS (e.g. Law et al. 2011). The first two of these are unsuitable for our study, because: 

\begin{itemize}
\item Linearity of $\chi$ vs. $\lambda^2$ does not imply Faraday simplicity, despite the converse being true.
\item $p$ is significantly affected by Ricean bias for faint sources. The optimal debiasing scheme is signal-dependent, as is the residual bias on the estimate of the true value of $p$ obtained using any given scheme (Simmons \& Stewart 1985). 
\end{itemize}

As part of a 1.1--1.4 GHz study into characterization of Faraday complexity, Sun et al. (2015) found that Faraday complexity arising from interference of two Faraday thin components could be detected more reliably by identifying non-sinusoidal variation in Stokes $q$ \& $u$ than by using any other existing method. However, no such advantage was evident for detection of complexity arising from Faraday thick components, and the authors conclude that over their limited bandwidth, all currently available methods are subject to type 2 (false negative) errors for detecting Faraday thick emission components. As we will show, the majority of Faraday complexity that we observe is probably of this type.

We proceeded by noting that our primary requirement is to make a binary classification of sources into Faraday complex / simple categories, and that we wish to employ conservative detection criteria in any case so as to minimise type 1 (false positive) errors. We therefore made a choice to detect complexity by searching for multiple components in the FDS resulting from RM synthesis \& {\sc rmclean}, which is both simple \& intuitive, has proven effective for wavelength coverage and spectral sampling comparable to our own (cf. Law et al. 2011), and at the same time, provides estimates of the peak and dispersion in Faraday depth of the emission from our sources.


Specifically, our method identifies complexity using the 2nd moment of the {\sc rmclean} component model, as described by Brown (2011). Denoting the {\sc rmclean} model as $F_M$ and the 2nd moment of this quantity as $\sigma_\phi$, it is calculated it as:

\begin{eqnarray}
\sigma_\phi&=&K^{-1}\sum\limits_{i=1}^{n}(\phi_i-\mu_\phi)^2 |F_M(\phi_i)|
\end{eqnarray}

where the normalization constant K is given by
\begin{eqnarray}
K=\sum\limits_{i=0}^{n}|F_M(\phi_i)|
\end{eqnarray}

and $\mu_\phi$, the first moment of the distribution, is given by
\begin{eqnarray}
\mu_\phi=K^{-1}\sum\limits_{i=0}^{n}\phi_i|F_M(\phi_i)|
\end{eqnarray}

where $n$ is the number of distinct $\phi$ values at which $F_M$ is non-zero, while $\phi_i$ is the Faraday depth of the $i$th Faraday depth vector component (i.e. the vector of $\phi$ values over which equation \ref{eq:RMs} was evaluated to calculate the FDS). $\sigma_\phi$ is a measure of the dispersion of the {\sc rmclean} model components; Faraday-simple sources will show zero or very little spread (i.e. they are dominated by polarized emission from a single Faraday depth), while Faraday-complex sources with multiple {\sc rmclean} components should reflect this as a larger value of $\sigma_\phi$. In Section 6.4.3, we show that this algorithm correctly selects the most heavily depolarising / repolarising sources as being Faraday complex, which acts as \emph{a posteriori} evidence that our method is effective. Further comparison of spectropolarimetric analysis techniques is beyond the scope of this work, but note that this is a topic of ongoing research (see Sun et al. 2015).

The practical implementation of our method requires choices to be made for two parameters. First, a threshold must be adopted for $\sigma_\phi$ above which a source is classified as Faraday-complex. We show that a value for this threshold is naturally suggested by our data in Section \ref{sec-compprev}. The second parameter is the {\sc rmclean} cutoff, which must be set low enough to detect faint emission components, while remaining high enough to render the false detection probability negligible. Since no formalism currently exists that describes the statistical behaviour of FDS containing multiple unresolved components, this is not a trivial problem. Instead we make use of the analytic formalism of Hales et al. (2012), which fully characterizes the detection significance of polarized emission in FDS when multiple unresolved components are absent (see also Macquart et al. 2012). The equations derived therein relate the polarized signal-to-noise ratio (S/N) of an emission component in a Faraday simple FDS to a Gaussian Equivalent Significance (GES) level --- i.e. a 3$\sigma$ GES detection for a non-Gaussian probability density function (PDF) is equivalent to a $3\sigma$ detection for a Gaussian PDF. If the {\sc rmclean} cutoff is set to a sufficiently high GES (i.e. one which ensures a negligible false detection probability), subsequent detection of multiple FDS components then implies complexity, even if the detection significance of the additional components cannot be formally calculated.

To provide a gauge of detection significance / degree of complexity, we calculated $\sigma_\phi$ at three different {\sc rmclean} cutoffs --- $6\sigma$ GES, $8\sigma$ GES \& $10\sigma$ GES --- then used the complexity classification assigned at each level as a proxy in this regard. For example, when we refer to a source as being `complex at the $8\sigma$ GES level', we mean it was detected as complex using an $8\sigma$ GES {\sc rmclean} cutoff (and therefore also a $6\sigma$ GES cutoff), but not using a $10\sigma$ GES cutoff. We then attribute a higher degree of complexity / a more robust detection to this source than to a $6\sigma$ GES complex detection, but vice versa for a $10\sigma$ GES detection. We discuss detection reliability under these {\sc rmclean} cutoffs in Appendix \ref{sec-discussion-soundness-cldepth}.
 
\section{Ancillary source characterization}\label{sec-additional} 

\subsection{Radio morphology}\label{sec-radmorph}

 Using the source finding image (see Section \ref{sec-srcfind}), we classified the Stokes $I$ morphology of our sources as: `unresolved' if they were well-fit by a Gaussian of the same dimensions as the restoring beam; `lobe/jet component' if they either a) possessed obvious radio lobe or jet morphology, or b) possessed a counterpart radio source within 3' of similar brightness (Hammond et al. 2012 and refs therein); `core-jet' if the total flux was dominated by an unresolved component in the presence of additional resolved components that were either a) substantially fainter than the core or b) radiated away from the core with linear morphology; or `extended' for resolved sources not meeting the above criteria. Our morphological classifications are presented in column 22 of Table \ref{tab:allsourcedata}. Note that these morphological classifications (made using the 15" resolution source finding image) are distinct from the resolution status (resolved / unresolved in the 1' resolution spectropolarimetric images) that we assigned to the sources for the purpose of spectropolarimetric data extraction in Section \ref{sec-extract}.

\subsection{Multiwavelength counterparts}\label{sec-mwxmatch}

 We crossmatched sources with counterparts in infrared (WISE; Wright et al. 2010), optical (SuperCOSMOS; Hambly et al. 2001), and ultraviolet (GALEX) images, and in the ROSAT All Sky Survey Bright Source Catalogue (RASS-BSC; Voges et al. 1999) and Faint Source Catalogue (RASS-FSC; Voges et al. 2000). In IR, optical and UV, we manually identified counterparts by overlaying contours from the source finding image onto the relevant survey images. We assigned a `match' when objects were present within 15" of the radio contour centroid (i.e. within the beam FWHM of the source finding image); a `match - off source' if either a) The radio source was extended, and a candidate counterpart source lay inside the 10\% radio flux contour, b) A second radio source was located within 3', and a candidate counterpart was located within 15" of the position of the flux centroid of the two sources (e.g. Best et al. 2005) or c) NASA Extragalactic Database (NED) queries revealed an existing association in the literature; or `no match' otherwise. The counterpart statuses in IR, optical and UV are listed in columns 23, 24 \& 25 of Table \ref{tab:allsourcedata} respectively. Our X-ray cross matches were made in a binary manner based on the distance to the nearest catalogued sources and the positional errors for the RASS-BSC and RASS-FSC (Parejko et al. 2008). We assigned a `match' if the radio emission centroid fell within 20" of a catalogued RASS-BSC source position or 40" of a RASS-FSC source position, and `no match' otherwise. The X-ray counterpart classifications are listed for each source in column 26 of Table \ref{tab:allsourcedata}.\
 
 We furthermore crossmatched our sample with redshift data in the literature using NED and a matching radius of $1'$ --- approximately the FWHM of the spectropolarimetric image restoring beam. We accepted 61 redshift matches for our sample, 21 of which were polarized sources. These are recorded in column 27 of Table \ref{tab:allsourcedata}. 
 
\section{Results}\label{sec-results}

\begin{figure}
\centering
\includegraphics[width=0.475\textwidth]{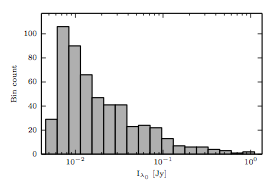}
\caption{Histogram of $I_{\lambda_0}$ for the sample.}
\label{fig:SandSI}
\end{figure}

Our final sample consists of 563 sources lying $<0.2^\circ$ from the nearest mosaic pointing centre and $>1^\circ$ from the core of Fornax A. A histogram of $I_{\lambda_0}$ for the sample is presented in Figure \ref{fig:SandSI}. The brightest source in the sample has $I_{\lambda_0} \approx 1.7$ Jy, while the median $I_{\lambda_0}$ is 13 mJy. 

\begin{turnpage} 
\begin{table*} 
\caption{(Columns 1--15) Selected quantities calculated / determined for polarized sources in our sample. Faraday complex sources are located above the horizontal break, with simple sources below it.} 
\label{tab:allsourcedata} 
\scriptsize 
\tabcolsep=0.09cm 
\begin{tabular}{c c c c c c c c c c c c c c c c} 
\hline 
\hline 
\multicolumn{1}{c}{(1)} & \multicolumn{1}{c}{(2)} & \multicolumn{1}{c}{(3)} & \multicolumn{1}{c}{(4)} & \multicolumn{1}{c}{(5)} & \multicolumn{1}{c}{(6)} & \multicolumn{1}{c}{(7)} & \multicolumn{1}{c}{(8)} & \multicolumn{1}{c}{(9)} & \multicolumn{1}{c}{(10)} & \multicolumn{1}{c}{(11)} & \multicolumn{1}{c}{(12)} & \multicolumn{1}{c}{(13)} & \multicolumn{1}{c}{(14)} & \multicolumn{1}{c}{(15)} 
\\ 
\multicolumn{1}{c}{Name} & \multicolumn{1}{c}{RA (J2000)} & \multicolumn{1}{c}{Dec (J2000)} & \multicolumn{1}{c}{PC sep} & \multicolumn{1}{c}{PC ang} & \multicolumn{1}{c}{rez} & \multicolumn{1}{c}{$\lambda_0$ (3 s.f.)} & \multicolumn{1}{c}{$I_{\lambda_0}$} & \multicolumn{1}{c}{$\Delta I_{\lambda_0}$} & \multicolumn{1}{c}{$\alpha_{\lambda_0}$} & \multicolumn{1}{c}{$\Delta\alpha_{\lambda_0}$} & \multicolumn{1}{c}{Comp} & \multicolumn{1}{c}{Cut GES} & \multicolumn{1}{c}{Weight} & \multicolumn{1}{c}{$\sigma_\phi$} 
\\ 
\multicolumn{1}{c}{} & \multicolumn{1}{c}{[H:M:S]} & \multicolumn{1}{c}{[D:M:S]} & \multicolumn{1}{c}{[Deg]} & \multicolumn{1}{c}{[Deg]} & \multicolumn{1}{c}{} & \multicolumn{1}{c}{[m]} & \multicolumn{1}{c}{[Jy]} & \multicolumn{1}{c}{[Jy]} & \multicolumn{1}{c}{} & \multicolumn{1}{c}{} & \multicolumn{1}{c}{} & \multicolumn{1}{c}{} & \multicolumn{1}{c}{} & \multicolumn{1}{c}{[rad m$^{-2}$]} 
\\ 
\hline 
    032006-362044 & 03:20:06.15 & -36:20:44.17 & 0.116 & 66.7 & u & 0.192 & 0.0184 & 0.0003 & -1.1 & 0.1 & c & $10\sigma$ & n & 188.13 \\ 
    032228-384841 & 03:22:28.01 & -38:48:41.21 & 0.170 & -173.8 & u & 0.189 & 0.899 & 0.002 & -0.41 & 0.01 & c & $10\sigma$ & n & 33.47 \\ 
    033019-365308 & 03:30:19.57 & -36:53:08.15 & 0.134 & -46.0 & r & 0.187 & 0.271 & 0.002 & -0.61 & 0.04 & c & $10\sigma$ & n & 18.45 \\ 
    033123-361041 & 03:31:23.49 & -36:10:41.76 & 0.057 & -132.7 & u & 0.190 & 0.1047 & 0.0004 & -0.73 & 0.03 & c & $6\sigma$ & n & 29.94 \\ 
    033147-332912 & 03:31:47.72 & -33:29:12.42 & 0.175 & -39.2 & u & 0.191 & 0.4491 & 0.0007 & -0.69 & 0.01 & c & $10\sigma$ & n & 69.25 \\ 
    033242-363645 & 03:32:42.20 & -36:36:45.94 & 0.133 & -51.1 & r & 0.183 & 0.0515 & 0.0008 & -0.8 & 0.1 & c & $10\sigma$ & c & 17.35 \\ 
    033329-384204 & 03:33:29.22 & -38:42:04.31 & 0.135 & -111.2 & r & 0.183 & 0.111 & 0.001 & -1.32 & 0.06 & c & $6\sigma$ & c & 22.37 \\ 
    033653-361606 & 03:36:53.98 & -36:16:06.81 & 0.152 & -25.1 & u & 0.186 & 0.5899 & 0.0008 & 0.26 & 0.01 & c & $8\sigma$ & n & 39.36 \\ 
    033725-375958 & 03:37:25.64 & -37:59:58.48 & 0.084 & 6.1 & r & 0.183 & 0.0678 & 0.0006 & -1.17 & 0.06 & c & $6\sigma$ & c & 29.19 \\ 
    033726-380229 & 03:37:26.72 & -38:02:29.68 & 0.044 & 16.6 & u & 0.182 & 0.0261 & 0.0006 & -1.6 & 0.2 & c & $6\sigma$ & c & 21.91 \\ 
    033754-351735 & 03:37:54.40 & -35:17:35.51 & 0.161 & 91.7 & r & 0.180 & 0.0025 & 0.0004 & -0.0 & 1.0 & c & $10\sigma$ & n & 121.59 \\ 
    033828-352659 & 03:38:28.98 & -35:26:59.41 & 0.160 & 44.1 & r & 0.192 & 0.285 & 0.004 & -1.62 & 0.09 & c & $6\sigma$ & n & 60.78 \\ 
    033829-352818 & 03:38:29.52 & -35:28:18.43 & 0.146 & 50.6 & r & 0.196 & 0.078 & 0.002 & -2.4 & 0.2 & c & $10\sigma$ & n & 70.51 \\ 
    033843-352335 & 03:38:43.34 & -35:23:35.92 & 0.112 & 177.7 & r & 0.186 & 0.545 & 0.002 & -0.78 & 0.03 & c & $10\sigma$ & n & 51.40 \\ 
    033848-352215 & 03:38:48.30 & -35:22:15.93 & 0.092 & 166.6 & r & 0.186 & 1.091 & 0.004 & -0.76 & 0.03 & c & $10\sigma$ & n & 40.56 \\ 
    034133-362252 & 03:41:33.82 & -36:22:52.01 & 0.094 & -89.3 & r & 0.183 & 0.609 & 0.001 & -0.8 & 0.02 & c & $10\sigma$ & c & 24.95 \\ 
    034202-361520 & 03:42:02.88 & -36:15:20.04 & 0.127 & 1.4 & u & 0.182 & 0.005 & 0.0004 & -3.5 & 1.1 & c & $8\sigma$ & c & 28.25 \\ 
    034205-370322 & 03:42:05.40 & -37:03:22.00 & 0.115 & -177.0 & u & 0.186 & 1.82 & 0.002 & -0.62 & 0.01 & c & $10\sigma$ & n & 21.41 \\ 
    034437-382640 & 03:44:37.63 & -38:26:40.85 & 0.198 & 34.8 & u & 0.182 & 0.333 & 0.001 & -0.58 & 0.03 & c & $6\sigma$ & c & 30.81 \\ 
    \tableline 
    031533-375052 & 03:15:33.31 & -37:50:52.65 & 0.100 & 148.7 & u & 0.189 & 0.0249 & 0.0003 & -1.0 & 0.1 & s & - & - & 0.05 \\ 
    031537-375056 & 03:15:37.39 & -37:50:56.39 & 0.108 & 142.9 & u & 0.188 & 0.02 & 0.0003 & -0.7 & 0.1 & s & - & - & 0.05 \\ 
    031538-342345 & 03:15:38.53 & -34:23:45.08 & 0.057 & -75.3 & u & 0.189 & 0.1209 & 0.0003 & -0.58 & 0.02 & s & - & - & 0.05 \\ 
    031551-364449 & 03:15:51.29 & -36:44:49.80 & 0.123 & 145.8 & u & 0.187 & 0.403 & 0.0006 & -0.7 & 0.01 & s & - & - & 0.05 \\ 
    031611-353540 & 03:16:11.63 & -35:35:40.83 & 0.118 & 124.4 & u & 0.190 & 0.0078 & 0.0002 & -0.7 & 0.2 & s & - & - & 0.00 \\ 
    031616-381438 & 03:16:16.65 & -38:14:38.45 & 0.140 & -50.4 & u & 0.188 & 0.0148 & 0.0003 & -0.4 & 0.2 & s & - & - & 0.05 \\ 
    031636-350112 & 03:16:36.43 & -35:01:12.34 & 0.165 & -104.0 & u & 0.188 & 0.0191 & 0.0003 & -0.2 & 0.1 & s & - & - & 0.00 \\ 
    031651-370212 & 03:16:51.64 & -37:02:12.84 & 0.160 & 131.8 & u & 0.190 & 0.0072 & 0.0002 & -0.9 & 0.3 & s & - & - & 0.00 \\ 
    031653-382609 & 03:16:53.59 & -38:26:09.01 & 0.104 & 172.7 & r & 0.191 & 0.0772 & 0.0009 & -1.01 & 0.07 & s & - & - & 0.02 \\ 
    031654-382434 & 03:16:54.99 & -38:24:34.06 & 0.078 & 166.9 & r & 0.191 & 0.0761 & 0.0008 & -1.0 & 0.07 & s & - & - & 0.00 \\ 
    031656-375122 & 03:16:56.74 & -37:51:22.80 & 0.082 & 177.3 & u & 0.187 & 0.1059 & 0.0003 & -0.47 & 0.02 & s & - & - & 0.00 \\ 
    031705-353441 & 03:17:05.08 & -35:34:41.31 & 0.059 & -131.7 & r & 0.190 & 0.0271 & 0.0003 & -0.47 & 0.07 & s & - & - & 0.05 \\ 
    031745-384734 & 03:17:45.23 & -38:47:34.83 & 0.177 & 170.4 & r & 0.192 & 0.032 & 0.001 & -1.1 & 0.2 & s & - & - & 0.00 \\ 
    031747-344234 & 03:17:47.85 & -34:42:34.35 & 0.086 & -92.5 & u & 0.195 & 0.008 & 0.0002 & -1.2 & 0.3 & s & - & - & 0.00 \\ 
    031756-375619 & 03:17:56.32 & -37:56:19.08 & 0.129 & 21.6 & u & 0.190 & 0.0522 & 0.0003 & -0.81 & 0.05 & s & - & - & 0.03 \\ 
    031817-382921 & 03:18:17.38 & -38:29:21.71 & 0.150 & -166.2 & u & 0.189 & 0.0947 & 0.0005 & -0.53 & 0.04 & s & - & - & 0.00 \\ 
    031826-360959 & 03:18:26.00 & -36:09:59.86 & 0.096 & -127.8 & u & 0.188 & 0.0385 & 0.0002 & -0.64 & 0.05 & s & - & - & 0.05 \\ 
    031840-352549 & 03:18:40.68 & -35:25:49.61 & 0.126 & -19.9 & r & 0.190 & 0.2011 & 0.0006 & -0.71 & 0.02 & s & - & - & 0.04 \\ 
    031850-350545 & 03:18:50.95 & -35:05:45.46 & 0.109 & -167.4 & u & 0.187 & 0.0626 & 0.0004 & -0.14 & 0.06 & s & - & - & 0.04 \\ 
    031852-380826 & 03:18:52.73 & -38:08:26.51 & 0.116 & -128.7 & u & 0.189 & 0.0842 & 0.0004 & -0.74 & 0.04 & s & - & - & 0.00 \\ 
    031858-332622 & 03:18:58.59 & -33:26:22.82 & 0.194 & 40.5 & u & 0.184 & 0.0131 & 0.0003 & 0.3 & 0.2 & s & - & - & 0.00 \\ 
    031922-340222 & 03:19:22.18 & -34:02:22.51 & 0.154 & -40.9 & u & 0.190 & 0.038 & 0.0002 & -0.65 & 0.04 & s & - & - & 0.00 \\ 
    031954-350558 & 03:19:54.86 & -35:05:58.84 & 0.163 & -128.3 & u & 0.190 & 0.0147 & 0.0003 & -0.5 & 0.2 & s & - & - & 0.00 \\ 
    032012-351244 & 03:20:12.96 & -35:12:44.05 & 0.119 & 58.8 & r & 0.188 & 0.069 & 0.0005 & -0.32 & 0.05 & s & - & - & 0.05 \\ 
    032015-351400 & 03:20:15.53 & -35:14:00.21 & 0.118 & 69.9 & r & 0.188 & 0.0685 & 0.0005 & -0.31 & 0.05 & s & - & - & 0.00 \\ 
    032039-352111 & 03:20:39.11 & -35:21:11.93 & 0.150 & -118.5 & u & 0.192 & 0.0294 & 0.0002 & -0.82 & 0.06 & s & - & - & 0.03 \\ 
    032043-343600 & 03:20:43.23 & -34:36:00.80 & 0.162 & 171.6 & u & 0.192 & 0.0596 & 0.0002 & -0.91 & 0.03 & s & - & - & 0.05 \\ 
    032044-382210 & 03:20:44.40 & -38:22:10.89 & 0.123 & 98.2 & r & 0.189 & 0.165 & 0.001 & -0.77 & 0.04 & s & - & - & 0.05 \\ 
    032046-382230 & 03:20:46.29 & -38:22:30.02 & 0.130 & 100.1 & r & 0.189 & 0.166 & 0.001 & -0.77 & 0.04 & s & - & - & 0.04 \\ 
    032046-383729 & 03:20:46.39 & -38:37:29.01 & 0.029 & -66.8 & u & 0.181 & 0.094 & 0.0004 & 0.57 & 0.04 & s & - & - & 0.05 \\ 
\end{tabular} 
\end{table*} 
\end{turnpage}    
   
\addtocounter{table}{-1}
\begin{turnpage} 
\begin{table*} 
\caption{(Table 1 cont., columns 1--15) Selected quantities calculated / determined for polarized sources in our sample.} 
\scriptsize 
\tabcolsep=0.09cm 
\begin{tabular}{c c c c c c c c c c c c c c c c} 
\hline 
\hline 
\multicolumn{1}{c}{(1)} & \multicolumn{1}{c}{(2)} & \multicolumn{1}{c}{(3)} & \multicolumn{1}{c}{(4)} & \multicolumn{1}{c}{(5)} & \multicolumn{1}{c}{(6)} & \multicolumn{1}{c}{(7)} & \multicolumn{1}{c}{(8)} & \multicolumn{1}{c}{(9)} & \multicolumn{1}{c}{(10)} & \multicolumn{1}{c}{(11)} & \multicolumn{1}{c}{(12)} & \multicolumn{1}{c}{(13)} & \multicolumn{1}{c}{(14)} & \multicolumn{1}{c}{(15)} 
\\ 
\multicolumn{1}{c}{Name} & \multicolumn{1}{c}{RA (J2000)} & \multicolumn{1}{c}{Dec (J2000)} & \multicolumn{1}{c}{PC sep} & \multicolumn{1}{c}{PC ang} & \multicolumn{1}{c}{rez} & \multicolumn{1}{c}{$\lambda_0$ (3 s.f.)} & \multicolumn{1}{c}{$I_{\lambda_0}$} & \multicolumn{1}{c}{$\Delta I_{\lambda_0}$} & \multicolumn{1}{c}{$\alpha_{\lambda_0}$} & \multicolumn{1}{c}{$\Delta\alpha_{\lambda_0}$} & \multicolumn{1}{c}{Comp} & \multicolumn{1}{c}{Cut GES} & \multicolumn{1}{c}{Weight} & \multicolumn{1}{c}{$\sigma_\phi$} 
\\ 
\multicolumn{1}{c}{} & \multicolumn{1}{c}{[H:M:S]} & \multicolumn{1}{c}{[D:M:S]} & \multicolumn{1}{c}{[Deg]} & \multicolumn{1}{c}{[Deg]} & \multicolumn{1}{c}{} & \multicolumn{1}{c}{[m]} & \multicolumn{1}{c}{[Jy]} & \multicolumn{1}{c}{[Jy]} & \multicolumn{1}{c}{} & \multicolumn{1}{c}{} & \multicolumn{1}{c}{} & \multicolumn{1}{c}{} & \multicolumn{1}{c}{} & \multicolumn{1}{c}{[rad m$^{-2}$]} 
\\ 
\hline    
    032124-341417 & 03:21:24.73 & -34:14:17.07 & 0.074 & -179.3 & r & 0.194 & 0.0385 & 0.0004 & -0.79 & 0.06 & s & - & - & 0.00 \\ 
    032125-340738 & 03:21:25.49 & -34:07:38.51 & 0.036 & 2.8 & r & 0.192 & 0.0318 & 0.0004 & -0.74 & 0.08 & s & - & - & 0.00 \\ 
    032126-341334 & 03:21:26.70 & -34:13:34.93 & 0.063 & 174.6 & r & 0.194 & 0.0375 & 0.0004 & -0.79 & 0.06 & s & - & - & 0.00 \\ 
    032132-340553 & 03:21:32.29 & -34:05:53.22 & 0.070 & 21.1 & u & 0.192 & 0.0793 & 0.0003 & -0.6 & 0.03 & s & - & - & 0.00 \\ 
    032213-345831 & 03:22:13.00 & -34:58:31.34 & 0.037 & 33.7 & u & 0.188 & 0.0558 & 0.0004 & -0.08 & 0.05 & s & - & - & 0.05 \\ 
    032315-384013 & 03:23:15.70 & -38:40:13.75 & 0.140 & 101.5 & u & 0.189 & 0.06 & 0.0009 & -0.9 & 0.1 & s & - & - & 0.04 \\ 
    032331-333314 & 03:23:31.18 & -33:33:14.54 & 0.118 & 61.3 & u & 0.190 & 0.0944 & 0.0005 & -0.69 & 0.04 & s & - & - & 0.05 \\ 
    032334-345532 & 03:23:34.73 & -34:55:32.47 & 0.089 & -15.4 & u & 0.191 & 0.0284 & 0.0004 & -0.7 & 0.1 & s & - & - & 0.02 \\ 
    032349-352401 & 03:23:49.08 & -35:24:01.83 & 0.169 & -129.1 & u & 0.183 & 0.0094 & 0.0002 & 0.7 & 0.2 & s & - & - & 0.00 \\ 
    032349-351059 & 03:23:49.61 & -35:10:59.49 & 0.171 & -49.7 & u & 0.197 & 0.0062 & 0.0004 & -1.8 & 0.5 & s & - & - & 0.00 \\ 
    032358-344500 & 03:23:58.19 & -34:45:00.25 & 0.110 & -98.1 & r & 0.190 & 0.014 & 0.0004 & -0.8 & 0.2 & s & - & - & 0.00 \\ 
    032358-340958 & 03:23:58.92 & -34:09:58.31 & 0.115 & -85.5 & r & 0.193 & 0.012 & 0.0003 & -1.0 & 0.2 & s & - & - & 0.04 \\ 
    032410-343927 & 03:24:10.67 & -34:39:27.08 & 0.102 & -40.8 & u & 0.192 & 0.0177 & 0.0003 & -1.0 & 0.1 & s & - & - & 0.00 \\ 
    032431-341910 & 03:24:31.45 & -34:19:10.65 & 0.144 & -178.9 & u & 0.191 & 0.0997 & 0.0003 & -0.73 & 0.03 & s & - & - & 0.00 \\ 
    032442-333555 & 03:24:42.59 & -33:35:55.04 & 0.033 & 58.6 & u & 0.191 & 0.0735 & 0.0003 & -0.52 & 0.04 & s & - & - & 0.00 \\ 
    032459-343906 & 03:24:59.87 & -34:39:06.24 & 0.131 & 51.0 & u & 0.186 & 0.0064 & 0.0002 & 0.0 & 0.3 & s & - & - & 0.04 \\ 
    032501-352200 & 03:25:01.46 & -35:22:00.54 & 0.136 & 122.6 & r & 0.192 & 0.0525 & 0.0004 & -0.7 & 0.05 & s & - & - & 0.00 \\ 
    032504-335859 & 03:25:04.36 & -33:58:59.61 & 0.101 & -147.7 & u & 0.190 & 0.0246 & 0.0003 & -0.4 & 0.1 & s & - & - & 0.00 \\ 
    032520-360357 & 03:25:20.91 & -36:03:57.75 & 0.074 & 23.1 & u & 0.191 & 0.0289 & 0.0003 & -0.84 & 0.08 & s & - & - & 0.00 \\ 
    032522-344644 & 03:25:22.30 & -34:46:44.89 & 0.150 & -105.8 & u & 0.191 & 0.025 & 0.0002 & -0.74 & 0.08 & s & - & - & 0.03 \\ 
    032532-353343 & 03:25:32.38 & -35:33:43.23 & 0.063 & 77.8 & u & 0.190 & 0.0572 & 0.0003 & -0.52 & 0.05 & s & - & - & 0.04 \\ 
    032546-333739 & 03:25:46.84 & -33:37:39.30 & 0.072 & -96.2 & u & 0.193 & 0.028 & 0.0003 & -1.0 & 0.1 & s & - & - & 0.00 \\ 
    032554-355641 & 03:25:54.57 & -35:56:41.54 & 0.091 & -166.1 & u & 0.192 & 0.0103 & 0.0003 & -0.9 & 0.2 & s & - & - & 0.05 \\ 
    032556-362527 & 03:25:56.85 & -36:25:27.22 & 0.012 & -135.7 & u & 0.199 & 0.0083 & 0.0007 & -2.4 & 0.7 & s & - & - & 0.00 \\ 
    032613-362142 & 03:26:13.00 & -36:21:42.87 & 0.071 & 40.4 & u & 0.186 & 0.0157 & 0.0002 & -0.0 & 0.1 & s & - & - & 0.00 \\ 
    032615-381131 & 03:26:15.06 & -38:11:31.73 & 0.121 & 145.1 & u & 0.188 & 0.0341 & 0.0004 & -0.7 & 0.1 & s & - & - & 0.00 \\ 
    032630-362643 & 03:26:30.02 & -36:26:43.55 & 0.107 & 106.2 & u & 0.186 & 0.0189 & 0.0003 & -0.0 & 0.1 & s & - & - & 0.02 \\ 
    032659-374856 & 03:26:59.56 & -37:48:56.90 & 0.052 & 91.1 & u & 0.187 & 0.0664 & 0.0004 & -0.22 & 0.05 & s & - & - & 0.04 \\ 
    032720-362938 & 03:27:20.95 & -36:29:38.08 & 0.090 & -147.0 & u & 0.188 & 0.0244 & 0.0002 & -0.24 & 0.07 & s & - & - & 0.00 \\ 
    032732-351518 & 03:27:32.23 & -35:15:18.22 & 0.049 & -22.1 & u & 0.191 & 0.0989 & 0.0005 & -1.07 & 0.04 & s & - & - & 0.05 \\ 
    032736-343304 & 03:27:36.59 & -34:33:04.57 & 0.178 & 120.8 & u & 0.190 & 0.0215 & 0.0002 & -0.62 & 0.09 & s & - & - & 0.05 \\ 
    032749-333337 & 03:27:49.49 & -33:33:37.91 & 0.069 & 26.9 & r & 0.191 & 0.1939 & 0.0008 & -0.97 & 0.03 & s & - & - & 0.05 \\ 
    032810-375009 & 03:28:10.66 & -37:50:09.75 & 0.042 & -117.2 & r & 0.189 & 0.0371 & 0.0006 & -0.6 & 0.1 & s & - & - & 0.02 \\ 
    032831-333858 & 03:28:31.63 & -33:38:58.13 & 0.147 & -100.1 & u & 0.193 & 0.06 & 0.0006 & -0.97 & 0.08 & s & - & - & 0.03 \\ 
    032834-351326 & 03:28:34.08 & -35:13:26.43 & 0.152 & -59.4 & r & 0.192 & 0.0127 & 0.0003 & -0.6 & 0.2 & s & - & - & 0.00 \\ 
    032849-334751 & 03:28:49.75 & -33:47:51.25 & 0.132 & 37.2 & u & 0.191 & 0.1508 & 0.0004 & -0.82 & 0.02 & s & - & - & 0.05 \\ 
    032908-373330 & 03:29:08.57 & -37:33:30.80 & 0.022 & -157.6 & u & 0.194 & 0.0099 & 0.0004 & -1.5 & 0.3 & s & - & - & 0.04 \\ 
    032918-351153 & 03:29:18.06 & -35:11:53.92 & 0.105 & 10.3 & u & 0.190 & 0.0367 & 0.0003 & -0.56 & 0.06 & s & - & - & 0.00 \\ 
    032918-360155 & 03:29:18.25 & -36:01:55.62 & 0.173 & 173.2 & r & 0.190 & 0.0781 & 0.0005 & -0.35 & 0.05 & s & - & - & 0.04 \\ 
    033002-360832 & 03:30:02.53 & -36:08:32.86 & 0.009 & 104.6 & r & 0.190 & 0.0687 & 0.0007 & -0.47 & 0.07 & s & - & - & 0.05 \\ 
    033005-354623 & 03:30:05.89 & -35:46:23.81 & 0.166 & -58.4 & u & 0.188 & 0.0875 & 0.0005 & -0.37 & 0.04 & s & - & - & 0.05 \\ 
    033008-365315 & 03:30:08.71 & -36:53:15.38 & 0.161 & -55.5 & r & 0.187 & 0.307 & 0.002 & -0.6 & 0.04 & s & - & - & 0.03 \\ 
    033020-355449 & 03:30:20.11 & -35:54:49.25 & 0.107 & -119.7 & r & 0.191 & 0.0268 & 0.0007 & -0.6 & 0.2 & s & - & - & 0.00 \\ 
    033057-341128 & 03:30:57.02 & -34:11:28.43 & 0.036 & 103.0 & r & 0.193 & 0.0307 & 0.0004 & -0.62 & 0.08 & s & - & - & 0.00 \\ 
    033109-380430 & 03:31:09.13 & -38:04:30.77 & 0.069 & 71.5 & u & 0.191 & 0.0115 & 0.0002 & -0.7 & 0.2 & s & - & - & 0.04 \\ 
    033123-372745 & 03:31:23.35 & -37:27:45.14 & 0.137 & 56.6 & u & 0.186 & 0.0121 & 0.0003 & 0.2 & 0.2 & s & - & - & 0.00 \\ 
    033138-342852 & 03:31:38.09 & -34:28:52.32 & 0.024 & 143.0 & u & 0.194 & 0.0213 & 0.0003 & -1.5 & 0.1 & s & - & - & 0.04 \\ 
    033201-342737 & 03:32:01.32 & -34:27:37.59 & 0.094 & 88.9 & u & 0.190 & 0.008 & 0.0003 & -0.6 & 0.3 & s & - & - & 0.04 \\ 
    033212-364144 & 03:32:12.49 & -36:41:44.63 & 0.120 & 88.6 & u & 0.197 & 0.0108 & 0.0004 & -1.9 & 0.3 & s & - & - & 0.00 \\ 
\end{tabular} 
\end{table*} 
\end{turnpage}    
   
\addtocounter{table}{-1}
\begin{turnpage} 
\begin{table*} 
\caption{(Table 1 cont., columns 1--15) Selected quantities calculated / determined for polarized sources in our sample.} 
\scriptsize 
\tabcolsep=0.09cm 
\begin{tabular}{c c c c c c c c c c c c c c c c} 
\hline 
\hline 
\multicolumn{1}{c}{(1)} & \multicolumn{1}{c}{(2)} & \multicolumn{1}{c}{(3)} & \multicolumn{1}{c}{(4)} & \multicolumn{1}{c}{(5)} & \multicolumn{1}{c}{(6)} & \multicolumn{1}{c}{(7)} & \multicolumn{1}{c}{(8)} & \multicolumn{1}{c}{(9)} & \multicolumn{1}{c}{(10)} & \multicolumn{1}{c}{(11)} & \multicolumn{1}{c}{(12)} & \multicolumn{1}{c}{(13)} & \multicolumn{1}{c}{(14)} & \multicolumn{1}{c}{(15)} 
\\ 
\multicolumn{1}{c}{Name} & \multicolumn{1}{c}{RA (J2000)} & \multicolumn{1}{c}{Dec (J2000)} & \multicolumn{1}{c}{PC sep} & \multicolumn{1}{c}{PC ang} & \multicolumn{1}{c}{rez} & \multicolumn{1}{c}{$\lambda_0$ (3 s.f.)} & \multicolumn{1}{c}{$I_{\lambda_0}$} & \multicolumn{1}{c}{$\Delta I_{\lambda_0}$} & \multicolumn{1}{c}{$\alpha_{\lambda_0}$} & \multicolumn{1}{c}{$\Delta\alpha_{\lambda_0}$} & \multicolumn{1}{c}{Comp} & \multicolumn{1}{c}{Cut GES} & \multicolumn{1}{c}{Weight} & \multicolumn{1}{c}{$\sigma_\phi$} 
\\ 
\multicolumn{1}{c}{} & \multicolumn{1}{c}{[H:M:S]} & \multicolumn{1}{c}{[D:M:S]} & \multicolumn{1}{c}{[Deg]} & \multicolumn{1}{c}{[Deg]} & \multicolumn{1}{c}{} & \multicolumn{1}{c}{[m]} & \multicolumn{1}{c}{[Jy]} & \multicolumn{1}{c}{[Jy]} & \multicolumn{1}{c}{} & \multicolumn{1}{c}{} & \multicolumn{1}{c}{} & \multicolumn{1}{c}{} & \multicolumn{1}{c}{} & \multicolumn{1}{c}{[rad m$^{-2}$]} 
\\ 
\hline    
    033214-382447 & 03:32:14.22 & -38:24:47.93 & 0.121 & 107.8 & u & 0.188 & 0.0574 & 0.0006 & -0.48 & 0.09 & s & - & - & 0.02 \\ 
    033218-363949 & 03:32:18.57 & -36:39:49.66 & 0.145 & 76.1 & u & 0.190 & 0.005 & 0.0004 & -0.6 & 0.7 & s & - & - & 0.03 \\ 
    033218-353815 & 03:32:18.70 & -35:38:15.44 & 0.158 & 111.3 & u & 0.190 & 0.0301 & 0.0003 & -0.54 & 0.07 & s & - & - & 0.00 \\ 
    033224-334116 & 03:32:24.04 & -33:41:16.34 & 0.067 & 166.6 & u & 0.195 & 0.0313 & 0.0003 & -1.29 & 0.09 & s & - & - & 0.05 \\ 
    033228-365620 & 03:32:28.37 & -36:56:20.00 & 0.040 & 14.2 & u & 0.189 & 0.0715 & 0.0006 & -0.63 & 0.07 & s & - & - & 0.04 \\ 
    033250-360829 & 03:32:50.81 & -36:08:29.13 & 0.071 & -93.1 & u & 0.189 & 0.0408 & 0.0008 & -0.4 & 0.2 & s & - & - & 0.00 \\ 
    033300-365535 & 03:33:00.44 & -36:55:35.82 & 0.127 & 66.5 & u & 0.188 & 0.0437 & 0.0004 & -0.29 & 0.07 & s & - & - & 0.00 \\ 
    033302-341137 & 03:33:02.67 & -34:11:37.75 & 0.146 & 94.8 & u & 0.185 & 0.0363 & 0.0005 & 0.3 & 0.1 & s & - & - & 0.00 \\ 
    033319-344546 & 03:33:19.81 & -34:45:46.40 & 0.125 & -101.3 & r & 0.189 & 0.0115 & 0.0003 & -0.6 & 0.2 & s & - & - & 0.00 \\ 
    033323-375247 & 03:33:23.22 & -37:52:47.92 & 0.069 & 160.2 & u & 0.192 & 0.1135 & 0.0004 & -1.05 & 0.03 & s & - & - & 0.03 \\ 
    033331-373806 & 03:33:31.53 & -37:38:06.62 & 0.148 & -133.3 & u & 0.193 & 0.0064 & 0.0003 & -1.1 & 0.3 & s & - & - & 0.00 \\ 
    033332-344206 & 03:33:32.15 & -34:42:06.41 & 0.088 & -65.5 & u & 0.192 & 0.0459 & 0.0003 & -0.99 & 0.06 & s & - & - & 0.05 \\ 
    033336-335206 & 03:33:36.47 & -33:52:06.32 & 0.108 & 72.5 & u & 0.188 & 0.0666 & 0.0004 & -0.42 & 0.04 & s & - & - & 0.05 \\ 
    033342-371804 & 03:33:42.18 & -37:18:04.66 & 0.102 & 116.5 & u & 0.196 & 0.0133 & 0.0004 & -1.7 & 0.2 & s & - & - & 0.03 \\ 
    033347-342225 & 03:33:47.11 & -34:22:25.22 & 0.160 & 57.3 & u & 0.189 & 0.0242 & 0.0003 & -0.6 & 0.1 & s & - & - & 0.04 \\ 
    033400-364316 & 03:34:00.37 & -36:43:16.75 & 0.159 & 99.0 & u & 0.188 & 0.0148 & 0.0003 & -0.3 & 0.2 & s & - & - & 0.00 \\ 
    033405-344329 & 03:34:05.36 & -34:43:29.22 & 0.036 & 67.6 & r & 0.188 & 0.0503 & 0.0006 & -0.26 & 0.08 & s & - & - & 0.04 \\ 
    033415-372542 & 03:34:15.40 & -37:25:42.32 & 0.112 & 19.4 & u & 0.184 & 0.285 & 0.0006 & 0.28 & 0.02 & s & - & - & 0.03 \\ 
    033417-341230 & 03:34:17.72 & -34:12:30.66 & 0.087 & 109.8 & u & 0.191 & 0.0398 & 0.0004 & -0.86 & 0.08 & s & - & - & 0.00 \\ 
    033423-363628 & 03:34:23.21 & -36:36:28.07 & 0.123 & -46.2 & u & 0.190 & 0.0225 & 0.0004 & -0.7 & 0.1 & s & - & - & 0.05 \\ 
    033423-341136 & 03:34:23.58 & -34:11:36.56 & 0.103 & 98.0 & u & 0.183 & 0.007 & 0.0002 & 0.5 & 0.2 & s & - & - & 0.00 \\ 
    033424-344435 & 03:34:24.77 & -34:44:35.78 & 0.100 & 92.8 & u & 0.188 & 0.0088 & 0.0003 & -0.5 & 0.3 & s & - & - & 0.00 \\ 
    033432-335605 & 03:34:32.68 & -33:56:05.25 & 0.045 & -145.7 & u & 0.187 & 0.0143 & 0.0003 & -0.2 & 0.2 & s & - & - & 0.04 \\ 
    033450-364735 & 03:34:50.13 & -36:47:35.91 & 0.100 & 179.4 & r & 0.191 & 0.1155 & 0.0006 & -0.72 & 0.03 & s & - & - & 0.04 \\ 
    033451-361019 & 03:34:51.13 & -36:10:19.33 & 0.039 & 163.3 & u & 0.193 & 0.0458 & 0.0005 & -1.18 & 0.09 & s & - & - & 0.03 \\ 
    033511-372737 & 03:35:11.15 & -37:27:37.17 & 0.123 & -55.7 & r & 0.192 & 0.0521 & 0.0007 & -1.03 & 0.09 & s & - & - & 0.05 \\ 
    033520-342803 & 03:35:20.33 & -34:28:03.30 & 0.133 & 94.6 & u & 0.188 & 0.0732 & 0.0005 & -0.58 & 0.06 & s & - & - & 0.00 \\ 
    033523-335523 & 03:35:23.57 & -33:55:23.85 & 0.153 & 99.7 & u & 0.189 & 0.029 & 0.0004 & -0.7 & 0.1 & s & - & - & 0.00 \\ 
    033529-362939 & 03:35:29.65 & -36:29:39.23 & 0.086 & -163.6 & u & 0.191 & 0.0295 & 0.0004 & -0.6 & 0.1 & s & - & - & 0.00 \\ 
    033555-350135 & 03:35:55.74 & -35:01:35.01 & 0.079 & -100.8 & u & 0.187 & 0.123 & 0.002 & -0.6 & 0.2 & s & - & - & 0.05 \\ 
    033558-332708 & 03:35:58.15 & -33:27:08.78 & 0.199 & 34.7 & u & 0.188 & 0.14 & 0.002 & -0.48 & 0.09 & s & - & - & 0.00 \\ 
    033559-340352 & 03:35:59.78 & -34:03:52.17 & 0.157 & 45.0 & u & 0.190 & 0.0326 & 0.0004 & -0.8 & 0.1 & s & - & - & 0.00 \\ 
    033601-342136 & 03:36:01.22 & -34:21:36.24 & 0.105 & -28.5 & u & 0.187 & 0.022 & 0.0004 & -0.4 & 0.1 & s & - & - & 0.00 \\ 
    033604-341342 & 03:36:04.38 & -34:13:42.39 & 0.137 & 112.8 & u & 0.189 & 0.0572 & 0.0005 & -0.61 & 0.07 & s & - & - & 0.00 \\ 
    033613-334714 & 03:36:13.85 & -33:47:14.38 & 0.106 & 1.0 & u & 0.189 & 0.0246 & 0.0003 & -0.7 & 0.1 & s & - & - & 0.05 \\ 
    033639-381057 & 03:36:39.26 & -38:10:57.14 & 0.174 & -124.8 & u & 0.192 & 0.0201 & 0.0006 & -1.1 & 0.2 & s & - & - & 0.00 \\ 
    033641-335828 & 03:36:41.75 & -33:58:28.66 & 0.128 & 129.7 & u & 0.190 & 0.0388 & 0.0003 & -0.83 & 0.07 & s & - & - & 0.04 \\ 
    033645-354229 & 03:36:45.86 & -35:42:29.67 & 0.161 & 148.6 & u & 0.184 & 0.0587 & 0.0004 & 0.02 & 0.06 & s & - & - & 0.04 \\ 
    033752-342219 & 03:37:52.76 & -34:22:19.77 & 0.075 & 7.9 & u & 0.190 & 0.1768 & 0.0004 & -0.82 & 0.02 & s & - & - & 0.00 \\ 
    033804-371130 & 03:38:04.08 & -37:11:30.90 & 0.050 & -9.6 & u & 0.190 & 0.0716 & 0.0005 & -0.52 & 0.06 & s & - & - & 0.04 \\ 
    033826-355129 & 03:38:26.48 & -35:51:29.71 & 0.067 & -105.1 & u & 0.191 & 0.0197 & 0.0003 & -0.9 & 0.1 & s & - & - & 0.00 \\ 
    033827-352540 & 03:38:27.79 & -35:25:40.52 & 0.154 & -161.7 & r & 0.192 & 0.081 & 0.002 & -1.7 & 0.1 & s & - & - & 0.04 \\ 
    033832-355108 & 03:38:32.82 & -35:51:08.72 & 0.045 & -105.1 & u & 0.188 & 0.0187 & 0.0005 & -0.4 & 0.2 & s & - & - & 0.03 \\ 
    033849-375103 & 03:38:49.28 & -37:51:03.72 & 0.138 & 111.5 & u & 0.193 & 0.2375 & 0.0006 & -1.01 & 0.02 & s & - & - & 0.00 \\ 
    033903-375437 & 03:39:03.71 & -37:54:37.29 & 0.166 & 2.7 & u & 0.190 & 0.0415 & 0.0005 & -0.78 & 0.09 & s & - & - & 0.05 \\ 
    033913-345302 & 03:39:13.98 & -34:53:02.59 & 0.123 & -22.0 & u & 0.194 & 0.0066 & 0.0002 & -1.4 & 0.3 & s & - & - & 0.00 \\ 
    033950-370006 & 03:39:50.99 & -37:00:06.30 & 0.140 & -111.5 & u & 0.187 & 0.0741 & 0.0006 & -0.16 & 0.06 & s & - & - & 0.05 \\ 
    034006-363545 & 03:40:06.19 & -36:35:45.48 & 0.119 & 48.3 & u & 0.191 & 0.2355 & 0.0007 & -1.17 & 0.02 & s & - & - & 0.04 \\ 
\end{tabular} 
\end{table*} 
\end{turnpage}    
   
\addtocounter{table}{-1}
\begin{turnpage} 
\begin{table*} 
\caption{(Table 1 cont., columns 1--15) Selected quantities calculated / determined for polarized sources in our sample.} 
\scriptsize 
\tabcolsep=0.09cm 
\begin{tabular}{c c c c c c c c c c c c c c c c} 
\hline 
\hline 
\multicolumn{1}{c}{(1)} & \multicolumn{1}{c}{(2)} & \multicolumn{1}{c}{(3)} & \multicolumn{1}{c}{(4)} & \multicolumn{1}{c}{(5)} & \multicolumn{1}{c}{(6)} & \multicolumn{1}{c}{(7)} & \multicolumn{1}{c}{(8)} & \multicolumn{1}{c}{(9)} & \multicolumn{1}{c}{(10)} & \multicolumn{1}{c}{(11)} & \multicolumn{1}{c}{(12)} & \multicolumn{1}{c}{(13)} & \multicolumn{1}{c}{(14)} & \multicolumn{1}{c}{(15)} 
\\ 
\multicolumn{1}{c}{Name} & \multicolumn{1}{c}{RA (J2000)} & \multicolumn{1}{c}{Dec (J2000)} & \multicolumn{1}{c}{PC sep} & \multicolumn{1}{c}{PC ang} & \multicolumn{1}{c}{rez} & \multicolumn{1}{c}{$\lambda_0$ (3 s.f.)} & \multicolumn{1}{c}{$I_{\lambda_0}$} & \multicolumn{1}{c}{$\Delta I_{\lambda_0}$} & \multicolumn{1}{c}{$\alpha_{\lambda_0}$} & \multicolumn{1}{c}{$\Delta\alpha_{\lambda_0}$} & \multicolumn{1}{c}{Comp} & \multicolumn{1}{c}{Cut GES} & \multicolumn{1}{c}{Weight} & \multicolumn{1}{c}{$\sigma_\phi$} 
\\ 
\multicolumn{1}{c}{} & \multicolumn{1}{c}{[H:M:S]} & \multicolumn{1}{c}{[D:M:S]} & \multicolumn{1}{c}{[Deg]} & \multicolumn{1}{c}{[Deg]} & \multicolumn{1}{c}{} & \multicolumn{1}{c}{[m]} & \multicolumn{1}{c}{[Jy]} & \multicolumn{1}{c}{[Jy]} & \multicolumn{1}{c}{} & \multicolumn{1}{c}{} & \multicolumn{1}{c}{} & \multicolumn{1}{c}{} & \multicolumn{1}{c}{} & \multicolumn{1}{c}{[rad m$^{-2}$]} 
\\ 
\hline    
    034008-372711 & 03:40:08.85 & -37:27:11.35 & 0.103 & -56.9 & u & 0.190 & 0.0595 & 0.0003 & -0.59 & 0.04 & s & - & - & 0.03 \\ 
    034009-332644 & 03:40:09.98 & -33:26:44.17 & 0.152 & 7.0 & u & 0.189 & 0.145 & 0.001 & -0.32 & 0.06 & s & - & - & 0.00 \\ 
    034017-375446 & 03:40:17.15 & -37:54:46.50 & 0.153 & 141.6 & u & 0.192 & 0.0224 & 0.0006 & -0.9 & 0.2 & s & - & - & 0.00 \\ 
    034033-374943 & 03:40:33.74 & -37:49:43.83 & 0.154 & 103.5 & u & 0.193 & 0.0283 & 0.0005 & -1.1 & 0.2 & s & - & - & 0.00 \\ 
    034042-340818 & 03:40:42.78 & -34:08:18.93 & 0.119 & 81.7 & u & 0.192 & 0.053 & 0.0006 & -1.08 & 0.09 & s & - & - & 0.00 \\ 
    034049-340903 & 03:40:49.58 & -34:09:03.48 & 0.141 & 88.1 & u & 0.191 & 0.4127 & 0.0005 & -0.89 & 0.01 & s & - & - & 0.04 \\ 
    034054-342252 & 03:40:54.43 & -34:22:52.23 & 0.051 & -12.5 & u & 0.187 & 0.0153 & 0.0003 & -0.2 & 0.2 & s & - & - & 0.00 \\ 
    034146-344840 & 03:41:46.87 & -34:48:40.95 & 0.106 & -179.8 & u & 0.192 & 0.0171 & 0.0003 & -1.0 & 0.2 & s & - & - & 0.00 \\ 
    034155-363653 & 03:41:55.21 & -36:36:53.45 & 0.140 & 68.4 & u & 0.187 & 0.0858 & 0.0007 & -0.28 & 0.07 & s & - & - & 0.00 \\ 
    034219-380805 & 03:42:19.09 & -38:08:05.60 & 0.076 & 177.8 & r & 0.182 & 0.0035 & 0.0005 & 0.3 & 0.9 & s & - & - & 0.00 \\ 
    034254-340503 & 03:42:54.97 & -34:05:03.01 & 0.089 & -54.1 & u & 0.177 & 0.0062 & 0.0003 & 1.5 & 0.4 & s & - & - & 0.00 \\ 
    034305-374023 & 03:43:05.08 & -37:40:23.47 & 0.101 & 1.2 & u & 0.192 & 0.0969 & 0.0004 & -0.87 & 0.04 & s & - & - & 0.04 \\ 
    034323-335144 & 03:43:23.71 & -33:51:44.37 & 0.197 & 90.1 & u & 0.190 & 0.3383 & 0.0006 & -0.74 & 0.01 & s & - & - & 0.04 \\ 
    034359-381408 & 03:43:59.30 & -38:14:08.75 & 0.187 & 58.7 & u & 0.187 & 0.0584 & 0.0007 & -0.6 & 0.1 & s & - & - & 0.05 \\ 
\end{tabular} 
\end{table*} 
\end{turnpage}

\addtocounter{table}{-1}
\begin{turnpage} 
\begin{table*} 
\caption{(Table 1 cont., columns 1, 16--27) Further quantities calculated / determined for polarized sources in our sample. Faraday complex sources are located above the horizontal break, with simple sources below it.} 
\scriptsize 
\tabcolsep=0.09cm 
\begin{tabular}{c c c c c c c c c c c c c c} 
\hline 
\hline 
\multicolumn{1}{c}{(1)} & \multicolumn{1}{c}{(16)}  & \multicolumn{1}{c}{(17)} &  \multicolumn{1}{c}{(18)} & \multicolumn{1}{c}{(19)} & \multicolumn{1}{c}{(20)} & \multicolumn{1}{c}{(21)} & \multicolumn{1}{c}{(22)} & \multicolumn{1}{c}{(23)} & \multicolumn{1}{c}{(24)} & \multicolumn{1}{c}{(25)} & \multicolumn{1}{c}{(26)} & \multicolumn{1}{c}{(27)}   
\\ 
\multicolumn{1}{c}{Name} & \multicolumn{1}{c}{$p$ cut} & \multicolumn{1}{c}{$p$} & \multicolumn{1}{c}{$\Delta p$} & \multicolumn{1}{c}{$p$ off} & \multicolumn{1}{c}{$\phi_{peak}$} & \multicolumn{1}{c}{$\Delta \phi_{peak}$} & \multicolumn{1}{c}{Morph} & \multicolumn{1}{c}{IR\_c} & \multicolumn{1}{c}{O\_c} & \multicolumn{1}{c}{UV\_c} & \multicolumn{1}{c}{X\_c} & \multicolumn{1}{c}{$z$} 
\\ 
\multicolumn{1}{c}{} & \multicolumn{1}{c}{[RM beam$^{-1}$]} & \multicolumn{1}{c}{[RM beam$^{-1}$]} & \multicolumn{1}{c}{[RM beam$^{-1}$]} & \multicolumn{1}{c}{[RM beam$^{-1}$]} & \multicolumn{1}{c}{[rad m$^{-2}$]} & \multicolumn{1}{c}{[rad m$^{-2}$]} & \multicolumn{1}{c}{} & \multicolumn{1}{c}{} & \multicolumn{1}{c}{} & \multicolumn{1}{c}{} & \multicolumn{1}{c}{} & \multicolumn{1}{c}{} 
\\ 
\hline 
    032006-362044 & 0.032 & 0.05 & 0.01 & 0.37 & -300.4 & 6.0 & cj & no & no & no & no & - \\ 
    032228-384841 & 0.004 & 0.0383 & 0.0003 & 0.44 & -6.0 & 0.9 & cj & yes & no & no & no & - \\ 
    033019-365308 & 0.006 & 0.068 & 0.0009 & 0.34 & 7.1 & 0.8 & lc & off & off & no & no & - \\ 
    033123-361041 & 0.008 & 0.017 & 0.001 & 0.52 & 23.8 & 5.0 & u & yes & no & no & no & - \\ 
    033147-332912 & 0.004 & 0.0385 & 0.0004 & 0.48 & 32.0 & 1.0 & u & yes & yes & no & no & - \\ 
    033242-363645 & 0.021 & 0.208 & 0.004 & 0.25 & 4.6 & 0.9 & lc & off & may & may & no & - \\ 
    033329-384204 & 0.017 & 0.046 & 0.002 & 0.50 & -8.0 & 3.0 & ext & no & no & no & no & 0.103 \\ 
    033653-361606 & 0.003 & 0.0066 & 0.0003 & 0.72 & 37.5 & 4.0 & u & yes & yes & no & no & 1.540 \\ 
    033725-375958 & 0.020 & 0.198 & 0.003 & 0.41 & 8.6 & 0.9 & lc & off & off & off & no & - \\ 
    033726-380229 & 0.038 & 0.08 & 0.008 & 0.55 & 11.6 & 4.0 & u & yes & no & no & no & - \\ 
    033754-351735 & 0.179 & 0.3 & 0.2 & 0.41 & 165.0 & 6.0 & cj & - & - & - & no & - \\ 
    033828-352659 & 0.009 & 0.0094 & 0.0008 & 0.32 & -34.2 & 9.0 & ext & yes & yes & yes & yes & 0.005 \\ 
    033829-352818 & 0.021 & 0.034 & 0.003 & 0.91 & 144.2 & 6.0 & lc & off & off & off & no & 0.112 \\ 
    033843-352335 & 0.005 & 0.0422 & 0.0004 & 0.49 & 58.8 & 1.0 & lc & off & off & off & no & 0.112 \\ 
    033848-352215 & 0.002 & 0.0523 & 0.0002 & 0.56 & 16.7 & 0.4 & lc & off & off & off & no & 0.113 \\ 
    034133-362252 & 0.004 & 0.0212 & 0.0003 & 0.50 & 18.0 & 2.0 & ext & off & off & no & no & - \\ 
    034202-361520 & 0.243 & 0.57 & 0.04 & 0.59 & 13.2 & 4.0 & ext & yes & yes & no & no & 0.197 \\ 
    034205-370322 & 0.002 & 0.05599 & 9e-05 & 0.07 & 2.0 & 0.3 & cj & yes & yes & yes & yes & 0.284 \\ 
    034437-382640 & 0.009 & 0.0657 & 0.0009 & 0.16 & -19.8 & 1.0 & u & off & off & no & no & - \\ 
    \tableline 
    031533-375052 & 0.033 & 0.084 & 0.007 & 0.45 & 22.4 & 4.0 & lc & yes & no & no & no & - \\ 
    031537-375056 & 0.045 & 0.096 & 0.008 & 0.48 & 11.6 & 4.0 & lc & off & no & no & no & - \\ 
    031538-342345 & 0.011 & 0.084 & 0.001 & 0.46 & 24.2 & 1.0 & ext & yes & yes & no & no & - \\ 
    031551-364449 & 0.004 & 0.0431 & 0.0004 & 0.35 & 12.4 & 0.8 & u & yes & yes & yes & no & - \\ 
    031611-353540 & 0.120 & 0.12 & 0.02 & 0.00 & 13.0 & 9.0 & lc & - & - & - & no & - \\ 
    031616-381438 & 0.099 & 0.16 & 0.01 & 0.33 & 20.6 & 6.0 & cj & no & no & no & no & - \\ 
    031636-350112 & 0.043 & 0.052 & 0.008 & 0.00 & 14.1 & 8.0 & u & may & no & no & no & - \\ 
    031651-370212 & 0.106 & 0.17 & 0.02 & 0.00 & 1.9 & 6.0 & u & - & - & - & no & - \\ 
    031653-382609 & 0.029 & 0.066 & 0.004 & 0.05 & 26.6 & 4.0 & lc & off & off & off & no & - \\ 
    031654-382434 & 0.029 & 0.065 & 0.004 & 0.00 & 25.8 & 4.0 & lc & off & no & off & no & - \\ 
    031656-375122 & 0.006 & 0.006 & 0.002 & 0.00 & 3.2 & 9.0 & u & yes & may & yes & no & - \\ 
    031705-353441 & 0.039 & 0.068 & 0.006 & 0.33 & 8.8 & 5.0 & cj & yes & no & no & no & - \\ 
    031745-384734 & 0.058 & 0.073 & 0.009 & 0.00 & 26.5 & 7.0 & ext & yes & no & may & no & - \\ 
    031747-344234 & 0.083 & 0.09 & 0.02 & 0.00 & 13.9 & 8.0 & u & - & - & - & no & - \\ 
    031756-375619 & 0.018 & 0.019 & 0.003 & 0.10 & 20.5 & 9.0 & u & yes & no & no & no & - \\ 
    031817-382921 & 0.013 & 0.019 & 0.002 & 0.00 & 32.4 & 6.0 & u & yes & yes & yes & no & - \\ 
    031826-360959 & 0.037 & 0.124 & 0.004 & 0.37 & 18.4 & 3.0 & u & no & yes & no & no & - \\ 
    031840-352549 & 0.009 & 0.0109 & 0.0008 & 0.27 & 70.1 & 7.0 & ext & off & no & no & no & - \\ 
    031850-350545 & 0.017 & 0.045 & 0.002 & 0.15 & 55.7 & 4.0 & u & yes & may & no & no & - \\ 
    031852-380826 & 0.016 & 0.026 & 0.002 & 0.00 & 20.1 & 6.0 & ext & may & no & no & no & - \\ 
    031858-332622 & 0.089 & 0.1 & 0.01 & 0.00 & 11.4 & 8.0 & u & nod & no & no & no & - \\ 
    031922-340222 & 0.018 & 0.04 & 0.004 & 0.00 & 20.0 & 4.0 & ext & yes & no & no & no & - \\ 
    031954-350558 & 0.050 & 0.1 & 0.01 & 0.00 & 26.9 & 5.0 & ext & no & may & no & no & - \\ 
    032012-351244 & 0.023 & 0.041 & 0.003 & 0.31 & 16.8 & 5.0 & ext & yes & yes & no & no & - \\ 
    032015-351400 & 0.023 & 0.041 & 0.003 & 0.00 & 17.4 & 5.0 & ext & yes & yes & no & no & - \\ 
    032039-352111 & 0.031 & 0.032 & 0.005 & 0.10 & 19.7 & 9.0 & u & no & no & no & no & - \\ 
    032043-343600 & 0.015 & 0.101 & 0.003 & 0.44 & 11.8 & 1.0 & cj & no & no & no & no & - \\ 
    032044-382210 & 0.016 & 0.046 & 0.001 & 0.43 & 8.0 & 3.0 & ext & off & off & off & no & - \\ 
    032046-382230 & 0.017 & 0.046 & 0.001 & 0.28 & 7.8 & 3.0 & ext & off & off & off & no & - \\ 
    032046-383729 & 0.014 & 0.035 & 0.002 & 0.29 & 19.3 & 4.0 & u & off & off & off & no & - \\ 
\end{tabular} 
\end{table*} 
\end{turnpage}     
    
\addtocounter{table}{-1}
\begin{turnpage} 
\begin{table*} 
\caption{(Table 1 cont., columns 1, 16--27) Further quantities calculated / determined for polarized sources in our sample} 
\scriptsize 
\tabcolsep=0.09cm 
\begin{tabular}{c c c c c c c c c c c c c c} 
\hline 
\hline 
\multicolumn{1}{c}{(1)} & \multicolumn{1}{c}{(16)}  & \multicolumn{1}{c}{(17)} &  \multicolumn{1}{c}{(18)} & \multicolumn{1}{c}{(19)} & \multicolumn{1}{c}{(20)} & \multicolumn{1}{c}{(21)} & \multicolumn{1}{c}{(22)} & \multicolumn{1}{c}{(23)} & \multicolumn{1}{c}{(24)} & \multicolumn{1}{c}{(25)} & \multicolumn{1}{c}{(26)} & \multicolumn{1}{c}{(27)}   
\\ 
\multicolumn{1}{c}{Name} & \multicolumn{1}{c}{$p$ cut} & \multicolumn{1}{c}{$p$} & \multicolumn{1}{c}{$\Delta p$} & \multicolumn{1}{c}{$p$ off} & \multicolumn{1}{c}{$\phi_{peak}$} & \multicolumn{1}{c}{$\Delta \phi_{peak}$} & \multicolumn{1}{c}{Morph} & \multicolumn{1}{c}{IR\_c} & \multicolumn{1}{c}{O\_c} & \multicolumn{1}{c}{UV\_c} & \multicolumn{1}{c}{X\_c} & \multicolumn{1}{c}{$z$}
\\ 
\multicolumn{1}{c}{} & \multicolumn{1}{c}{[RM beam$^{-1}$]} & \multicolumn{1}{c}{[RM beam$^{-1}$]} & \multicolumn{1}{c}{[RM beam$^{-1}$]} & \multicolumn{1}{c}{[RM beam$^{-1}$]} & \multicolumn{1}{c}{[rad m$^{-2}$]} & \multicolumn{1}{c}{[rad m$^{-2}$]} & \multicolumn{1}{c}{} & \multicolumn{1}{c}{} & \multicolumn{1}{c}{} & \multicolumn{1}{c}{} & \multicolumn{1}{c}{} & \multicolumn{1}{c}{}
\\ 
\hline     
    032124-341417 & 0.040 & 0.093 & 0.005 & 0.00 & 11.0 & 4.0 & cj & off & off & no & no & 0.212 \\ 
    032125-340738 & 0.044 & 0.049 & 0.005 & 0.00 & 9.9 & 8.0 & lc & no & no & no & no & 0.214 \\ 
    032126-341334 & 0.041 & 0.092 & 0.005 & 0.00 & 10.8 & 4.0 & cj & yes & yes & yes & no & 0.212 \\ 
    032132-340553 & 0.009 & 0.012 & 0.002 & 0.00 & -4.2 & 7.0 & cj & yes & yes & yes & no & 0.109 \\ 
    032213-345831 & 0.020 & 0.055 & 0.003 & 0.40 & 7.6 & 3.0 & u & yes & yes & yes & no & - \\ 
    032315-384013 & 0.021 & 0.053 & 0.004 & 0.24 & -2.4 & 4.0 & u & yes & may & may & no & - \\ 
    032331-333314 & 0.018 & 0.052 & 0.002 & 0.43 & 5.6 & 3.0 & u & may & no & no & no & - \\ 
    032334-345532 & 0.031 & 0.05 & 0.005 & 0.07 & 6.8 & 6.0 & ext & yes & no & no & no & - \\ 
    032349-352401 & 0.072 & 0.08 & 0.02 & 0.00 & 5.5 & 8.0 & u & yes & yes & yes & no & - \\ 
    032349-351059 & 0.138 & 0.14 & 0.03 & 0.00 & 26.8 & 9.0 & u & - & - & - & no & - \\ 
    032358-344500 & 0.073 & 0.09 & 0.02 & 0.00 & 11.5 & 8.0 & lc & off & off & no & no & - \\ 
    032358-340958 & 0.083 & 0.15 & 0.01 & 0.23 & 9.2 & 5.0 & cj & - & - & - & no & - \\ 
    032410-343927 & 0.045 & 0.062 & 0.009 & 0.00 & 10.9 & 7.0 & u & yes & yes & no & no & - \\ 
    032431-341910 & 0.010 & 0.019 & 0.002 & 0.00 & 11.1 & 5.0 & u & may & may & may & no & - \\ 
    032442-333555 & 0.012 & 0.024 & 0.002 & 0.00 & 7.8 & 5.0 & ext & yes & yes & yes & no & - \\ 
    032459-343906 & 0.116 & 0.19 & 0.02 & 0.26 & 8.9 & 6.0 & u & - & - & - & no & 0.108 \\ 
    032501-352200 & 0.026 & 0.041 & 0.003 & 0.00 & 9.5 & 6.0 & cj & may & no & nod & no & - \\ 
    032504-335859 & 0.033 & 0.054 & 0.006 & 0.00 & 9.3 & 6.0 & u & yes & may & no & no & - \\ 
    032520-360357 & 0.023 & 0.027 & 0.006 & 0.00 & 19.4 & 8.0 & u & may & no & no & no & - \\ 
    032522-344644 & 0.025 & 0.025 & 0.006 & 0.10 & 6.4 & 9.0 & u & no & no & no & no & - \\ 
    032532-353343 & 0.025 & 0.057 & 0.003 & 0.27 & -1.3 & 4.0 & u & yes & yes & nod & no & - \\ 
    032546-333739 & 0.056 & 0.153 & 0.006 & 0.00 & 14.6 & 3.0 & ext & no & no & no & no & - \\ 
    032554-355641 & 0.058 & 0.1 & 0.01 & 0.40 & 7.5 & 5.0 & u & - & - & - & no & - \\ 
    032556-362527 & 0.092 & 0.12 & 0.03 & 0.00 & -17.9 & 7.0 & u & - & - & - & no & - \\ 
    032613-362142 & 0.058 & 0.06 & 0.01 & 0.00 & -12.5 & 9.0 & u & no & no & no & no & - \\ 
    032615-381131 & 0.042 & 0.083 & 0.006 & 0.00 & 9.4 & 5.0 & ext & no & no & no & no & - \\ 
    032630-362643 & 0.031 & 0.07 & 0.008 & 0.05 & -4.9 & 4.0 & u & yes & yes & yes & no & - \\ 
    032659-374856 & 0.011 & 0.017 & 0.002 & 0.27 & 19.3 & 6.0 & u & yes & yes & no & no & - \\ 
    032720-362938 & 0.034 & 0.041 & 0.006 & 0.00 & -1.9 & 8.0 & u & yes & no & no & no & - \\ 
    032732-351518 & 0.010 & 0.019 & 0.002 & 0.48 & 1.9 & 5.0 & u & may & no & nod & no & - \\ 
    032736-343304 & 0.052 & 0.1 & 0.008 & 0.39 & 14.7 & 5.0 & lc & may & no & no & no & - \\ 
    032749-333337 & 0.009 & 0.073 & 0.001 & 0.28 & 20.4 & 1.0 & ext & off & no & no & no & - \\ 
    032810-375009 & 0.033 & 0.081 & 0.005 & 0.04 & 7.8 & 4.0 & ext & no & no & no & no & - \\ 
    032831-333858 & 0.016 & 0.017 & 0.003 & 0.10 & 17.8 & 9.0 & u & yes & no & no & no & - \\ 
    032834-351326 & 0.071 & 0.08 & 0.02 & 0.00 & 3.4 & 8.0 & cj & - & - & - & no & - \\ 
    032849-334751 & 0.006 & 0.02 & 0.001 & 0.39 & 16.9 & 3.0 & u & yes & no & may & no & - \\ 
    032908-373330 & 0.082 & 0.12 & 0.02 & 0.26 & -4.4 & 7.0 & u & no & no & no & no & - \\ 
    032918-351153 & 0.025 & 0.051 & 0.004 & 0.00 & 23.6 & 5.0 & u & off & off & off & no & - \\ 
    032918-360155 & 0.014 & 0.08 & 0.002 & 0.16 & 5.9 & 2.0 & ext & no & no & no & no & - \\ 
    033002-360832 & 0.016 & 0.105 & 0.002 & 0.40 & 2.2 & 1.0 & ext & off & off & off & no & - \\ 
    033005-354623 & 0.013 & 0.033 & 0.002 & 0.49 & 2.3 & 4.0 & u & no & no & no & no & - \\ 
    033008-365315 & 0.005 & 0.0609 & 0.0008 & 0.08 & -1.2 & 0.8 & lc & off & off & no & no & - \\ 
    033020-355449 & 0.041 & 0.058 & 0.006 & 0.00 & -10.3 & 7.0 & ext & no & no & nod & no & - \\ 
    033057-341128 & 0.030 & 0.043 & 0.007 & 0.00 & 35.4 & 6.0 & ext & may & no & no & no & - \\ 
    033109-380430 & 0.083 & 0.11 & 0.02 & 0.17 & -4.3 & 7.0 & ext & may & may & may & no & - \\ 
    033123-372745 & 0.061 & 0.07 & 0.01 & 0.00 & 12.0 & 8.0 & ext & yes & yes & no & no & - \\ 
    033138-342852 & 0.044 & 0.051 & 0.008 & 0.19 & 22.5 & 8.0 & u & no & no & no & no & - \\ 
    033201-342737 & 0.106 & 0.13 & 0.02 & 0.19 & 69.6 & 8.0 & u & yes & yes & no & no & - \\ 
    033212-364144 & 0.066 & 0.08 & 0.02 & 0.00 & -29.5 & 7.0 & ext & no & no & no & no & - \\ 
\end{tabular} 
\end{table*} 
\end{turnpage}     

\addtocounter{table}{-1}
\begin{turnpage} 
\begin{table*} 
\caption{(Table 1 cont., columns 1, 16--27) Further quantities calculated / determined for polarized sources in our sample} 
\scriptsize 
\tabcolsep=0.09cm 
\begin{tabular}{c c c c c c c c c c c c c c} 
\hline 
\hline 
\multicolumn{1}{c}{(1)} & \multicolumn{1}{c}{(16)}  & \multicolumn{1}{c}{(17)} &  \multicolumn{1}{c}{(18)} & \multicolumn{1}{c}{(19)} & \multicolumn{1}{c}{(20)} & \multicolumn{1}{c}{(21)} & \multicolumn{1}{c}{(22)} & \multicolumn{1}{c}{(23)} & \multicolumn{1}{c}{(24)} & \multicolumn{1}{c}{(25)} & \multicolumn{1}{c}{(26)} & \multicolumn{1}{c}{(27)} 
\\ 
\multicolumn{1}{c}{Name} & \multicolumn{1}{c}{$p$ cut} & \multicolumn{1}{c}{$p$} & \multicolumn{1}{c}{$\Delta p$} & \multicolumn{1}{c}{$p$ off} & \multicolumn{1}{c}{$\phi_{peak}$} & \multicolumn{1}{c}{$\Delta \phi_{peak}$} & \multicolumn{1}{c}{Morph} & \multicolumn{1}{c}{IR\_c} & \multicolumn{1}{c}{O\_c} & \multicolumn{1}{c}{UV\_c} & \multicolumn{1}{c}{X\_c} & \multicolumn{1}{c}{$z$} 
\\ 
\multicolumn{1}{c}{} & \multicolumn{1}{c}{[RM beam$^{-1}$]} & \multicolumn{1}{c}{[RM beam$^{-1}$]} & \multicolumn{1}{c}{[RM beam$^{-1}$]} & \multicolumn{1}{c}{[RM beam$^{-1}$]} & \multicolumn{1}{c}{[rad m$^{-2}$]} & \multicolumn{1}{c}{[rad m$^{-2}$]} & \multicolumn{1}{c}{} & \multicolumn{1}{c}{} & \multicolumn{1}{c}{} & \multicolumn{1}{c}{} & \multicolumn{1}{c}{} & \multicolumn{1}{c}{} 
\\ 
\hline     
    033214-382447 & 0.023 & 0.046 & 0.003 & 0.05 & -6.0 & 5.0 & u & no & no & no & no & - \\ 
    033218-363949 & 0.171 & 0.22 & 0.03 & 0.08 & 16.6 & 7.0 & u & no & no & no & no & - \\ 
    033218-353815 & 0.029 & 0.038 & 0.005 & 0.00 & 15.2 & 7.0 & u & no & no & no & no & - \\ 
    033224-334116 & 0.033 & 0.062 & 0.006 & 0.32 & -2.1 & 5.0 & u & no & no & no & no & - \\ 
    033228-365620 & 0.012 & 0.072 & 0.002 & 0.19 & 0.4 & 2.0 & u & no & no & no & no & - \\ 
    033250-360829 & 0.035 & 0.11 & 0.004 & 0.00 & -0.8 & 3.0 & u & yes & yes & yes & no & - \\ 
    033300-365535 & 0.020 & 0.023 & 0.003 & 0.00 & 26.3 & 8.0 & u & off & off & off & no & - \\ 
    033302-341137 & 0.018 & 0.019 & 0.004 & 0.00 & 24.8 & 9.0 & u & yes & yes & yes & no & - \\ 
    033319-344546 & 0.059 & 0.06 & 0.01 & 0.00 & 39.1 & 9.0 & ext & - & - & - & no & - \\ 
    033323-375247 & 0.011 & 0.033 & 0.002 & 0.08 & -8.6 & 3.0 & u & yes & no & no & no & - \\ 
    033331-373806 & 0.107 & 0.11 & 0.03 & 0.00 & 7.9 & 9.0 & cj & no & no & no & no & - \\ 
    033332-344206 & 0.026 & 0.077 & 0.004 & 0.31 & 12.7 & 3.0 & u & yes & no & no & no & - \\ 
    033336-335206 & 0.018 & 0.052 & 0.002 & 0.40 & 25.8 & 3.0 & u & yes & yes & yes & no & - \\ 
    033342-371804 & 0.063 & 0.15 & 0.01 & 0.10 & -0.5 & 4.0 & u & no & no & no & no & - \\ 
    033347-342225 & 0.043 & 0.094 & 0.007 & 0.18 & 13.2 & 4.0 & u & yes & yes & no & no & - \\ 
    033400-364316 & 0.061 & 0.1 & 0.01 & 0.00 & 9.2 & 5.0 & u & no & no & no & no & - \\ 
    033405-344329 & 0.026 & 0.125 & 0.003 & 0.24 & 18.2 & 2.0 & ext & yes & yes & no & no & - \\ 
    033415-372542 & 0.004 & 0.021 & 0.0005 & 0.10 & 7.1 & 2.0 & u & yes & yes & yes & yes & - \\ 
    033417-341230 & 0.027 & 0.06 & 0.004 & 0.00 & 18.5 & 4.0 & ext & off & off & no & no & - \\ 
    033423-363628 & 0.052 & 0.097 & 0.007 & 0.39 & -7.2 & 5.0 & ext & may & no & no & no & - \\ 
    033423-341136 & 0.134 & 0.18 & 0.02 & 0.00 & 25.7 & 7.0 & u & - & - & - & no & - \\ 
    033424-344435 & 0.089 & 0.1 & 0.02 & 0.00 & 28.7 & 9.0 & cj & - & - & - & no & - \\ 
    033432-335605 & 0.066 & 0.08 & 0.01 & 0.19 & 11.3 & 7.0 & u & yes & no & no & no & - \\ 
    033450-364735 & 0.013 & 0.099 & 0.001 & 0.18 & 1.6 & 1.0 & cj & no & no & no & no & - \\ 
    033451-361019 & 0.020 & 0.036 & 0.004 & 0.13 & -2.2 & 5.0 & u & may & no & no & no & - \\ 
    033511-372737 & 0.024 & 0.116 & 0.003 & 0.40 & 0.8 & 2.0 & ext & yes & yes & no & no & - \\ 
    033520-342803 & 0.011 & 0.017 & 0.002 & 0.00 & 31.4 & 6.0 & ext & may & yes & yes & no & - \\ 
    033523-335523 & 0.037 & 0.059 & 0.005 & 0.00 & 18.6 & 6.0 & u & no & no & no & no & - \\ 
    033529-362939 & 0.045 & 0.134 & 0.006 & 0.00 & 3.8 & 3.0 & ext & off & no & off & no & - \\ 
    033555-350135 & 0.010 & 0.028 & 0.001 & 0.42 & 19.2 & 3.0 & u & may & no & no & no & - \\ 
    033558-332708 & 0.012 & 0.032 & 0.001 & 0.00 & 26.8 & 4.0 & u & yes & no & no & no & - \\ 
    033559-340352 & 0.028 & 0.053 & 0.005 & 0.00 & 26.7 & 5.0 & u & may & may & no & no & - \\ 
    033601-342136 & 0.040 & 0.055 & 0.007 & 0.00 & 18.4 & 7.0 & u & yes & yes & no & no & - \\ 
    033604-341342 & 0.018 & 0.037 & 0.003 & 0.00 & 19.3 & 4.0 & u & yes & no & no & no & - \\ 
    033613-334714 & 0.047 & 0.123 & 0.006 & 0.35 & 25.3 & 4.0 & ext & yes & no & no & no & - \\ 
    033639-381057 & 0.049 & 0.06 & 0.01 & 0.00 & -11.0 & 8.0 & u & off & off & off & no & - \\ 
    033641-335828 & 0.020 & 0.029 & 0.004 & 0.25 & 27.0 & 6.0 & u & yes & no & no & no & - \\ 
    033645-354229 & 0.016 & 0.02 & 0.003 & 0.27 & -3.7 & 7.0 & u & yes & yes & yes & no & 1.568 \\ 
    033752-342219 & 0.005 & 0.008 & 0.0009 & 0.00 & 38.5 & 6.0 & u & yes & yes & yes & no & - \\ 
    033804-371130 & 0.012 & 0.078 & 0.002 & 0.19 & 6.4 & 1.0 & u & may & yes & yes & no & - \\ 
    033826-355129 & 0.056 & 0.12 & 0.009 & 0.00 & 24.3 & 4.0 & u & off & off & off & no & 0.305 \\ 
    033827-352540 & 0.021 & 0.048 & 0.003 & 0.18 & 51.2 & 4.0 & lc & off & off & off & no & 0.006 \\ 
    033832-355108 & 0.041 & 0.05 & 0.009 & 0.09 & 36.6 & 8.0 & lc & no & no & no & no & 0.305 \\ 
    033849-375103 & 0.004 & 0.041 & 0.0008 & 0.00 & 8.3 & 0.9 & ext & may & may & no & no & - \\ 
    033903-375437 & 0.035 & 0.081 & 0.005 & 0.37 & 8.4 & 4.0 & u & off & off & off & no & - \\ 
    033913-345302 & 0.130 & 0.18 & 0.03 & 0.00 & 29.8 & 7.0 & u & - & - & - & no & 0.103 \\ 
    033950-370006 & 0.016 & 0.036 & 0.002 & 0.39 & 3.6 & 4.0 & u & yes & yes & yes & no & 0.141 \\ 
    034006-363545 & 0.004 & 0.037 & 0.0008 & 0.18 & 3.8 & 1.0 & ext & yes & no & no & no & - \\ 
\end{tabular} 
\end{table*} 
\end{turnpage}     
    
\addtocounter{table}{-1}
\begin{turnpage} 
\begin{table*} 
\caption{(Table 1 cont., columns 1, 16--27) Further quantities calculated / determined for polarized sources in our sample} 
\label{tab:allsourcedatab4} 
\scriptsize 
\tabcolsep=0.09cm 
\begin{tabular}{c c c c c c c c c c c c c c} 
\hline 
\hline 
\multicolumn{1}{c}{(1)} & \multicolumn{1}{c}{(16)}  & \multicolumn{1}{c}{(17)} &  \multicolumn{1}{c}{(18)} & \multicolumn{1}{c}{(19)} & \multicolumn{1}{c}{(20)} & \multicolumn{1}{c}{(21)} & \multicolumn{1}{c}{(22)} & \multicolumn{1}{c}{(23)} & \multicolumn{1}{c}{(24)} & \multicolumn{1}{c}{(25)} & \multicolumn{1}{c}{(26)} & \multicolumn{1}{c}{(27)} 
\\ 
\multicolumn{1}{c}{Name} & \multicolumn{1}{c}{$p$ cut} & \multicolumn{1}{c}{$p$} & \multicolumn{1}{c}{$\Delta p$} & \multicolumn{1}{c}{$p$ off} & \multicolumn{1}{c}{$\phi_{peak}$} & \multicolumn{1}{c}{$\Delta \phi_{peak}$} & \multicolumn{1}{c}{Morph} & \multicolumn{1}{c}{IR\_c} & \multicolumn{1}{c}{O\_c} & \multicolumn{1}{c}{UV\_c} & \multicolumn{1}{c}{X\_c} & \multicolumn{1}{c}{$z$} 
\\ 
\multicolumn{1}{c}{} & \multicolumn{1}{c}{[RM beam$^{-1}$]} & \multicolumn{1}{c}{[RM beam$^{-1}$]} & \multicolumn{1}{c}{[RM beam$^{-1}$]} & \multicolumn{1}{c}{[RM beam$^{-1}$]} & \multicolumn{1}{c}{[rad m$^{-2}$]} & \multicolumn{1}{c}{[rad m$^{-2}$]} & \multicolumn{1}{c}{} & \multicolumn{1}{c}{} & \multicolumn{1}{c}{} & \multicolumn{1}{c}{} & \multicolumn{1}{c}{} & \multicolumn{1}{c}{} 
\\ 
\hline     
    034008-372711 & 0.016 & 0.033 & 0.003 & 0.11 & -3.8 & 4.0 & ext & may & off & off & no & - \\ 
    034009-332644 & 0.009 & 0.017 & 0.001 & 0.00 & 20.8 & 5.0 & u & yes & no & no & no & - \\ 
    034017-375446 & 0.059 & 0.133 & 0.008 & 0.00 & 9.6 & 4.0 & ext & may & may & no & no & - \\ 
    034033-374943 & 0.038 & 0.046 & 0.006 & 0.00 & 8.7 & 8.0 & u & yes & yes & no & no & 0.042 \\ 
    034042-340818 & 0.019 & 0.037 & 0.003 & 0.00 & 22.9 & 5.0 & lc & may & may & no & no & - \\ 
    034049-340903 & 0.003 & 0.0295 & 0.0004 & 0.21 & 32.0 & 0.9 & lc & may & may & no & no & - \\ 
    034054-342252 & 0.056 & 0.11 & 0.01 & 0.00 & 22.7 & 5.0 & u & no & no & no & no & - \\ 
    034146-344840 & 0.051 & 0.052 & 0.009 & 0.00 & 17.6 & 9.0 & cj & may & no & no & no & - \\ 
    034155-363653 & 0.011 & 0.014 & 0.002 & 0.00 & 38.0 & 7.0 & u & yes & no & no & no & - \\ 
    034219-380805 & 0.169 & 0.2 & 0.1 & 0.00 & -6.3 & 9.0 & ext & - & - & - & no & - \\ 
    034254-340503 & 0.115 & 0.15 & 0.03 & 0.00 & 26.9 & 7.0 & u & no & no & no & no & - \\ 
    034305-374023 & 0.014 & 0.041 & 0.002 & 0.22 & -20.2 & 3.0 & u & yes & no & no & no & - \\ 
    034323-335144 & 0.004 & 0.0549 & 0.0006 & 0.24 & 21.0 & 0.7 & u & yes & yes & yes & no & - \\ 
    034359-381408 & 0.035 & 0.084 & 0.004 & 0.39 & -6.6 & 4.0 & u & may & no & no & no & - \\ 
\end{tabular} 
\end{table*} 
\end{turnpage}

\subsection{Faraday complexity / polarization classification of the sample}\label{sec-compprev}

 In this section we present the Faraday complexity classifications obtained by applying the methods described in Sections \ref{sec-rmsynth} \& \ref{sec-compcat}. Initially we describe the results obtained using a $6\sigma$ GES {\sc rmclean} cutoff and natural weighting in RM synthesis (defined as $\mathit{w}_j=\sigma_{qu,j}^{-2}$ in eqns \ref{eq:RMs} \& \ref{eq:lSQ_0}). We then explain how these classifications are affected when the {\sc rmclean} cutoff or RM synthesis weighting scheme is changed in Sections \ref{sec-cleaneffect} \& \ref{sec-weighteffect}.  
 
We classified as unpolarized the 403 of 563 sources for which max($|$FDS$|$) $<$ {\sc rmclean} cutoff ($6\sigma$ GES). The remaining 160 were classified as polarized, yielding $\sim 5$ polarized sources / sq. degree. This value is consistent with estimates of polarized source density in the literature (e.g. Stil et al. 2014), given the median $6\sigma$ GES {\sc rmclean} cutoff we obtain is equivalent to $P\approx 0.6$ mJy beam$^{-1}$ (rmsf beam)$^{-1}$. 

\begin{figure}
\includegraphics[width=0.475\textwidth]{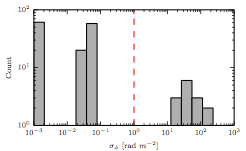}
\caption{Histogram of $\sigma_\phi$ for the sample after applying RM synthesis \& {\sc rmclean} with natural weighting and a $6\sigma$ GES {\sc rmclean} cutoff. We have altered $\sigma_\phi = 0$ values to $\sigma_\phi = 0.001$ rad m$^{-2}$ to fit on the logarithmic $x$-axis. We classify sources as Faraday-simple or complex based on the value of $\sigma_\phi$ as described in the main text, placing the distinguishing threshold at 1 rad m$^{-2}$ (red dashed line).}
\label{fig:phimom2_hist}
\end{figure}

We calculated $\sigma_\phi$ for each polarized source (recorded in column 15 of Table \ref{tab:allsourcedata}) and present a histogram of the resulting values in Figure \ref{fig:phimom2_hist}. Three populations are evident: The first consists of sources with $\sigma_\phi = 0$ --- i.e. all {\sc rmclean} components lie at precisely the same Faraday depth. The second at $\sigma_\phi \approx 0.1 $ rad m$^{-2}$ occurs when all {\sc rmclean} components are detected in two adjacent bins in $\phi$ space (separated by 0.1 rad m$^{-2}$), with only a small minority of components in the second bin. We classified the 146 sources comprising both this and the $\sigma_\phi = 0$ population together as Faraday-simple. 

The third population contains 14 sources (12\% of the polarized source sample), with $\sigma_\phi$ values in the range $10 - 300 $ rad m$^{-2}$. A clear division exists between the first two populations and the third, which suggests a $\sigma_\phi$ threshold to use for distinguishing between Faraday simple and Faraday complex sources. We classified this last population of sources as Faraday complex, noting here that the Faraday classifications thereby obtained agree well with the depolarization characteristics of sample sources presented in Section \ref{sec-observables-depolrepol}. 

\subsubsection{Effect of {\sc rmclean} cutoff threshold on classifications}\label{sec-cleaneffect}

Repeating the analysis of the preceding section using $8\sigma$ GES and $10\sigma$ GES {\sc rmclean} cutoffs, we obtained the results recorded in Table \ref{tab:natwei}. The {\sc rmclean} cutoff has a moderate effect on the polarization / Faraday complexity classifications: The number of complex and simple detections decreases as the {\sc rmclean} cutoff level is increased, while the number of unpolarized sources increases. This is caused by emission components progressively falling below the {\sc rmclean} cutoff and thus not being detected / used in the calculation of $\sigma_\phi$. Despite this, the majority of complex detections remain classified as such regardless of the {\sc rmclean} cutoff adopted, implying the additional polarized components in these sources are detected with high significance. 
 
 \begin{center}
 \begin{table}
 \caption{Number of sources in each Faraday category as a function of {\sc rmclean} cutoff GES level using natural weighting in RM synthesis.}
    \begin{tabular}{| l | l | l | l |}
    \hline
    Faraday classification & $6\sigma$ GES & $8\sigma$ GES & $10\sigma$ GES \\ \hline
    Unpolarized & 403 & 436 & 456\\ \hline
    Simple & 146 & 115 & 98 \\ \hline
    Complex & 14 & 12 & 9 \\
    \hline
    \end{tabular}
\label{tab:natwei}
\end{table}
\end{center}
 
\subsubsection{Effect of RM synthesis weighting on classifications}\label{sec-weighteffect}

The weighting scheme used in RM synthesis can affect sensitivity to Faraday complexity. In our observations, primary beam attenuation causes a small but systematic increase in the uncertainty of ($q$,$u$) data for off-axis sources towards higher frequencies. Under these circumstances, natural weighting will decrease sensitivity to depolarizing signals. We therefore repeated the analysis of the previous two sections using constant weighting (i.e. $\mathit{w}_j=1$ in eqns \ref{eq:RMs} \& \ref{eq:lSQ_0}). The resulting classifications are presented in Table \ref{tab:uniwei}, and we compare the classifications made under the natural and constant weighting schemes in Table \ref{tab:sets}. The majority of $10\sigma$ GES detections are unaffected by the weighting scheme employed. At $6\sigma$ GES however, only a single source is classified as complex under both schemes, with the majority of complex detections occurring when constant weighting is used. Regardless of the cause of this effect, it is clear that choosing between the two weighting schemes may affect our ability to detect specific types of complexity (e.g. strongly depolarizing signals). Thus, we consider sources to be complex if they are detected as such under either natural \emph{or} constant weighting. This results in a total of 19 sources being classified as complex using a $6\sigma$ GES {\sc rmclean} cutoff. 

 \begin{center}
 \begin{table}
 \caption{Number of sources in each Faraday category as a function of {\sc rmclean} cutoff GES level using constant weighting in RM synthesis.}
    \begin{tabular}{| l | l | l | l |}
    \hline
    Faraday classification & $6\sigma$ GES & $8\sigma$ GES & $10\sigma$ GES \\ \hline
    Unpolarized & 400 & 438 & 461\\ \hline
    Simple & 145 & 114 & 93 \\ \hline
    Complex & 18 & 11 & 9 \\
    \hline
  \end{tabular}
\label{tab:uniwei}
\end{table}
\end{center}

 \begin{center}
 \begin{table}
 \caption{Number of elements in the union, intersection and differences of sets comprising the Faraday complex source detections using constant (C) \& natural (N) RM synthesis weighting schemes, and which are detected at the specified GES {\sc rmclean} level but not higher (e.g. sources detected as complex using a $10\sigma$ GES cutoff are also detected as such using a $6\sigma$ GES cutoff, but are not counted in the $6\sigma$ GES column).}
    \begin{tabular}{| l | l | l | l | l |}
    \hline
    Set & $6\sigma$ GES & $8\sigma$ GES & $10\sigma$ GES\\ \hline
    $|C\cup N|$ & 7 & 1 & 11 \\ \hline
    $|C\cap N|$ & 1 & 1 & 7\\ \hline
    $|N\setminus C|$  & 1 & 0 & 2 \\ \hline
    $|C\setminus N|$  & 5 & 0 & 2\\
    \hline
    \end{tabular}
\label{tab:sets}
\end{table}
\end{center}

\subsection{Spectropolarimetric data for selected sources}\label{sec-indiv}

We now present the spectropolarimetric data for all sources classified as complex, regardless of the weighting scheme and {\sc rmclean} cutoff under which they were detected as such. We also include data for selected Faraday simple and unpolarized sources. The complexity classification of each source is noted in the figure captions. The data are presented in 7 panels. The uppermost panel (panel a) shows the result of applying RM synthesis to the source data. It includes plots of the magnitude of the FDS, the {\sc rmclean} component distribution and the {\sc rmclean} cutoff level as described in the figure captions. To more clearly show the multiple {\sc rmclean} components characteristic of complex sources, we include an inset panel zoomed on $|\phi|<750$ rad m$^{-2}$ and $0<|$FDS$|<3\times$ the {\sc rmclean} cutoff. Panel b) shows Stokes $I$($\lambda^2$) with its fitted model. Panels c) and e) show the Stokes q($\lambda^2$) \& u($\lambda^2$) and Stokes v($\lambda^2$) spectra respectively. Panel d) plots fractional polarization against $\lambda^2$, along with the best fit depolarization models described in Section \ref{sec-observables-depolrepol}. Panel f) plots polarization angle vs. $\lambda^2$.

\begin{figure*}[htpb]
\centering
\includegraphics[width=0.8\textwidth]{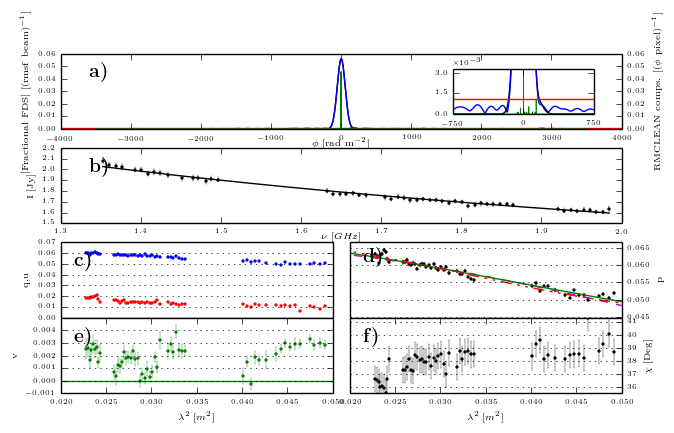}
\caption{$10\sigma$ GES complex detection. Source: 034205-370322. \textbf{a)} blue curve $\equiv$ $|FDS|$, red line $\equiv$ {\sc rmclean} cutoff, green lines $\equiv$ {\sc rmclean} component model, black curve $\equiv$ $|FDS-${\sc rmclean} residuals$|$. Inset: Zoom on detail in panel a). \textbf{b)} Total intensity spectrum + model fit. \textbf{c)} Stokes q (red) \& u (blue) spectra. \textbf{d)} $p$ spectrum + best-fit depolarization models from Section \ref{sec-observables-depolrepol}. Red dot-dot-dash line $\equiv$ double component model, green line $\equiv$ Tribble model, magenta dashed line $\equiv$ Burn model. \textbf{e)} Stokes v spectrum. Green line indicates Stokes v$=0$. \textbf{f)} Polarization angle.}
\label{fig:034205-370322}
\end{figure*}

\begin{figure*}[htpb]
\centering
\includegraphics[width=0.8\textwidth]{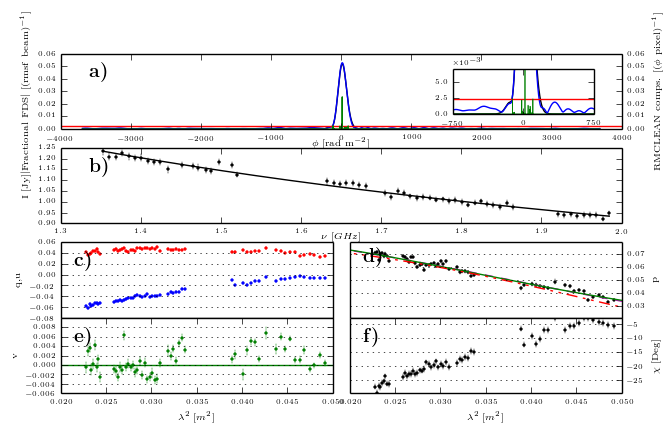}
\caption{$10\sigma$ GES complex detection. Plots as described in Figure \ref{fig:034205-370322} caption. Source: 033848-352215}
\label{fig:033848-352215}
\end{figure*}
 
\begin{figure*}[htpb]
\centering
\includegraphics[width=0.8\textwidth]{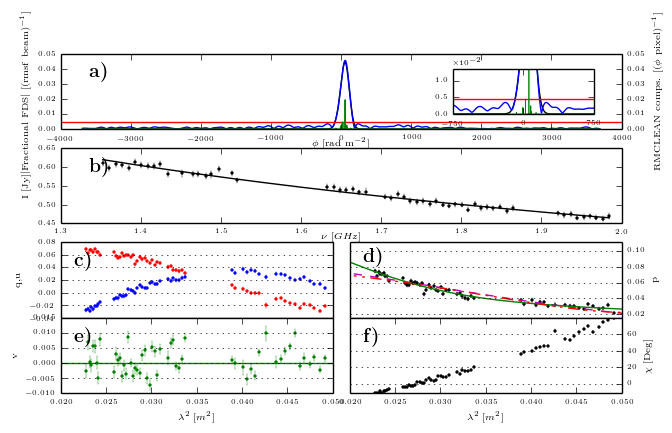}
\caption{$10\sigma$ GES complex detection. Plots as described in Figure \ref{fig:034205-370322} caption. Source: 033843-352335}
\label{fig:033843-352335}
\end{figure*}

\begin{figure*}[htpb]
\centering
\includegraphics[width=0.8\textwidth]{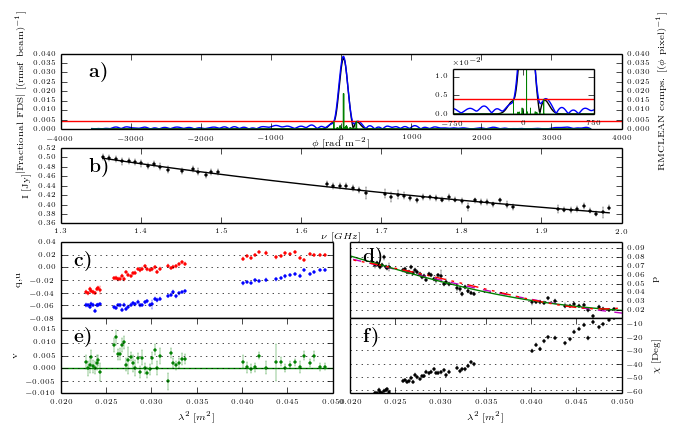}
\caption{$10\sigma$ GES complex detection. Plots as described in Figure \ref{fig:034205-370322} caption. Source: 033147-332912}
\label{fig:033147-332912}
\end{figure*}

\begin{figure*}[htpb]
\centering
\includegraphics[width=0.8\textwidth]{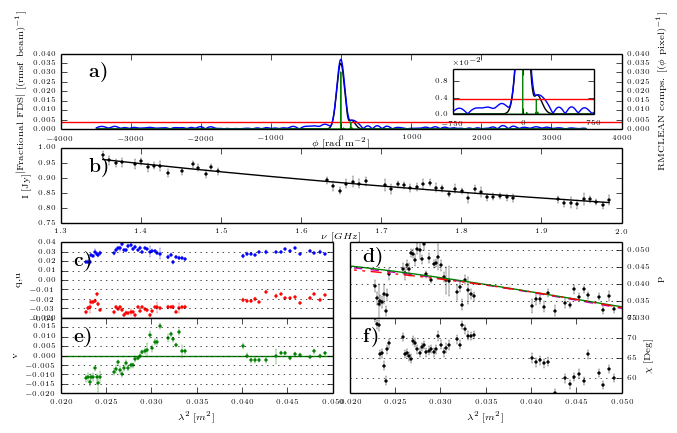}
\caption{$10\sigma$ GES complex detection. Plots as described in Figure \ref{fig:034205-370322} caption. Source: 032228-384841}
\label{fig:032228-384841}
\end{figure*}

\begin{figure*}[htpb]
\centering
\includegraphics[width=0.8\textwidth]{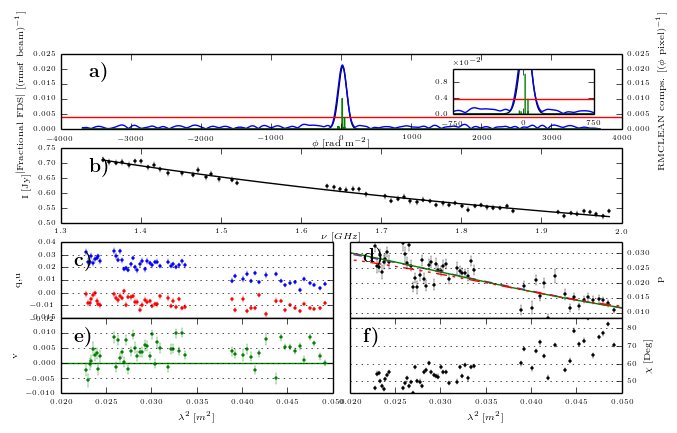}
\caption{$10\sigma$ GES complex detection. Plots as described in Figure \ref{fig:034205-370322} caption. Source: 034133-362252}
\label{fig:034133-362252}
\end{figure*}
  
\begin{figure*}[htpb]
\centering
\includegraphics[width=0.8\textwidth]{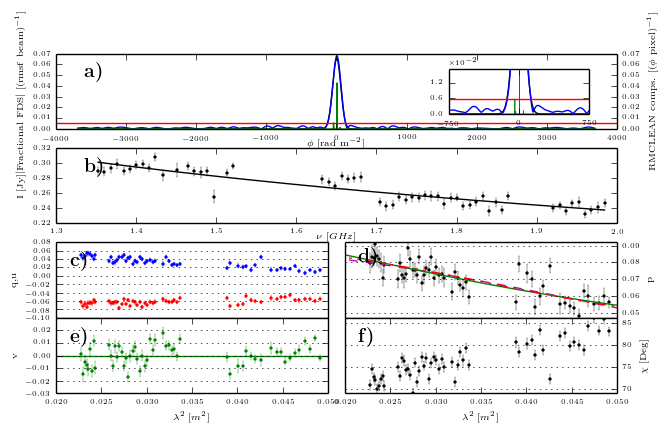}
\caption{$10\sigma$ GES complex detection. Plots as described in Figure \ref{fig:034205-370322} caption. Source: 033019-365308}
\label{fig:033019-365308}
\end{figure*}

\begin{figure*}[htpb]
\centering
\includegraphics[width=0.8\textwidth]{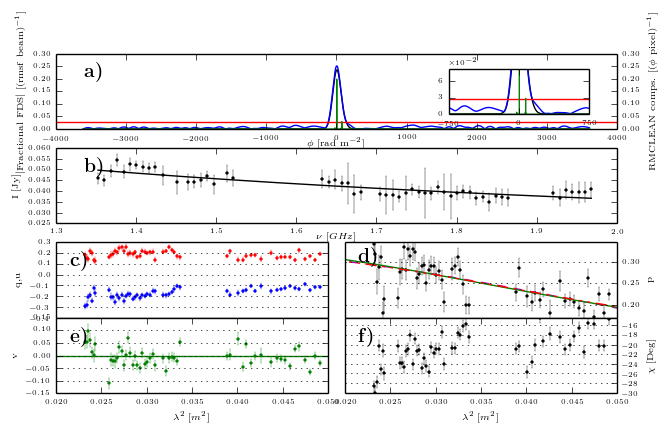}
\caption{$10\sigma$ GES complex detection. Plots as described in Figure \ref{fig:034205-370322} caption. Source: 033242-363645}
\label{fig:033242-363645}
\end{figure*}
 
\begin{figure*}[htpb]
\centering
\includegraphics[width=0.8\textwidth]{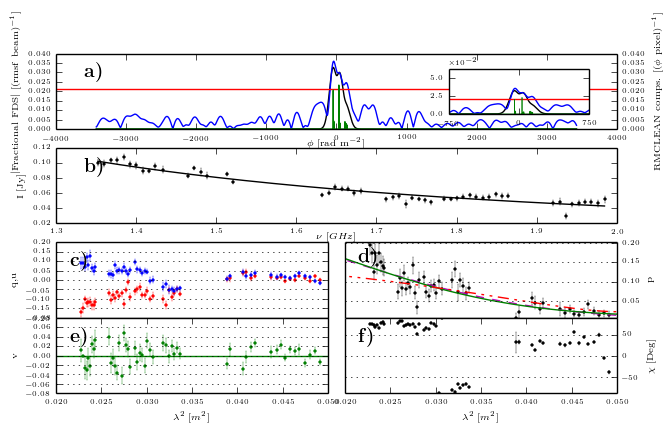}
\caption{$10\sigma$ GES complex detection. Plots as described in Figure \ref{fig:034205-370322} caption. Source: 033829-352818}
\label{fig:033829-352818}
\end{figure*}

\begin{figure*}[htpb]
\centering
\includegraphics[width=0.8\textwidth]{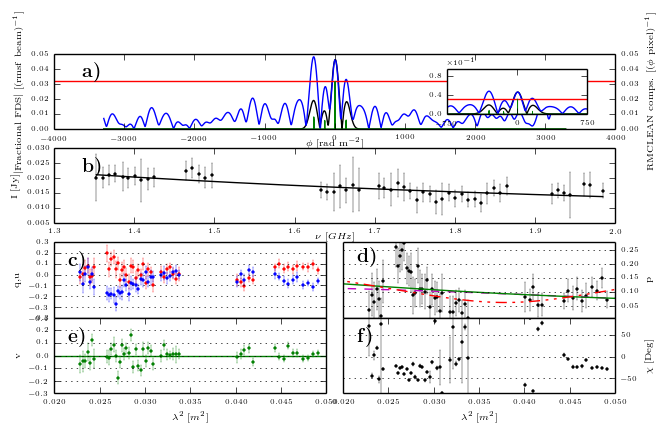}
\caption{$10\sigma$ GES complex detection. Plots as described in Figure \ref{fig:034205-370322} caption. Source: 032006-362044}
\label{fig:032006-362044}
\end{figure*}
 
\begin{figure*}[htpb]
\centering
\includegraphics[width=0.8\textwidth]{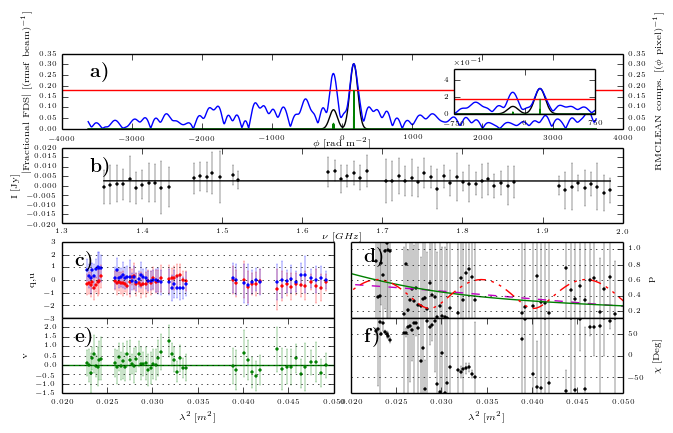}
\caption{$10\sigma$ GES complex detection. Plots as described in Figure \ref{fig:034205-370322} caption. Source: 033754-351735}
\label{fig:033754-351735}
\end{figure*}

\begin{figure*}[htpb]
\centering
\includegraphics[width=0.8\textwidth]{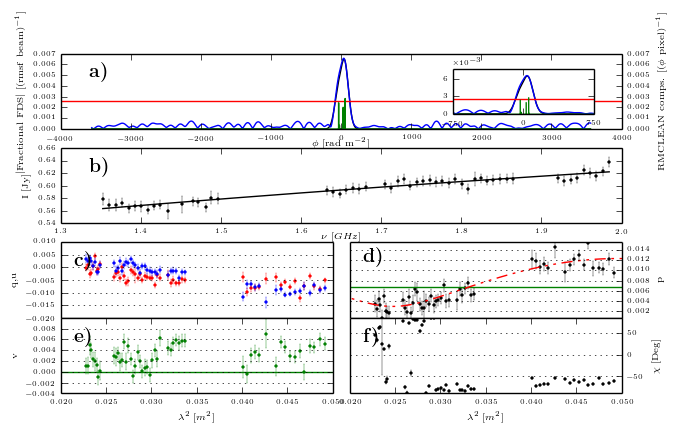}
\caption{$8\sigma$ GES complex detection. Plots as described in Figure \ref{fig:034205-370322} caption. Source: 033653-361606}
\label{fig:033653-361606}
\end{figure*}

\begin{figure*}[htpb]
\centering
\includegraphics[width=0.8\textwidth]{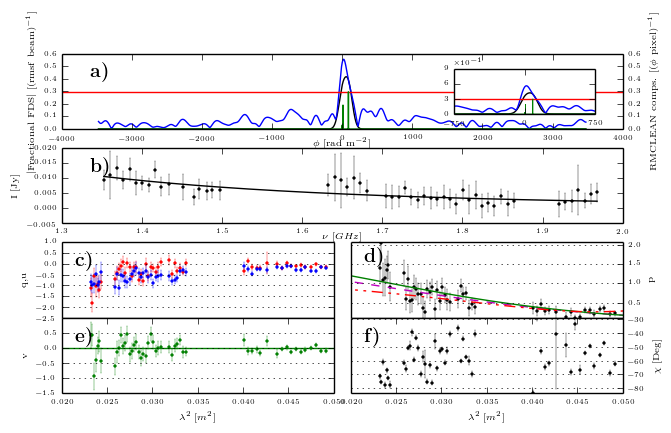}
\caption{$8\sigma$ GES complex detection. Plots as described in Figure \ref{fig:034205-370322} caption. Source: 034202-361520}
\label{fig:034202-361520}
\end{figure*}

\begin{figure*}[htpb]
\centering
\includegraphics[width=0.8\textwidth]{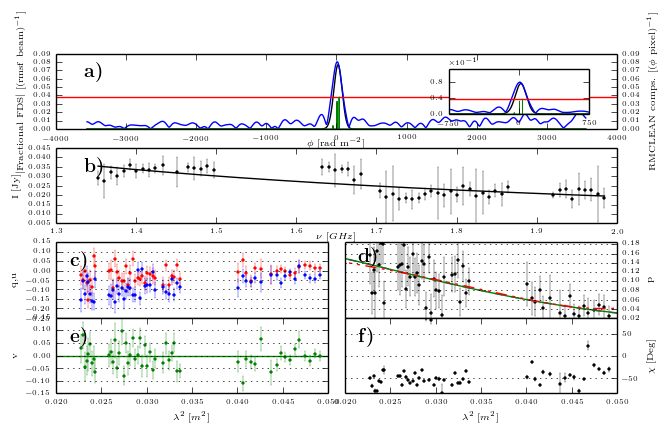}
\caption{$6\sigma$ GES complex detection. Plots as described in Figure \ref{fig:034205-370322} caption. Source: 033726-380229}
\label{fig:033726-380229}
\end{figure*}

\begin{figure*}[htpb]
\centering
\includegraphics[width=0.8\textwidth]{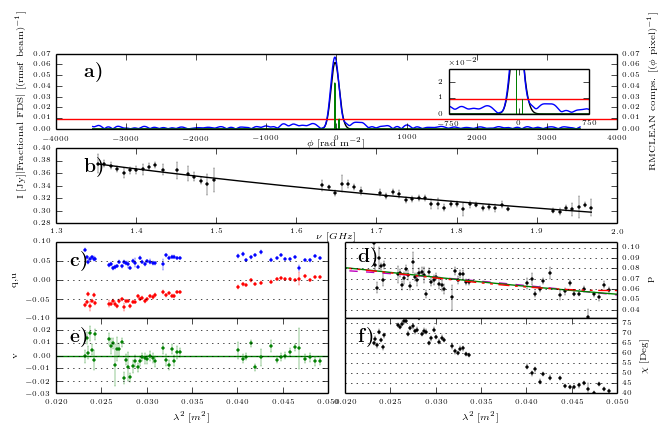}
\caption{$6\sigma$ GES complex detection. Plots as described in Figure \ref{fig:034205-370322} caption. Source: 034437-382640}
\label{fig:034437-382640}
\end{figure*}

\begin{figure*}[htpb]
\centering
\includegraphics[width=0.8\textwidth]{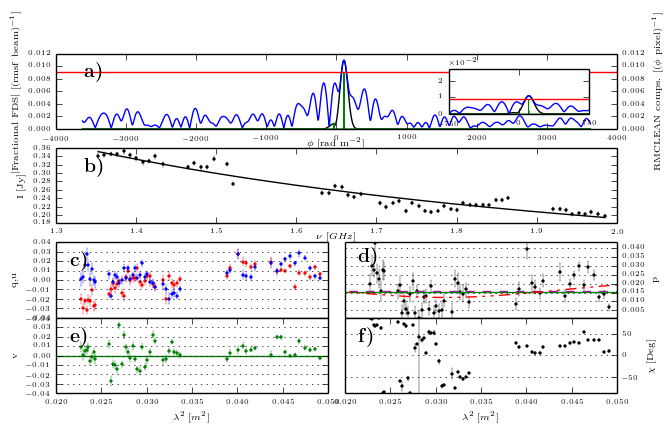}
\caption{$6\sigma$ GES complex detection. Plots as described in Figure \ref{fig:034205-370322} caption. Source: 033828-352659} 
\label{fig:033828-352659}
\end{figure*}

\begin{figure*}[htpb]
\centering
\includegraphics[width=0.8\textwidth]{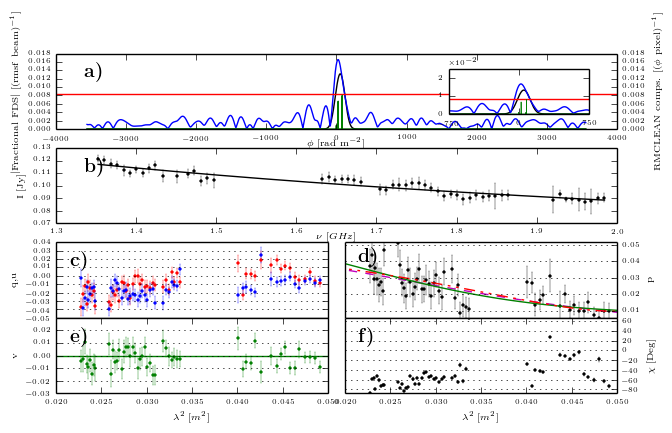}
\caption{$6\sigma$ GES complex detection. Plots as described in Figure \ref{fig:034205-370322} caption. Source: 033123-361041}
\label{fig:033123-361041}
\end{figure*}
 
\begin{figure*}[htpb]
\centering
\includegraphics[width=0.8\textwidth]{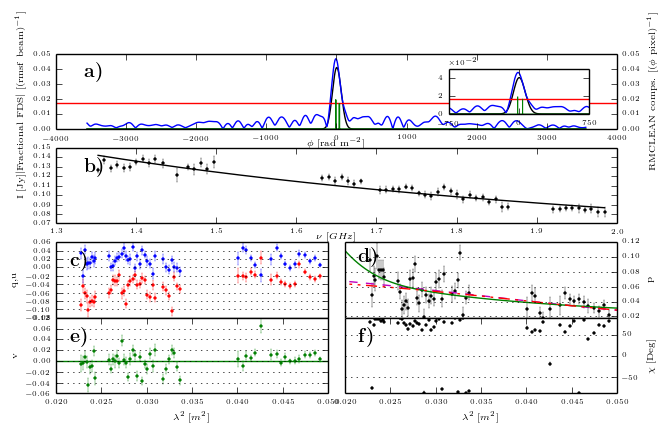}
\caption{$6\sigma$ GES complex detection. Plots as described in Figure \ref{fig:034205-370322} caption. Source: 033329-384204}
\label{fig:033329-384204}
\end{figure*}


 \begin{figure*}[htpb]
 \centering
\includegraphics[width=0.8\textwidth]{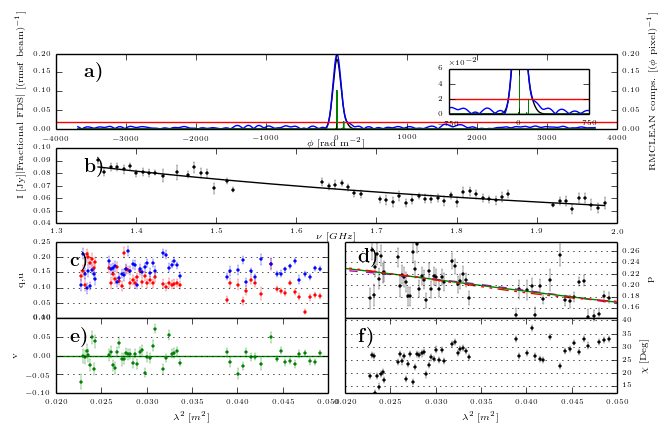}
\caption{$6\sigma$ GES complex detection. Plots as described in Figure \ref{fig:034205-370322} caption. Source: 033725-375958}
\label{fig:033725-375958}
\end{figure*}

\begin{figure*}[htpb]
\centering
\includegraphics[width=0.8\textwidth]{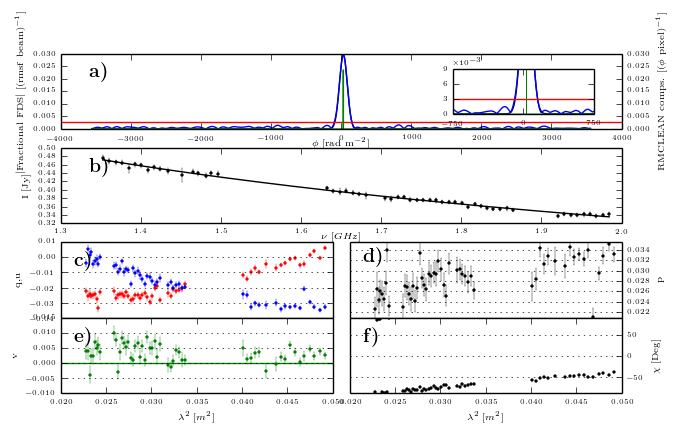}
\caption{Faraday simple detection. Plots as described in Figure \ref{fig:034205-370322} caption. Source: 034049-340903}
\label{fig:034049-340903}
\end{figure*}
  
\begin{figure*}[htpb]
\centering
\includegraphics[width=0.8\textwidth]{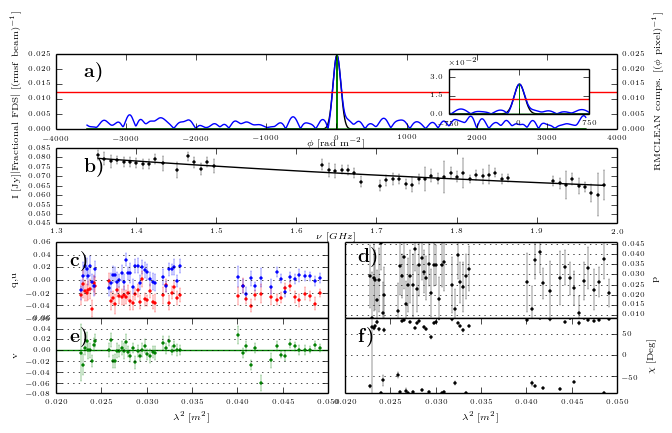}
\caption{Faraday simple detection. Plots as described in Figure \ref{fig:034205-370322} caption. Source: 032442-333555}
\label{fig:032442-333555}
\end{figure*}

\begin{figure*}[htpb]
\centering
\includegraphics[width=0.8\textwidth]{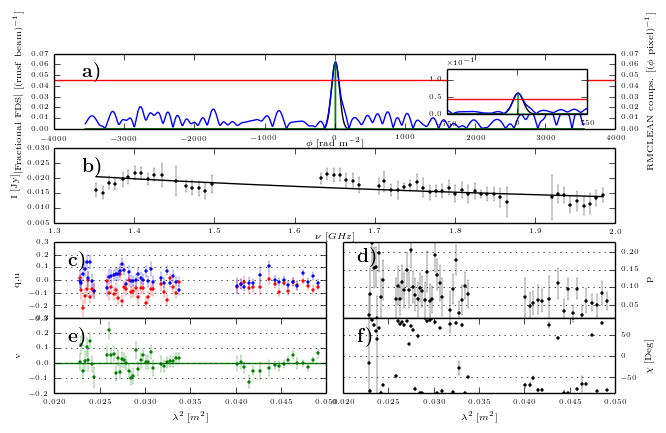}
\caption{Faraday simple detection. Plots as described in Figure \ref{fig:034205-370322} caption. Source: 032410-343927}
\label{fig:032410-343927}
\end{figure*}
  
\begin{figure*}[htpb]
\centering
\includegraphics[width=0.8\textwidth]{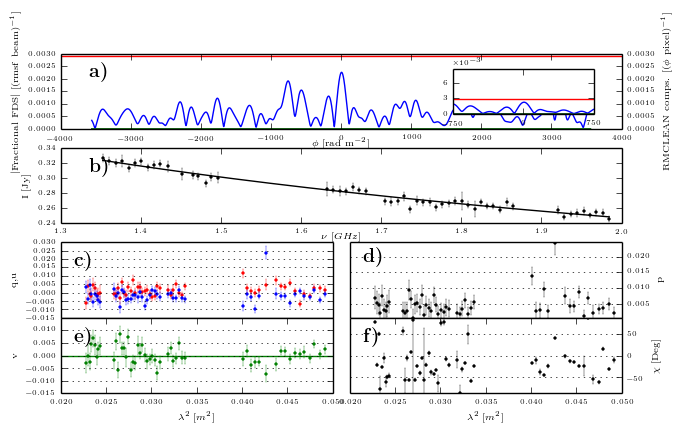}
\caption{Unpolarized source. Plots as described in Figure \ref{fig:034205-370322} caption. Source: 033336-360823}
\label{fig:033336-360823}
\end{figure*}

\begin{figure*}[htpb]
\centering
\includegraphics[width=0.8\textwidth]{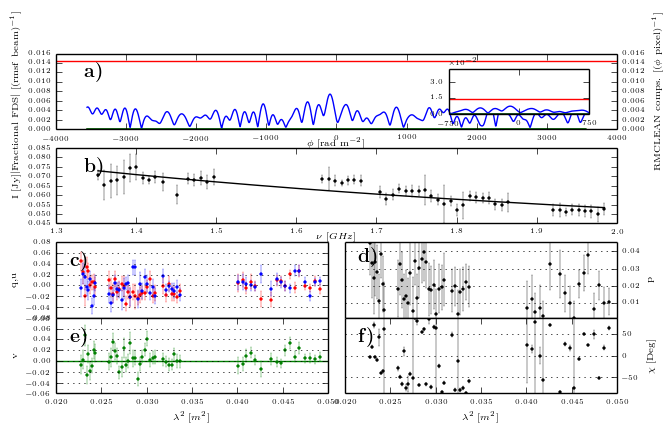}
\caption{Unpolarized source. Plots as described in Figure \ref{fig:034205-370322} caption. Source: 033554-365949}
\label{fig:033554-365949}
\end{figure*}
 
\begin{figure*}[htpb]
\centering
\includegraphics[width=0.8\textwidth]{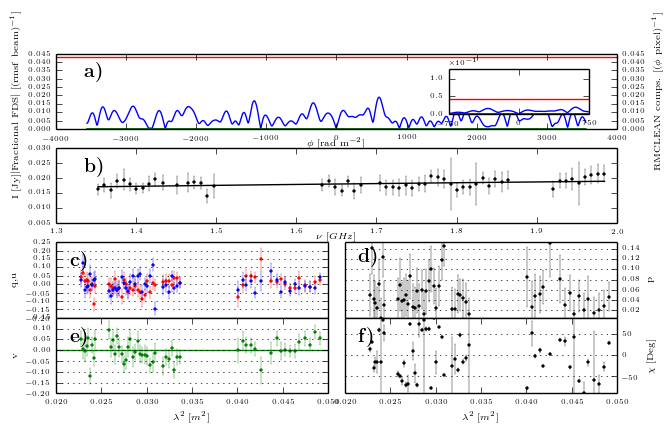}
\caption{Unpolarized source. Plots as described in Figure \ref{fig:034205-370322} caption. Source: 031827-341833}
\label{fig:031827-341833}
\end{figure*}

\subsection{Relationship between complexity and non-polarimetric source properties}\label{sec-additional-nonpol}

In the following sections we examine several non-polarimetric properties of the sample with the aim of identifying characteristics that distinguish Faraday complex sources.

\subsubsection{Source flux density}\label{sec-observables-bright}

\begin{figure}[ht]
\includegraphics[width=0.475\textwidth]{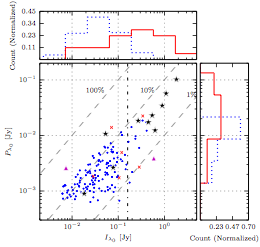}
\caption{Scatter plot of monochromatic polarized vs. total intensities (see main text) for polarized sample sources. Objects classified as Faraday-simple are plotted as blue dots, while Faraday-complex objects are plotted as black stars stars ($10\sigma$ GES detections), purple triangles ($8\sigma$ GES) or red crosses ($6\sigma$ GES). In all cases, sources are assigned the highest detection significance attained under analysis using each RM synthesis weighting scheme. Diagonal dashed grey lines represent lines of constant fractional polarization and are labelled in \%. Above and to the right of the plot are normalized histograms of the source distributions projected onto each axis. In each case, the blue dotted lines are the histograms for Faraday-simple sources, while the red solid lines are the histograms for Faraday-complex objects. The vertical black dot-dashed line indicates the integrated flux density that divides the bright source sample (BSS) from the faint source sample (FSS) as defined in Section \ref{sec-fcvsfluxdens}.}
\label{fig:Faradaypropsvsflux+polint}
\end{figure}

We calculated the monochromatic polarized intensity $P_{\lambda_{0}}$ for each source by multiplying $I_{\lambda_{0}}$ by max($|$FDS$|$). In Figure \ref{fig:Faradaypropsvsflux+polint}, we plot $P_{\lambda_{0}}$ vs. $I_{\lambda_{0}}$, along with histograms of simple and complex source counts (normalized) projected onto each axis. The plot markers indicate the highest degree of complexity or polarization attained by a source under the different weighting schemes and {\sc rmclean} cutoff levels adopted (see caption) --- a convention we use through the remainder of this paper. It is evident that the likelihood of a source being assigned a complex classification is strongly dependent on both its total and polarized intensity: 100\% of the brightest seven sources in both $P_{\lambda_{0}}$ \& $I_{\lambda_{0}}$ are detected as Faraday complex, but the projected $P_{\lambda_{0}}$ \& $I_{\lambda_{0}}$ histograms show that the ratio of complex to simple detection counts drops off rapidly moving to lower integrated flux densities. 

\subsubsection{Spectral index}\label{sec-observables-SI}

\begin{figure}
\includegraphics[width=0.475\textwidth]{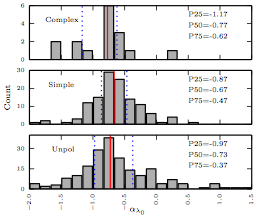}
\caption{Histograms of $\alpha_{\lambda_0}$ for sample sources with $I_{\lambda_{0}} > 10$ mJy, split according to Faraday complexity. Vertical red lines indicate the sample median (the `P50' value in text inset), while the blue dotted lines represent the 25th (P25) and 75th (P75) percentiles. Note that one complex source with $\alpha_{\lambda_0}=-2.4\pm0.2$ extends beyond the x-axis limit in the top panel.}
\label{fig:SIofSources}
\end{figure}

Figure \ref{fig:SIofSources} shows histograms of $\alpha_{\lambda_{0}}$ ($I(\nu)\propto\nu^\alpha$) for the sources in each polarization / complexity category possessing $I_{\lambda_{0}} > 10$ mJy (the large uncertainties of fainter sources obscure the shape of the distributions). Complex sources (upper panel) possess $\alpha_{\lambda_{0}}$ between -2.4 and 0.26 --- a range comparable to that of both the simple and unpolarized distributions (middle \& lower panels). The majority of the complex detections have steep $\alpha_{\lambda_{0}}$: 90\% have $\alpha_{\lambda_{0}}<-0.5$, with the median, 25th and 75th percentiles of the distribution equal to -0.77, -1.17 and -0.67 respectively. The inverted spectral index of the complex source 033653-361606 ($\alpha_{\lambda_{0}}=0.26$) is a notable exception. The distribution percentiles all lie steeper than the equivalents for the simple and unpolarized distributions (see inset text on figure), but the differences do not achieve a robust level of statistical significance. 

\subsubsection{Radio morphology}\label{sec-smorph}
 
Table \ref{tab:morph} presents the numbers and percentages of complex, simple and unpolarized sources possessing the morphological types defined in Section \ref{sec-radmorph}. Complex sources are almost evenly split between the unresolved, lobe / jet component, core-jet and extended morphological types, implying that Faraday complexity is not uniquely associated with specific source morphologies. At the same time, $\sim3/4$ of the complex sources are morphologically resolved compared to only $\sim1/4$ of the unpolarized sources and $\sim2/5$ of the simple sources. This suggests an association between complexity and apparent source size, but degeneracies between morphology, source brightness and Faraday complexity in our sample prevent a deeper analysis of this and other apparent morphological differences between the polarization / complexity categories.

 \begin{center}
 \begin{table}
    \caption{Number and percentage of complex, simple \& unpolarized sources with the morphological types defined in Section \ref{sec-radmorph}.}
    \begin{tabular}{| l | l | l | l | l | l | l | l | l | l |}
    \hline
     & \multicolumn{2}{c|}{Unresolved} & \multicolumn{2}{c|}{Lobe / jet} & \multicolumn{2}{c|}{Core-jet} & \multicolumn{2}{c|}{Extended} \\
    \hline
     & \multicolumn{1}{c|}{\#} & \multicolumn{1}{c|}{\%} & \multicolumn{1}{c|}{\#} & \multicolumn{1}{c|}{\%} & \multicolumn{1}{c|}{\#} & \multicolumn{1}{c|}{\%} & \multicolumn{1}{c|}{\#} & \multicolumn{1}{c|}{\%}\\
    \hline
   Complex & 5 & 26 & 6 & 32 & 4 & 21 & 4 & 21  \\
   Simple & 79 & 56 & 13 & 9 & 13 & 9 & 36 & 26\\ 
   Unpolarized & 305 & 76 & 28 & 7 & 10 & 2 & 60 & 15\\
   \hline
    \end{tabular}
   \label{tab:morph}
    \end{table}
\end{center}

\subsubsection{Multiwavelength counterparts}\label{sec-fcvsfluxdensb}

\begin{figure}
\includegraphics[width=0.475\textwidth]{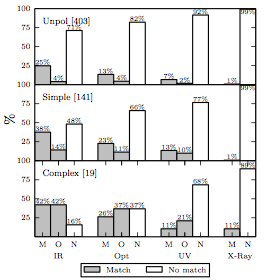}
\caption{Percentage of sources with an on-source counterpart match (M), an off-source match (O), and no match (N) as described in Section \ref{sec-fcvsfluxdensb}, broken down by Faraday category and wavelength. The number of sources in each Faraday category is listed in square brackets next to the main axes labels.}
\label{fig:CPstats}
\end{figure}

In Figure \ref{fig:CPstats} we divide the sample according to Faraday complexity, then for each wavelength at which cross matches were made in Section \ref{sec-mwxmatch}, we plot the percentage of sources that were assigned on-source / off-source / no match counterpart statuses. The main trend visible is that the proportion of non-detections progressively increases from complex to simple to unpolarized sources. This probably follows trivially from the fact that polarized emission, complexity and counterparts are all more readily detectable for brighter sources. When only the detections are considered, a trend is apparent in the ratio of off-source to on-source detections, which increases going from unpolarized to simple to complex sources. When counterparts were detected for complex sources, 50\%, 59\% \& 66\% (IR, optical \& UV) of the time they are not spatially coincident with the radio source. In contrast, the proportion of on-source counterparts for Faraday simple and unpolarized sources is greater than the proportion of off-source counterparts at all wavelengths --- in most cases substantially so. This may represent indirect evidence of a link between complexity and extended radio structure. We found no evidence to suggest that Faraday complexity was linked to the presence or absence of counterparts at specific wavelengths.

\subsubsection{Redshift}\label{redshiftdata}

 In Table \ref{tab:rs} we present the median value of crossmatched redshifts as a function of Faraday category. The overlapping 66\% confidence intervals for the median of each category demonstrate that any relationship between Faraday complexity and redshift must be substantially weaker than can be detected with our limited number of redshift crossmatches. 

 \begin{center}
 \begin{table}
    \caption{Number of redshift matches and median redshift for crossmatched sources in our sample, split by Faraday classification. Upper and lower bounds on the 66\% confidence intervals for the median are also shown.}
    \begin{tabular}{| l | l | l | l |}
    \hline
     & Complex & Simple & Unpolarized \\ \hline
    \# matches & 8 & 12 &  41  \\ \hline
    Median $^{66\% ~\text{CI}~\text{upper}}_{66\% ~\text{CI}~\text{lower}}$ & $0.11 _{0.11} ^{0.28}$ & $0.18 _{0.14} ^{0.21}$ & $0.12 _{0.11} ^{0.19}$ \\ \hline
    \end{tabular}
    \label{tab:rs}
    \end{table}
\end{center}

 \subsubsection{Location in the mosaic field}\label{sec-srcdistinfield}

\begin{figure}
\includegraphics[width=0.475\textwidth]{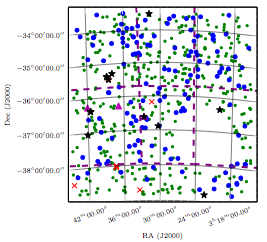}
\caption{Positions of sources detected in the field. The plotting markers follow the conventions of Figure \ref{fig:Faradaypropsvsflux+polint}. The patch without sources is the rejection region around Fornax A. The purple dashed lines delimit the approximate boundaries of the seven separately observed and calibrated submosaics (Section \ref{sec-observations}).}
\label{fig:Faradaypropsvsmosposish}
\end{figure}

\begin{figure}
\includegraphics[width=0.475\textwidth]{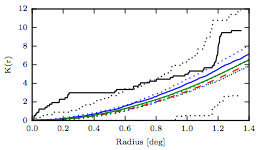}
\caption{Plots of Ripley's K function $K(r)$ vs. angular separation radius (solid lines, see main text) for positions of complex (black) simple (blue) \& unpolarized (green) sources in the field. The expectation (red dashed line) and 99\% confidence intervals for complete spatial randomness (dotted lines) are also shown.}
\label{fig:ripleysKfunction}
\end{figure}

In Figure \ref{fig:Faradaypropsvsmosposish} we plot the spatial location of complex, simple \& unpolarized sources in the mosaic field. We analysed their spatial distribution using Ripley's K function $K(r)$ (Ripley 1976), which computes the mean number of sources lying closer than $r$ degrees to other sources in a sample, normalized by the mean spatial source density. This can be compared to the expectation for complete spatial randomness (CSR), which is simply $\pi r^2$. Our results are shown in Figure \ref{fig:ripleysKfunction}, where we plot $K(r)$ for the complex, simple and unpolarized source samples, as well as 99\% confidence intervals for $K(r)$ under the assumption of CSR. The plots show that $K(r)$ lies substantially above the 99\% confidence interval for complex sources for scales $0^\circ$ -- $0.6^\circ$, indicating clustering on these scales at $>99$\% confidence. A particularly prominent clustering of complex sources is located at $\sim$ 03h38m -35d25m. By comparison, the  unpolarized and simple sources show deviations from CSR that are either small and/or statistically insignificant.

\subsection{Relationship between complexity and polarimetric source properties}\label{sec-additional-pol}

We now examine several polarimetric properties of the sample sources, with the aim of determining the characteristics of complex sources as well as how these characteristics distinguish complex sources from simple ones.

\subsubsection{Structure in complex Faraday Dispersion Spectra}\label{sec-observables-FDSstruc}

For 14 out of 19 complex sources, the additional polarized emission components in the FDS are unresolved (i.e. the {\sc rmclean} components are clustered within the width of the {\sc rmclean} restoring beam --- $\approx120$ rad m$^{-2}$. See Fig. \ref{fig:033848-352215} for example). For five complex sources --- 033147-332912 (Fig. \ref{fig:033147-332912}), 032228-384841 (Fig. \ref{fig:032228-384841}), 033829-352818 (Fig. \ref{fig:033829-352818}), 032006-362044 (Fig. \ref{fig:032006-362044}) \& 033754-351735 (Fig. \ref{fig:033754-351735}) --- multiple components are either fully or partially resolved in the FDS. In either case, the additional components typically contribute between 15\% and 60\% (mean $\approx45\%$) of the total polarized flux, comparable to the 10\%--70\% range reported by Law et al. (2011). 

\subsubsection{Faraday depth}\label{sec-observables-FDcomparo}

Here we determine whether the Faraday depth at which a source's emission peaks is correlated with its Faraday complexity classification. The existence of such a correlation would indicate that complexity is generated in regions possessing anisotropic ordered $\boldsymbol{B}$ field components --- for example, jets in AGN. 

\begin{figure}
\includegraphics[width=0.475\textwidth]{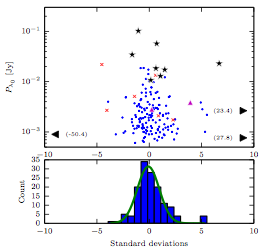}
\caption{\emph{Top:} $P_{\lambda_{0}}$ vs. the standardized residuals of $\phi$ for polarized sources in our sample (see text). Objects are plotted using the conventions adopted in Figure \ref{fig:Faradaypropsvsflux+polint}, with the exception that sources off the scale are plotted with triangular arrows and accompanying text indicating position on the $x$-axis. \emph{Bottom:} Histogram of the standardized residuals of the simple sources with accompanying Gaussian model fit (green line). The mean and standard deviation of this fitted Gaussian are $-0.06\pm0.08$ and $1.09\pm0.08$ respectively.}
\label{fig:CompFDs}
\end{figure}

We first calculate the residual peak Faraday depth of each source ($\phi_{res,peak}$) by subtracting the smooth Galactic contribution ($\phi_{gal,peak}$) to $\phi_{peak}$, which is well fit by the following  paraboloid:

\begin{eqnarray}
\phi_{gal,peak}(\text{RA}_{rel},\text{Dec}_{rel}) = 0.792(\text{RA}_{rel}-0.149)^2 + \dots \nonumber \\
1.040(\text{Dec}_{rel}+2.052)^2 + 1.6331 ~\text{rad m$^{-2}$} \nonumber \\
\label{eq:galscreen}
\end{eqnarray}

where, with RA and Dec in J2000 coordinates:\\
\\
$\text{RA}_{rel} = \text{RA} - 52.385^\circ$ and $\text{Dec}_{rel} = \text{Dec} + 35.793^\circ$\\
\\
The residual Faraday depth of each source is then calculated as $\phi_{res,peak} = \phi_{peak} - \phi_{gal,peak}$. It is difficult to use $\phi_{res,peak}$ directly to compare the dispersion of $\phi_{res,peak}$ (not to be confused with $\sigma_\phi$) for simple and complex sources, because the dominant contribution to this dispersion is signal dependent. Instead we calculate the standardized residual (SR) of $\phi_{res,peak}$ for each source. This is defined as:

\begin{eqnarray}
\text{SR} = \phi_{res,peak} / \sigma_{ex+err}
\label{eq:standardizedresid}
\end{eqnarray}
\\

where

\begin{eqnarray}
\sigma_{ex+err}^2 = \sigma_{err}^2 + \sigma_{ex}^2 
\label{eq:varcontrib}
\end{eqnarray}
\\

and $\sigma_{err}^2$ \& $\sigma_{ex}^2$ are the contributions to $\sigma_{ex+err}$ from measurement error and the mean extragalactic contribution to the Faraday depth of a radio source respectively. We use $\sigma_{ex}^2 = 49$ rad$^2$m$^{-4}$ (Schnitzeler 2010). If the value adopted for $\sigma_{ex}^2$ and the function describing the Galactic foreground are both accurate, SR should be Gaussian distributed with $\mu=0$ and $\sigma=1$. We confirm that this is the case, finding $\mu=-0.06\pm0.08$ and $\sigma=1.09\pm0.08$ for the best fit Gaussian to the histogram of SR data (both plotted in the bottom panel of Figure \ref{fig:CompFDs}).
  
 In the upper panel of Figure \ref{fig:CompFDs}, we plot $P_{\lambda_{0}}$ vs. SR so the variance in SR can be easily seen. Most sources, both complex and simple, are located within $1$--$2\sigma$ of zero and thus show no statistically significant enhancement in $\phi_{res,peak}$. However, several Faraday complex sources \emph{do} show statistically significant enhancements in $\phi_{res,peak}$. While no Faraday simple sources lie $>5.5\sigma$ from the mean, four complex sources have $\phi_{res,peak}$ values between $6.5\sigma$ and $50\sigma$. This is clearly statistically unlikely, even allowing for the slight non-normality of the SR distribution (bottom panel). The names and $|\phi_{peak}|$ values of these sources are 033843-352335 (Fig. \ref{fig:033843-352335}): 59 rad m$^{-2}$, 033829-352818 (Fig. \ref{fig:033829-352818}): 145 rad m$^{-2}$, 033754-351735 (Fig. \ref{fig:033754-351735}): 165 rad m$^{-2}$ \& 032006-362044 (Fig. \ref{fig:032006-362044}): 300 rad m$^{-2}$. We note that the first three of these sources are among the prominent cluster of complex sources identified in Section \ref{sec-srcdistinfield}.
  
\subsubsection{Depolarization / repolarization}\label{sec-observables-depolrepol}

Here we determine the extent to which Faraday complexity is associated with either depolarization or repolarization, which broadly constrains the possible mechanisms through which complexity is generated. For depolarizing sources, we then fit (existing) depolarization models to $p(\lambda^2)$ to obtain additional information about the depolarization behaviour. 

\begin{figure}
\includegraphics[width=0.475\textwidth]{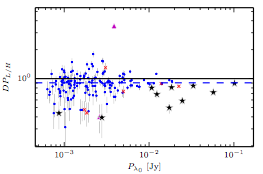}
\caption{Plot of the depolarization ratio $\text{DP}_{\text{L}/\text{H}}$ vs. $P_{\lambda_{0}}$. The $\text{DP}_{\text{L}/\text{H}}$ ratio was calculated as described in the text. The plotting markers follow the conventions of Figure \ref{fig:Faradaypropsvsflux+polint}. The black line represents no change in polarization over the band. The blue dashed line is the sample median for the simple sources.}
\label{fig:CompDepol}
\end{figure}

First we identify depolarizing / repolarizing sources by generating two sets of channels $L=\{i|\lambda_i>\lambda_0\}$ \& $H=\{i|\lambda_i<\lambda_0\}$ and calculating the following depolarization ratio: 

 \begin{equation}
\text{DP}_{\text{L}/\text{H}}=\frac{\sum\limits_{i\in L} p_{i,corr}/|L|}{\sum\limits_{i\in H} p_{i,corr}/|H|} 
\label{eq:depol}
 \end{equation}

 where $p_{i,corr}=\sqrt{p_{i}^2-\sigma_{qu,i}^2}$ is the bias-corrected polarization in channel $i$ (see Simmons \& Stewart 1985), $|L|$ \& $|H|$ are the number of channels in the sets $L$ \& $H$, and the subscript L/H denotes that this is a ratio of the lower frequency band to the higher frequency band. We plot $\text{DP}_{\text{L}/\text{H}}$ against $P_{\lambda_{0}}$ in Figure \ref{fig:CompDepol}, indicating the complexity / polarization category of each source with the plot markers (see caption) and the median value of $\text{DP}_{\text{L}/\text{H}}$ for Faraday simple sources with a blue dashed line. 

The simple sources are located in an envelope clustered around a median of 0.9 which broadens as the S/N drops with decreasing $P_{\lambda_{0}}$. The offset from $\text{DP}_{\text{L}/\text{H}}=1$ is statistically significant but caused by the mean residual bias from our polarization bias correction. The complex sources are mostly cleanly separated from the simple sources in the $P_{\lambda_{0}}$--$\text{DP}_{\text{L}/\text{H}}$ plane, being found primarily along the high $P_{\lambda_{0}}$, low $\text{DP}_{\text{L}/\text{H}}$ (i.e. higher S/N, stronger depolarization) edge of the source distribution. The observed agreement between the presence / lack of depolarization or repolarization and the assigned complex / simple Faraday classification represents \emph{a posteriori} evidence that our complexity classification algorithm is effective at distinguishing genuinely complex sources from simple ones.

18 (of 19 total) complex sources have $\text{DP}_{\text{L}/\text{H}}<1$ and 17 have $\text{DP}_{\text{L}/\text{H}}<0.9$ (i.e. the simple source median). The complexity in our sample is therefore predominantly a result of source depolarization. The single exception is the repolarizing source 033653-361606 (Fig. \ref{fig:033653-361606}). 

Information about the physical nature of depolarizing Faraday screens can be obtained by modelling the behaviour of $p(\lambda^2)$. We consider three depolarization models that, in combination, are capable of generating most of the $p(\lambda^2)$ behaviours that we observe. They are:\\

1) The Burn (1966) foreground screen, in which depolarization results from a large number of unresolved cells in the telescope beam:

\begin{eqnarray}
p(\lambda) = p_{[\lambda=0]}e^{-2\sigma_{RM}^2\lambda^4}
\label{eqn-BurnDepol}
\end{eqnarray}

2) A Tribble (1991, 1992) screen, with $N$ independent Faraday depth cells in the synthesized beam area (where $N\propto t^2/s_0^2$, $s_0$ is the cell scale length, and the synthesized beam $\text{FWHM}=2t\sqrt{\text{ln}2}$):

\begin{eqnarray}
p^2(\lambda) = p_{[\lambda=0]}^2\frac{1-\text{exp}(1-s_0^2/2t^2-4\sigma_{RM}^2\lambda^4)}{1+8\sigma_{RM}^2\lambda^4t^2/s_0^2}\dots\\ \nonumber
+p_{[\lambda=0]}^2\text{exp}(1-s_0^2/2t^2-4\sigma_{RM}^2\lambda^4)
\end{eqnarray}

3) Interference of polarized components at two separate Faraday depths:

\begin{eqnarray}
p(\lambda) = |p_1e^{2i\chi_1+\phi_1\lambda^2} + p_2e^{2i\chi_2+\phi_2\lambda^2}|
\end{eqnarray}

In the equations above, $p(\lambda)$ is the fractional polarization amplitude at wavelength $\lambda$, $p_{[\lambda=0]}$ is the fractional polarization at $\lambda=0$, $\sigma_{RM}$ is the dispersion in RM across a source due to a foreground Faraday screen, and $p_i$, $\phi_i$ and $\chi_i$ are the fractional polarization amplitude, Faraday depth and polarization angle of the $i$th polarized emission component. 

We used the `Emcee' sampler package (Foreman-Mackay et al. 2013) to find maximum likelihood values and errors for parameters in these models when fit to $p(\lambda^2)$. For the Tribble model, single-resolution observations cannot break the degeneracy between $\sigma_{RM}$ and $s_0/t$ (Tribble 1991). We present maximum likelihood values for these parameters (without errors) regardless to show that the Tribble model provides a demonstrably superior reproduction of $p(\lambda^2)$ for some sources. We list best-fit parameter values and $\bar{\chi}^2$ in Table \ref{tab:depolfits}. We also record the value of the Akaike information criterion (AIC; Akaike 1974), where AIC $=2k-2\text{ln}(L)$, $k=$ no. of model parameters, and $L$ is the maximum of the likelihood distribution returned by Emcee. Briefly, the AIC is a model selection criteria that is optimal in the sense that, for two models 1 \& 2 in which model 1 has the lowest AIC value, model 2 is $\text{exp}((\text{AIC}_1-\text{AIC}_2)/2)$ times as likely as model 1 to minimize the information lost by either underfitting or overfitting the data. We consider model 1 significantly favored when $\text{AIC}_1+5<\text{AIC}_2$. For a model where this is true against both other models, we list its AIC and $\bar{\chi}^2$ values in underlined bold text in Table \ref{tab:depolfits}. The best fit obtained with each model is overplotted on the $p(\lambda^2)$ plots presented in Section \ref{sec-indiv}.

\begin{table*}
\caption{Reduced $\chi^2$ and values for selected best-fit parameters for the Burn, Tribble and two polarized component model fits to $p(\lambda^2)$ for each complex source as described in the text. Fit parameters are bolded where specific models provide a clearly superior AIC value, as described in the main text.}
\scriptsize
\tabcolsep=0.09cm
\begin{tabular}{|c|c|c|c|c|c|c|c|c|c|c|c|c|c|c|}
\hline
 & \multicolumn{4}{ |c| }{Burn} & \multicolumn{5}{ c }{2 comp.} & \multicolumn{5}{ |c| }{Tribble} \\
 \hline
\multicolumn{1}{ |c| }{Src. Name} & \multicolumn{1}{ c| }{$\bar{\chi}^2$} & \multicolumn{1}{ c| }{AIC} & \multicolumn{1}{ c| }{$p_0$} & \multicolumn{1}{ c| }{$\sigma_{RM}$} & \multicolumn{1}{ c| }{$\bar{\chi}^2$} & \multicolumn{1}{ c| }{AIC} & \multicolumn{1}{ c| }{$p_1$} & \multicolumn{1}{ c| }{$p_2$} & \multicolumn{1}{ c| }{$|\phi_1-\phi_2|$} & \multicolumn{1}{ c| }{$\bar{\chi}^2$} & \multicolumn{1}{ c| }{AIC} & \multicolumn{1}{ c| }{$p_0$} & \multicolumn{1}{ c| }{$\sigma_{RM}$} & \multicolumn{1}{ c| }{$s_0/t$}\\
  & & & & \multicolumn{1}{ c| }{rad m$^{-2}$} & & & & & \multicolumn{1}{ c| }{rad m$^{-2}$} & & & & \multicolumn{1}{ c| }{rad m$^{-2}$} & \\
\hline
 032006-362044 & 1.1 & -257.0 & 0.12$\pm$0.02 & 9$\pm$3 & 1.1 & -238.7 & 0.12$\pm$0.03 & 0.13$\pm$0.02 & 41$\pm$1 & 1.1 & -232.9 & 0.16 & 28.0 & 2.1\\
 032228-384841 & 10.6 & -137.1 & 0.0470$\pm$0.0005 & 8.2$\pm$0.2 & 9.4 & -229.9 & 0.023$\pm$0.008 & 0.023$\pm$0.008 & 16$\pm$2 &  8.9 & -228.8 & 0.047 & 8.7 & 0.56\\
 033019-365308 & 1.8 & -496.0 & 0.089$\pm$0.002 & 10.2$\pm$0.4 & 1.8 & -492.9 & 0.021$\pm$0.004 & 0.071$\pm$0.004 & 27$\pm$4 & 1.8 & -496.0 & 0.099 & 51.0 & 4.7\\
 033123-361041 & 0.9 & -475.4 & 0.045$\pm$0.005 & 18$\pm$1 & 1.1 & -469.8 & 0.019$\pm$0.006 & 0.019$\pm$0.006 & 29$\pm$3 & 1.0 & -473.4 & 0.065 & 26.0 & 0.6\\
 033147-332912 & 3.5 & -471.3 & 0.106$\pm$0.002 & 19.3$\pm$0.2 & 4.0 & -449.9 & 0.0586$\pm$0.0007 & 0.0370$\pm$0.0006 & 33.5$\pm$0.3 & \textbf{\underline{2.9}} & \textbf{\underline{-503.0}} & 0.13 & 24.0 & 0.51\\
 033242-363645 & 2.5 & -310.9 & 0.329$\pm$0.009 & 10.4$\pm$0.5 & 2.6 & -306.5 & 0.06$\pm$0.1 & 0.2$\pm$0.1 & 31$\pm$40 & 2.5 & -277.6 & 0.3 & 13 & 1\\
 033329-384204 & 2.7 & -347.1 & 0.079$\pm$0.005 & 14.1$\pm$0.9 & 2.8 & -342.2 & 0.048$\pm$0.009 & 0.03$\pm$0.01 & 27$\pm$5 & 2.6 & -350.3 & 1.2 & 55.0 & 0.21\\
 033653-361606 & 8.6 & -257.3 & 0.007$\pm$0.02 & 0.009$\pm$5 & \textbf{\underline{1.1}} & \textbf{\underline{-644.5}} & 0.0046$\pm$0.0003 & 0.0073$\pm$0.0003 & 64$\pm$1 & 8.8 & -255.0 & 0.0068 & 0.31 & 15.0\\
 033725-375958 & 3.2 & -298.9 & 0.239$\pm$0.005 & 8.4$\pm$0.4 & 3.3 & -295.2 & 0.20$\pm$0.01 & 0.05$\pm$0.02 & 24$\pm$6 & 3.2 & -297.4 & 0.25 & 58.0 & 7.6\\
 033726-380229 & 0.7 & -312.8 & 0.20$\pm$0.03 & 20$\pm$2 & 0.9 & -308.4 & 0.08$\pm$0.02 & 0.08$\pm$0.02 & 31$\pm$3 & 0.7 & -310.5 & 0.2 & 19.0 & 0.13\\
 033754-351735 & 0.2 & 3.6 & 19$\pm$30 & 60$\pm$700 & 0.1 & 7.1 & 0.4245$\pm$0.001 & 0.1812$\pm$0.001 & 272$\pm$1 & 0.1 & 5.3 & 1.7 & 35.0 & 0.68\\
 033828-352659 & 5.7 & -273.1 & 0.0156$\pm$0.0001 & 0.03$\pm$0.01 & \textbf{\underline{5.4}} & \textbf{\underline{-327.8}} & 0.019$\pm$0.001 & 0.007$\pm$0.002 & 49$\pm$3 & 5.9 & -298.7 & 0.016 & 1.9 & 8.0\\
 033829-352818 & 1.8 & -331.8 & 0.25$\pm$0.02 & 24.2$\pm$1 & 2.1 & -319.2 & 0.076$\pm$0.005 & 0.094$\pm$0.004 & 33.9$\pm$0.7 & 1.8 & -330.9 & 0.28 & 26.0 & 0.19\\
 033843-352335 & 5.3 & -412.5 & 0.090$\pm$0.001 & 16.9$\pm$0.2 & 4.6 & -452.5 & 0.0320$\pm$0.0006 & 0.0597$\pm$0.0006 & 34.2$\pm$0.3 & \textbf{\underline{2.8}} & \textbf{\underline{-546.9}} & 0.15 & 30.0 & 0.73\\
 033848-352215 & \textbf{\underline{6.1}} & \textbf{\underline{-444.5}} & 0.0843$\pm$0.0004 & 13.39$\pm$0.07 & 6.7 & -422.2 & 0.057$\pm$0.001 & 0.0259$\pm$0.0006 & 27.2$\pm$0.6 & 6.9 & -410.6 & 0.088 & 16.0 & 0.75\\
 034133-362252 & 4.1 & -487.4 & 0.0358$\pm$0.0009 & 15.1$\pm$0.3 & 4.2 & -485.4 & 0.024$\pm$0.001 & 0.0117$\pm$0.0005 & 30$\pm$1 & 4.1 & -484.7 & 0.036 & 16.0 & 0.47\\
 034202-361520 & 1.0 & -116.7 & 1.7$\pm$0.2 & 22$\pm$1 & 1.0 & -115.5 & 0.58$\pm$0.06 & 0.80$\pm$0.05 & 34.6$\pm$0.8 & 1.2 & -107.3 & 1.6 & 23.0 & 0.51\\
 034205-370322 & 5.5 & -530.3 & 0.0656$\pm$0.0002 & 7.63$\pm$0.06 & \textbf{\underline{3.6}} & \textbf{\underline{-629.3}} & 0.0602$\pm$0.0003 & 0.0096$\pm$0.0002 & 29.9$\pm$0.6 & 4.6 & -580.1 & 0.068 & 56.0 & 8.4\\
 034437-382640 & 3.4 & -388.8 & 0.084$\pm$0.002 & 9.4$\pm$0.4 & 3.5 & -389.4 & 0.074$\pm$0.003 & 0.017$\pm$0.002 & 31$\pm$3 & 3.4 & -391.0 & 0.091 & 30.0 & 3.1\\
\hline
\end{tabular}
\label{tab:depolfits}
\end{table*}

The results show that a variety of depolarizing behaviours are present. The sources 032228-384841 (Fig. \ref{fig:032228-384841}), 032006-362044 (Fig. \ref{fig:032006-362044}) and possibly 033829-352818 (Fig. \ref{fig:033829-352818}) show complicated oscillatory depolarization that the models we fit to the data are not capable of reproducing. We do not attempt to model this behaviour here, but note that doing so would require $>2$ polarized components or even Faraday-thick components. The Tribble model is strongly favored for the sources 033147-332912 (Fig. \ref{fig:033147-332912}) \& 033843-352335 (Fig. \ref{fig:033843-352335}), implying Faraday screens structured on angular scales not much smaller than the sources themselves. Finally, the core-dominated source 033653-361606 (Fig. \ref{fig:033653-361606}) cannot be fit with depolarization models; two Faraday thin components are required to fit its observed repolarization.

We generally observe $\sigma_\phi$ in the range 0--25 rad m$^{-2}$ for sources best fit by the Burn model and $|\phi_1-\phi_2|$ in the range 20--50 rad m$^{-2}$ for those best fit by a double component model. While the model parameters are poorly constrained for the Tribble model fits, we note that it requires $\sigma_\phi$ at least as large as the Burn model, but generally larger depending on the factor $s_0/t$. 

\subsection{Summary of results from Section \ref{sec-results}}\label{sec-results-summary}

We have classified the Faraday complexity of the 160 polarized radio sources in our sample, detecting complexity in 19 ($\sim12$\% of polarized sources). The number of complex detections is only mildly dependent on the {\sc rmclean} cutoff and RM synthesis weighting scheme employed (Section \ref{sec-compprev}). We analysed the non-polarimetric properties of the sample and found that the ratio of the number of complex to simple detections was strongly dependent on source brightness (Section \ref{sec-observables-bright}). The spectral indices of complex sources span approximately the same range as the simple and unpolarized sources. However, 90\% are steeper than $\alpha<-0.5$ (Section \ref{sec-observables-SI}). The complex sources are not uniquely associated with a specific morphological type. However, $\sim3/4$ of complex sources are partially resolved on 15" scales, compared with only 44\% and 34\% of the simple and unpolarized sources respectively (Section \ref{sec-smorph}). In terms of multiwavelength counterparts, we detect more counterparts for complex sources than for simple or unpolarized sources and find that these counterparts are more often located near to, rather than cospatial with, the radio sources themselves when compared with simple and unpolarized sources (Section \ref{sec-fcvsfluxdensb}). We were unable to find any evidence that the redshifts of complex sources differ from that of simple or unpolarized sources (Section \ref{redshiftdata}). In Section \ref{sec-srcdistinfield}, we showed that the complex sources in the field are clustered on scales of 0--0.6$^\circ$ at $>99$\% confidence. For the polarimetric source properties, we found that the Faraday depths of simple sources do not significantly exceed that expected on the basis of estimates of the mean contributions from the sources themselves and the Galactic ISM. While most complex sources are similar in this regard, some complex sources do show a significant enhancement in $\phi$ (Section \ref{sec-observables-FDcomparo}). With one exception, the complex sources show net depolarization. This depolarization behaviour is often well characterized by either Burn, Tribble or 2-component depolarization models, but several sources show oscillatory depolarization which would require more sophisticated modelling (Section \ref{sec-observables-depolrepol}). 

\subsection{Tabulated quantities for the polarized sources in our sample}\label{sec-maintable}

The analysis presented in Section \ref{sec-results} generated the following quantities that we record (for the polarized sources only) in Table \ref{tab:allsourcedata}: 
\\
\\
 \emph{Column 1:} Source name\\
 \emph{Columns 2--3:} Source position in J2000 coordinates\\
 \emph{Column 4--5:} Radial and azimuthal position of the source relative to the nearest mosaic point phase centre\\
 \emph{Column 6:} Resolved or unresolved at the 90"$\times$45" resolution of spectropolarimetric images?\\
 \emph{Column 7:} $\lambda_0$ to 3 sig. figs.\\
 \emph{Columns 8--9:} Value of $I_{model}$ evaluated at $\lambda_0$ and its associated error\\
 \emph{Columns 10--11:} Value of $\alpha$ evaluated at ${\lambda_0}$ and its associated error \\
 \emph{Column 12:} Complexity categorization of the source\\
 \emph{Column 13:} The highest {\sc rmclean} cutoff GES level at which the source appears complex\\
 \emph{Column 14:} The weighting scheme under which a complex source attains it highest value of $\sigma_\phi$\\
 \emph{Columns 15:} Calculated value of $\sigma_\phi$\\
 \emph{Column 16:} The {\sc rmclean} cutoff value to 3 sig. figs.\\
 \emph{Columns 17--18:} The amplitude of the FDS at $\phi_{peak}$ and its associated error\\
 \emph{Column 19:} The fractional contribution of off-peak {\sc rmclean} components --- i.e. those {\sc rmclean} components found in $\phi$ bins apart from that in which the majority of components were found --- to the total polarized flux\\
 \emph{Columns 20--21:} The Faraday depth at which the FDS is maximum, and its associated error\\
 \emph{Column 22:} The Stokes $I$ morphology of the radio source at 15" resolution\\
 \emph{Columns 23--26:} The counterpart status of a source in IR, optical, UV and X-rays\\
 \emph{Columns 27:} Crossmatched redshift\\
\\

\section{Discussion}\label{sec-discussion} 

\subsection{Prevalence of Faraday complex objects}\label{sec-fcvsfluxdens}

In Section \ref{sec-observables-bright}, we showed that the ratio of complex to simple source detections depends strongly on source brightness. The cause of this effect has significant implications for any attempt to estimate the underlying prevalence of Faraday complexity in the radio source population. 

We argue that the complexity of bright sources is not caused by polarization leakage in Appendix \ref{sec-discussion-soundness}. The transition from predominantly complex to simple source detections must therefore either reflect a genuine change in Faraday complexity with source brightness (which would be consistent with previous claims of anti-correlations between 1.4 GHz fractional polarization and total Stokes I intensity made by authors such as Mesa et al. 2002, Tucci et al. 2004, Subrahmanyan et al. 2010, Stil et al. 2014 \& Banfield et al. 2011, 2014), or be caused by the effect that signal-to-noise (S/N) has on detection likelihood. To distinguish between these possibilities, we have conducted an experiment to observe how decreasing the S/N level of a subset of our sample affected their Faraday classification. For this subset we selected sources with $I_{\lambda_{0}}>160$ mJy yielding 26 objects: 10 simple, 10 complex, and 6 unpolarized. We call this our `Bright Source Sample' (BSS) while sources with $I_{\lambda_{0}}<160$ mJy fall in the `Faint Source Sample' (FSS). We decreased the S/N of the BSS sources by adding Gaussian noise to the channelized ($q$,$u$) data in quadrature, then re-scaling the flux densities such that the measured, band-averaged value of $\sigma_{qu}$ was the same after the procedure as it had been prior. This procedure results in the polarized signal being decreased relative to our band-averaged observational sensitivity, so we refer to it henceforth as `dimming'. 

We split the range of $I_{\lambda_{0}}$ spanned by the FSS into 20 bins of equal logarithmic width. For each bin, we dimmed each of the 26 BSS sources such that their rescaled Stokes $I$ flux fell into the target bin. We repeated the entire procedure 10 times, resulting in 26 sources $\times$ 20 flux intervals $\times$ 10 iterations $= 5200$ total simulated sources spanning the FSS integrated flux range. Each of these sources was analyzed and classified using RM synthesis, {\sc rmclean} and $\sigma_\phi$. The result is an empirically-derived probability distribution for the Faraday classification of a source population as a function of $I_{\lambda_{0}}$, under the assumption that the population consists entirely of fainter versions of the BSS sample. 

\begin{figure}
\includegraphics[width=0.5\textwidth]{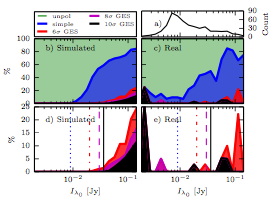}
\caption{Results of the source dimming experiment described in Section \ref{sec-fcvsfluxdens}. The proportion of objects that were classified as belonging to a given Faraday complexity category is plotted against integrated flux density. Plots in the same rows share y-axes, and plots in the same columns share $x$-axes. The color conventions are: $10\sigma$ GES complex detections (black), $8\sigma$ GES complex detections (magenta), $6\sigma$ GES complex detections (red), or Faraday simple (blue) and unpolarized (green). Panels \emph{b)} and \emph{c)} show the proportional breakdown into assigned Faraday classification of the simulated and real source populations respectively. Panels \emph{d)} and \emph{e)} show the proportional contribution of the various complex source categorizations to the polarized source sample in the simulated and real samples respectively. In both of these panels the vertical black solid lines, magenta dashed lines, red dot-dashed lines and blue dotted lines indicate the lowest integrated flux at which $10\sigma$ GES complex sources, $8\sigma$ GES complex sources, $6\sigma$ GES complex sources and simple polarized sources respectively are detected in the simulated source population. The number of sources in each flux bin in the real source population is plotted in panel \emph{a)}.}
\label{fig:FakeSrcs}
\end{figure}

The results are presented in Figure \ref{fig:FakeSrcs}. For the simulated sources (panel b), the stacked area plot demonstrates that decreasing S/N causes the proportion of both polarized and complex detections to drop. Though expected, this S/N effect has important implications for our analysis: While just under half the BSS sources are $10 \sigma$ GES complex detections, they are only detected 10\% of the time or less when dimmed below $\sim 100$ mJy, and go almost completely undetected below $\sim 40$ mJy. Only $\sim7\%$ \& $\sim20\%$ of our total sample sits above these two fluxes respectively. We therefore have insufficient S/N to detect complexity in up to 80\% of our 563 sample objects, assuming it is present at the level typical of BSS sources. 

We used an exact multinomial test to evaluate whether the Faraday classifications in our real data (i.e. the FSS) behaved in a manner consistent with the S/N effect observed in the simulated data. We defined the multinomial categories to be the `simple', `$6\sigma$ GES', `$8\sigma$ GES' \& `$10\sigma$ GES' Faraday classifications, and the probability mass function over these categories to be the proportion of the total source count in each flux bin falling in each category in the simulated detection data. Whenever this probability was zero, we manually assigned a probability of $1/260$ --- i.e. the inverse of the number of simulated Faraday classifications of dimmed sources in any given flux bin. We find that for flux bins where $I_{\lambda_{0}}>25$ mJy, the multinomial p-values fall in the range 0.12--1, meaning the drop off in complex source detections in the FSS is statistically indistinguishable from that caused by decreasing S/N in the simulated data. Given this fact and that the simulations predict a strong S/N dependence, it is likely that our detection of complex sources is S/N-limited. It follows that complex sources may be significantly more prevalent than the $\sim12\%$ of polarized sources that we have been able to classify as such.

Conversely, differences between the FSS and simulated data are apparent for $I_{\lambda_{0}}<25$ mJy. In panels d) \& e) of Figure \ref{fig:FakeSrcs}, we plot the percentage contribution of the complex Faraday categories to the total polarized source count for the simulated sample and FSS (respectively). Vertical lines on these plots (see figure caption) indicate lower $I_{\lambda_{0}}$ detection limits in the simulated data of 35, 30, 20 and 9 mJy for the $10\sigma$, $8\sigma$, $6\sigma$ GES complex and simple sources respectively. Yet in the FSS (panel e), $10\sigma$ GES complex detections occur at 20 mJy \& 2 mJy in spite of these limits, as does the $8\sigma$ GES detection at 7 mJy. We obtain p-values in these bins of 0.03, $2\times10^{-7}$, and 0.003, implying the counts are inconsistent with a purely S/N effect. This may point to a change in the nature of complexity of fainter sources, but sensitive observations would be required to confirm this.

\subsection{The physical origin of Faraday complexity}\label{sec-screendiscussion}

In the remaining sections we discuss the possible physical origins of Faraday complexity in our sample. We begin by pointing out the general implications of our results for the source(s) of complexity in our sample, which motivates the deeper discussion that follows.

Our most general result is that 18 of our 19 complex sources show depolarization (Section \ref{sec-observables-depolrepol}). Most of these sources show smooth depolarization that is broadly consistent with depolarization by Burn (1966) \& Tribble (1991)-type foreground screens (Section \ref{sec-observables-depolrepol}). In Section \ref{sec-srcdistinfield} we found that complex sources were spatially clustered on 0--0.6$^\circ$ scales. There are three possible explanations for this result, namely that: 1) sources with internally generated complexity are physically clustered, 2) sources are physically clustered within a medium that induces complexity, or 3) complexity is induced in otherwise unremarkable polarized background sources by a physically unrelated, spatially structured foreground screen. We suggest that our failure to uncover strong associations between complexity and source characteristics such as specific morphological type (Section \ref{sec-smorph}), enhancements in $\phi$ (Section \ref{sec-observables-FDcomparo}), redshift (Section \ref{redshiftdata}) or multiwavelength counterparts (Section \ref{sec-fcvsfluxdensb}) is most consistent with sources not being closely physically related to complexity-inducing Faraday screens, thus favouring interpretations b) \& c) over a) above. This idea is further supported by the fact that most of our complex detections are bright (Section \ref{sec-observables-bright}), $3/4$ are resolved (Section \ref{sec-smorph}), and 90\% have $\alpha<-0.5$ (Section \ref{sec-observables-SI}) --- all indications of physically extended and / or nearby sources -- since sources with larger angular diameter will intercept more depolarizing cells in remote Faraday screens. When considered together, we suggest that this evidence points to the predominant source of complexity in our sample being depolarization by large scale foreground screens that are physically unrelated and spatially remote from the sources themselves. In the following two sections we examine the leading candidates for screens of this type --- the Galactic interstellar medium (ISM) and intracluster medium (ICM) in galaxy clusters --- before going on to consider other possible complexity-inducing mechanisms.

\subsubsection{Galactic foregrounds}\label{sec-screendiscussion-galfore}

The Galactic ISM is thought to generate the majority of net Faraday rotation experienced by extragalactic sources at $\sim1.4$ GHz (e.g.  Oppermann  et  al.  2015). If this material possesses Faraday structure on sufficiently small scales, it could also act as a depolarizing foreground screen. 

First we test whether pervasive, turbulent structure in the Galactic ISM could form such a screen by calculating the second-order, one-dimensional structure function of $\phi_{peak}$ (e.g. Haverkorn et al. 2006):

 \begin{equation}
\text{SF}_{\theta} = 2\sigma_{\phi,ISM}^2(\theta) = \langle [ \phi_{peak}(\theta) - \phi_{peak}(\theta+\delta\theta)]^2 \rangle 
\label{eq:strucfunc}
 \end{equation}
 
where the variance associated with measurement uncertainty was removed following Appendix A of Haverkorn et al. (2004).

\begin{figure}
\includegraphics[width=0.475\textwidth]{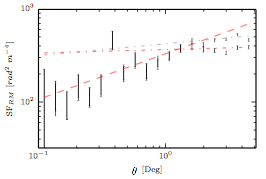}
\caption{Second order structure function for both $\phi_{peak}$ (grey error bars) and $\phi_{res,peak}$ (black error bars) in the field (i.e. $\phi_{peak}$ both before and after Galactic background subtraction; see Section \ref{sec-observables-FDcomparo}). The best fit power law model for $\theta < 1.5^\circ$ is shown as a red dashed line, while the fits for data at $\theta > 1.5^\circ$ both before and after Galactic background subtraction are shown as grey and red dot-dashed lines respectively.}
\label{fig:STfuncFTBFzoom}
\end{figure}

$\text{SF}_{\theta}$ for the field is shown in Figure \ref{fig:STfuncFTBFzoom}. We fitted separate power law models to the data above and below the break apparent at $\theta \approx 1.5^\circ$ as described in the figure caption, but data at $\theta > 1.5^\circ$ are not relevant here so we discuss them no further. Twice the variance of the astronomical signal at $\theta=1^\circ$ is $\sim 300$ rad$^2$m$^{-4}$. Subtracting the estimated intrinsic source contribution of $49$ rad$^2$m$^{-4}$ (Schnitzeler 2010) provides an upper limit to the Galactic contribution of $2\sigma_{\phi,ISM}^2=202$ rad$^2$m$^{-4}$. The best-fit slope to the data at $\theta < 1.5^\circ$ is $0.49 \pm 0.1$. Assuming this slope characterizes the turbulent cascade to scales less than the angular diameter of a typical source, we can extrapolate $\sigma_{\phi,ISM}^2$ down using:\\

 \begin{equation}
2\sigma_{\phi,ISM}^2(\theta) = 202 \times 10^{0.49\log \theta} ~\text{rad}^2 \text{m}^{-4}
\label{eq:extrap_variance}
 \end{equation}
 
The depolarization behaviour can be calculated if the scale size of the turbulent cells is known. Our results in Section \ref{sec-observables-depolrepol} show that a Burn depolarization model (Eqn. \ref{eqn-BurnDepol}) describes $p(\lambda^2)$ well for a number of our complex sources. Burn depolarization requires $\geq10$ turbulent cells across a source (Tribble 1991), or turbulent cell scales of 1.5" -- 30" for sources in our sample. From Eqn. \ref{eq:extrap_variance}, $\sigma_{\phi,ISM} \leq$ 1 -- 3 rad m$^{-2}$ at these scales. Given this, and considering the $\sigma_{\phi}$ values required to fit the depolarization behaviour of complex sources in our sample (Section \ref{sec-observables-depolrepol}, Table \ref{tab:depolfits}), enhancements of $>3\sigma_{\phi,ISM}$ in the Galactic screen are required. At most a few variations of this magnitude would be expected to randomly coincide with the 160 sight-lines towards our polarized objects, meaning a pervasive Burn screen is unlikely to be responsible for the complexity we observe. Relaxing the Burn criterion of $>10$ depolarizing cells across the source renders the pervasive screen scenario even less plausible, because the required value of $\sigma_{\phi,ISM}$ generally increases more rapidly than $\text{SF}_{\theta}$ for Tribble-type depolarization. 

\begin{figure}
\includegraphics[width=0.5\textwidth]{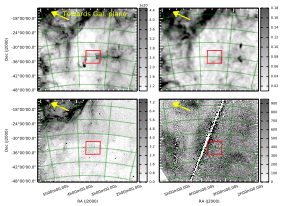}
\caption{The footprint of the mosaic (red rectangle) overlaid on images of (clockwise from top left) GASS H I column density (cm$^{-2})$, Planck spectral brightness (mJy sr$^{-1}$), ROSAT X-rays (counts s$^{-1}$ arcmin$^{-1}$ ) and WHAM + SHASSA H$\alpha$ photon flux (R). The yellow arrows indicate the direction to the Galactic plane along a line of constant Galactic longitude. The mosaic sits at the high latitude limits of the Galactic supershell GSH 238+00+09. Features probably associated with this supershell can be seen in both emission and absorption in the maps, including a prominent `Y'-shaped feature seen in emission in the H I \& Planck images and absorption in X-rays.}
\label{fig:multiWave}
\end{figure}

\begin{figure}
\includegraphics[width=0.425\textwidth]{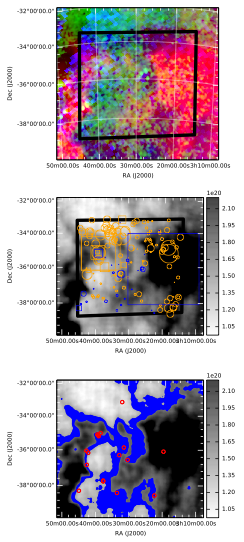}
\caption{\emph{Top:} 3-color map of ISM structure in the field. The black line delimits the mosaic field edge, while red, green and blue colors correspond to GASS H I column density [cm$^{-2}$], WHAM + SHASSA H$\alpha$ photon flux [R] \& ROSAT 1/4 keV X-ray [counts s$^{-1}$ arcmin$^{-1}$] respectively (see main text). \emph{Middle:} GASS H I column density with Faraday depth / complexity information derived from our work overlaid. Blue markers represent negative RMs, yellow markers represent positive RMs, square markers indicate complex sources while circular markers indicate simple sources. The diameter of the markers is proportional to $|\phi_{peak}|$. \emph{Bottom:} H I column density map, blue-filled between contours at $1.4\times10^{20}$ cm$^{-2}$ and $1.65\times10^{20}$ cm$^{-2}$. The positions of depolarizing complex sources are indicated with red circular plot markers.}
\label{fig:Field}
\end{figure}

The remaining possibility is that localized ISM structures induce Faraday complexity. The mosaic field spans the Galactic coordinates $-58^\circ < b < -52^\circ$, $233^\circ  < l < 243^\circ$. This is a sight line across the long axis of the Local Bubble, through a tunnel connecting the Local Bubble with the supershell GSH 238+00+09 (Lallement et al. 2003, Heiles 1998), then out into the lower Galactic halo. ISM structures along this LOS are shown in Figure \ref{fig:multiWave} as revealed in H I column density (GASS; McClure-Griffiths et al. 2009), mm (Planck all sky dust model; Abergel et al. 2013), X-rays (ROSAT 1/4 keV soft X-rays; Snowden et al. 1995) and H$\alpha$ photon flux (WHAM+SHASSA; Reynolds et al. 1998, Gaustad et al. 2001, Finkbeiner 2003). Structure is present at each of these wavelengths, some of which is probably associated with high Galactic latitude regions of GSH 238+00+09. In Figure \ref{fig:Field} (top panel) we present a 3-color multiwavelength map of the most prominent ISM structure in the mosaic field: X-ray emission (blue), H I column density (red) \& H$\alpha$ emission (green). The X-ray emission suffuses the field, increasing in strength towards the Northeast, while knotty tendrils of H I \& H$\alpha$ appear to be anti-correlated in the regions they occupy --- most notably where a wedge of H$\alpha$ running North-South through the middle of the field divides two regions of higher H I column density to the East and West. 

Galactic H I and H$\alpha$ have both previously been linked to RM structure (e.g. Leahy 1987, Heiles \& Haverkorn 2012, Foster et al. 2013). Our data also appear to show some relationship between $\phi_{peak}$ and H I / H$\alpha$: When $\phi_{peak}$ is plotted on the GASS H I data (Figure \ref{fig:Field}, middle panel), sources with large $|\phi_{peak}|$ tend to be found in the immediate vicinity of large H I column densities, while those with small $|\phi_{peak}|$ tend to be found in low H I column density / H$\alpha$ dominated regions (by comparison with Figure \ref{fig:Field}, top panel). In addition, the complex sources with enhanced $\phi_{peak,res}$ (Section \ref{sec-observables-FDcomparo}) each lie on or very close to regions of high H I column density. 

\begin{figure}
\includegraphics[width=0.425\textwidth]{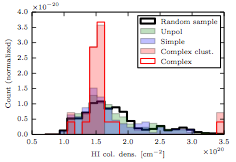}
\caption{Normalized histograms of H I column densities at the location of randomly sampled (black), unpolarized (green), simple (blue) \& complex (red line, no fill/unshaded) sources, as well as complex cluster centroids (red filled/shaded; sample described in text) for our sample. Note the change in color convention from previous plots. All but four (three) of the complex sources (complex cluster centroids) sit in the range $1.4\times10^{20}$ cm$^{-2}$ -- $1.65\times10^{20}$ cm$^{-2}$, while the distribution of randomly sampled, simple \& unpolarized sources are much broader.}
\label{fig:compOnH IHist}
\end{figure}

It therefore seems likely that Faraday rotation of extragalactic sources is related in some way to the presence or absence of Galactic H I / H$\alpha$ structures along the LOS. If this is the case, these same structures might also act as complexity-inducing Faraday screens. We tested this idea by extracting the H I column density at 1000 randomly-selected positions in the field, and at the locations of unpolarized, simple \& complex sources. We plot histograms of the resulting values in Figure \ref{fig:compOnH IHist}. While the random (black), unpolarized (green) \& simple (blue) distributions are all visually similar, the complex distribution (red line, no fill/unshaded) is clearly narrower and more peaked. We quantified the differences between the distributions using 2-sample K-S tests, the results of which are presented in Table \ref{tab:KStestH I}. We obtain D (i.e. the maximum difference between the two normalized empirical cumulative distribution functions) \& p (i.e. the probability of obtaining D given the null hypothesis) values of 0.42 \& $2\times10^{-3}$ respectively for comparison of the complex and random distributions and a similar result for comparison of the complex and unpolarized distributions. This is a large and statistically significant difference. No such difference is evident for the unpolarized vs. random comparison (D$=0.04$ \& $p=0.62$). For the simple vs. complex, the D value is reasonably large at 0.32, while for the simple vs. random it is quite small at (D$=0.12$). The statistical significance is weak in both cases however, with a $p$ value of $\sim0.04$. These results imply that Faraday complexity is preferentially associated with a particular range in H I column density. The results for the simple sources are curious --- we interpret these later in this section.

Before proceeding further however, we address two issues that could generate a false result. First, some complex sources are clustered on scales below the $16'$ resolution of the GASS image. While complex sources will obviously cluster around H I structures if these structures are in fact responsible for inducing complexity, the clustering will cause the D \& p-values to be in error if this is not the case. To address this, we redid the analysis after having replaced the individual column densities of sources located within 16' of one another with a single value for the cluster as a whole (extracted at the positional centroid of the source locations). The H I column densities of isolated sources were extracted as before. The histogram of the resulting values is plotted on Figure \ref{fig:compOnH IHist} as a red step plot (filled/shaded), which clearly shows the same peaked shape as before. We obtain new K-S test D \& p values of 0.43 \& $1.6\times10^{-2}$ (listed in Table \ref{tab:KStestH I} as `Comp. Clust.') --- i.e. the result is confirmed at $\sim$ 99\% confidence, with the D value even increasing slightly. Second, we verified that the results were not generated by chance alignments with H I structure. We took the positions of the random points and complex sources, shifted them by up to 10 degrees in random directions, then redid our analysis. After 50 such trials, we obtained a mean K-S test D value of 0.28 with a corresponding p-value of 0.32, while the maximum D value obtained was 0.35 with a corresponding p-value of 0.05. These D values (p-values) are substantially lower (higher) than the results we obtain for the unshifted data.

\begin{table}
\footnotesize
\tablefontsize{\footnotesize}
\caption{2-sample K-S test statistics and associated p-values comparing the H I column densities of different source categories, as referred to in the text. The complex category excludes data for the sole repolarizing source in the complex sample. The `Comp. Clust.' category refers to the H I column densities extracted from the positional centroid of source clusters lying within $16'$ of one another, or the position of complex sources for unclustered sources. The simple and unpolarized samples comprise the H I column densities extracted at the positions of the simple and unpolarized source sample used throughout the paper. The random sample is explained in the main body of the text. Bolded table entries highlight entries with D values $> 0.35$ \& p-values $< 0.02$.}
\tabcolsep=0.09cm
\begin{tabular}{| l | l l  | l  l  | l  l  |}
\hline
  & \multicolumn{2}{  c | }{Random} & \multicolumn{2}{ c| }{Unpolarized} & \multicolumn{2}{ |c| }{Simple} \\
 \hline
  & \multicolumn{1}{  c  }{D} & \multicolumn{1}{ c |}{ p } & \multicolumn{1}{ c }{D} & \multicolumn{1}{ c }{p} & \multicolumn{1}{| c }{ D} & \multicolumn{1}{ c | }{p} \\
 \hline
Complex & \textbf{0.42} & \textbf{0.002} & \textbf{0.43} & \textbf{0.002} & 0.32 & 0.04  \\
Comp. Clust. & \textbf{0.43} & \textbf{0.016} & \textbf{0.44} & \textbf{0.015} & 0.33 & 0.14 \\ 
Random & & & 0.04 & 0.62  & 0.12 & $0.04$  \\ 
Unpolarized & & & & & 0.13 & 0.06 \\  
\hline
\end{tabular}
\label{tab:KStestH I}
\end{table}

Thus, a relationship between H I column density and complexity in our field seems likely. However, rather than being a direct causal relationship, we suggest that the $1.4\times10^{20}$ -- $1.65\times10^{20} ~\text{cm}^{-2}$ H I column density range is characteristic of / acts as a proxy for regions in which complexity-inducing Faraday screens are present. To show how this range relates to the broader H I structure in the field, in Figure \ref{fig:Field} (bottom panel) we display the GASS H I column density map, then overplot contours at H I column densities of $1.4\times10^{20}$ \& $1.65\times10^{20} ~\text{cm}^{-2}$ with a blue fill applied between these values. We then overplot the positions of complex sources. While the blue-filled region occupies $<35\%$ of the map area, 15 of 19 (79\%) of the complex sources fall within it. A further two sources fall just outside it, both in terms of spatial proximity on the map and the value of H I column density at the source position. We propose the following speculative scenario: The blue-filled region in Figure \ref{fig:Field} (bottom panel) generally traces the edges of the knotty H I tendrils, and transition regions between H I and H$\alpha$-dominated regions of the mosaic field (Fig. \ref{fig:Field} top panel). We suppose that the soft X-ray emission apparent in the field (possibly originating from the supershell GSH 238+00+09) is ionizing the H I material and forming turbulent structure on small scales along its boundaries (e.g. Leahy 1987, Hennebelle \& Audit 2007, Gritschneder et al. 2009, Ntormousi et al. 2011), or possibly a warm partially ionized phase of the ISM (e.g. Heiles \& Haverkorn 2012, Foster et al. 2013). We can make a rough estimate of the magnitude of $\phi$ that could be produced in this way following Heiles \& Haverkorn (2012):

\begin{eqnarray}
\phi &=& 0.81n_eB_{||}L \nonumber \\
&=& 26N_{e,20}B_{||}  \nonumber \\
&\approx& 26x_eN_{H I,20}B_{||} \\
\label{eq:MWscreenRM}  \nonumber
\end{eqnarray}

where $n_e$ is electron density in cm$^{-3}$, $B_{||}$ is the LOS field strength in $\mu$G, $N_{e,20}$ is the electron column density in units of $10^{20}$cm$^{-2}$, $N_{H I,20}$ is the H I column density in units of $10^{20}$cm$^{-2}$ and $x_e$ is the ionization fraction of the H I cloud. The observed H I column density shows a difference in $N_{H I,20}$ between enhanced and sparse regions of $\sim1$, and we assume $B_{||} = 6~\mu$G. For an $x_e$ typical of the cold neutral medium (CNM) of 0.01, the region is only capable of generating $\phi\sim1$ rad m$^{-2}$. However, for an almost fully ionized fraction of $x_e=0.9$, $\phi\sim150$ rad m$^{-2}$. It is therefore plausible that magnetized turbulence in the vicinity of the H I clouds or even smooth gradients in ionization fraction or $B_{||}$ could readily produce the $\phi$ structure required to cause complexity we observe. 

Based on this physical picture, we now suggest reasons why we obtained only weak statistical significance for the simple vs. random and simple vs. complex HI column density comparison (Table \ref{tab:KStestH I}). One explanation might be that all of the simple sources in our field in the vicinity of ionization fronts are in fact complex, and with better S/N observations would be detected as such (i.e. see Section \ref{sec-fcvsfluxdens}). The newly obtained classification would then reveal no differences in the simple vs. unpolarized vs. random HI column densities, while the complex sample would further distinguish itself on this basis. Another possibility is that small scale turbulent structure in the ionization front means that sources intercepting these regions are only statistically more likely to become complex --- some simple sources may intercept gaps or relatively uniform regions in the foreground screen and remain simple. In this scenario, the spatial distribution of complex sources would closely trace turbulent ionization fronts, while simple sources would be randomly distributed in space apart from a (possibly small) relative under density in the vicinity of the ionization front structure. In either case the simple sources will differ from both the random and complex distributions, but more subtly than the random and complex distributions differ from one another.

We caution that these results presented in this section should be considered as being indicative at this stage. Our claim can be directly tested by making use of the high spatial density of polarized sources that will be detected by upcoming spectropolarimetric surveys. By analogy with the rotation measure grids that these surveys will generate and use to study large scale structure in Galactic and other Faraday screens, our findings mean that `depolarization grids' might allow ionization interfaces in the ISM to be studied in unprecedented detail. 

\subsubsection{Magnetized intercluster medium}\label{sec-screendiscussion-cluster}

 Galaxy clusters contain a hot, magnetized intracluster medium (ICM) that can impose Faraday structure on embedded sources. Our mosaic field contains the galaxy-poor Fornax cluster ($z\approx0.004$), a background cluster at $z=0.1$ (Hilker et al. 1999), and possibly a third cluster at $z>0.3$ (Scharf et al. 2005). Based on their redshift and impact parameter, the complex sources 033829-352818 (Fig. \ref{fig:033829-352818}), 033828-352659 (Fig. \ref{fig:033828-352659}), 033848-352215 (Fig. \ref{fig:033848-352215}) \& 033843-352335 (Fig. \ref{fig:033843-352335}) probably reside in the first two clusters, while the source 033754-351735 (Fig. \ref{fig:033754-351735}) has a small projected distance from the clusters but lacks a redshift to confirm membership. Furthermore, 033843- 352335, 033829-352818 \& 033754-351735 are among the sources that we identified as possessing large residual $\phi_{peak}$ in Section \ref{sec-observables-FDcomparo} (with $\phi_{peak}$ values of 59 rad m$^{-2}$, 145 rad m$^{-2}$ \& 165 rad m$^{-2}$ respectively), which is consistent with them being embedded in a Faraday active ICM. 

While there is little information in the literature about the background clusters, we can estimate the Faraday dispersion $\sigma_{\phi,clust}$ generated by the Fornax Cluster ICM using a modified version of the Felten (1996) model. The original model assumes a turbulent magnetized cluster medium with constant field strength, a single characteristic length scale for the turbulent eddies and a $\beta$ profile for electron density. The modification incorporates the effect of non-constant $B$ field strength for which $B \propto n_e^\gamma$ (Dolag et al. 2001). The relevant equation is:

 \begin{equation}
\sigma_{\phi,clust} = \frac{RB_0n_0r_c^{0.5}l^{0.5}}{[1+(r/r_c)^2]^{(6\zeta-1)/4}}\sqrt{\frac{\Gamma(3\zeta-0.5)}{\Gamma(3\zeta)}}
\label{eq:clusterRM}
 \end{equation}

where 

 \begin{equation}
\zeta = \beta(1+\gamma)
\label{eq:clusterRMextra}
 \end{equation}
 
\begin{figure}
\includegraphics[width=0.475\textwidth]{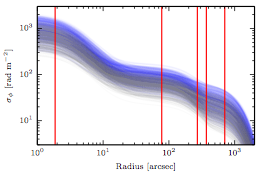}
\caption{Plot of the estimated $\sigma_{\phi,clust}$ that could be generated by the Fornax cluster as a function of cluster radius, generated as described in the main text. The blue lines are for $R=624$ (source behind cluster) and the grey lines are for $R=441$ (source embedded in cluster). The vertical red lines indicate the radial angular separation of sources from the core of the cD galaxy NGC 1399.}
\label{fig:fornClusterSigmaFD}
\end{figure}

In these equations, $R$ is a factor that depends on the location of the radio source relative to the Faraday rotating plasma (e.g. $R=624$ for sources behind the cluster; $R=441$ for centrally embedded sources), $B$ is the mean magnetic field strength in $\mu$G, $n_0$ is the electron density at the cluster centre in cm$^{-3}$, $l$ is an effective length scale of a cell in the turbulent ICM, $r_c$ is the scale radius of the cluster and $r$ is the radius of observation (all in kpc), $\Gamma$ is the Gamma function, and $\beta$ is a constant that sets the shape of the radial profile of $n_e$. We adopt observationally-derived values for $n_0$, $\beta$, and $r_c$ of the central cooling core, galactic and cluster contributions (Paolillo et al. 2002), summing the contributions from each. We then evaluate the model 500 times for $R = 441$, $B_0$ selected randomly in the interval [0.5, 2] $\mu$G, $l$ selected randomly in the interval [2, 10]  kpc (e.g. Murgia et al. 2004 and refs therein), and $\gamma$ selected randomly in the interval [0.5, 1.1] (Dolag et al. 2001), and then repeat this for $R = 629$. The $\sigma_\phi$ profiles in Figure \ref{fig:fornClusterSigmaFD} are the result.

Based on these calculations, $\sigma_{\phi,clust}$ is sufficient or in excess of that required to generate the complexity observed in the relevant sources, assuming the inner scale of turbulent cells in the ICM is $\ll$ the linear extent of embedded sources (or the apparent linear extent of background sources). At the distance of the Fornax Cluster the angular scale is 0.083 kpc / arcsec, which corresponds to a linear extent of embedded sources of $\sim5$ kpc. Estimates for the inner scale of turbulent structure in a typical ICM range from about 0.1 to a few kpc (e.g. Vacca et al. 2012). On this basis, some degree of either Burn or Tribble-type depolarization behaviour is likely to occur, which would be consistent with the depolarization behaviour we observe in the cluster-associated sources in Section \ref{sec-observables-depolrepol}. Thus, the ICM of clusters in our field probably contribute to Faraday complexity, but with the current observations we are unable to determine the relative extent of the contribution made by this mechanism versus that which was proposed in Section \ref{sec-screendiscussion-galfore}. Finally, we note that depolarizing Faraday screens might also occur at interfaces between the source and the cluster ICM (e.g. Rudnick \& Blundell 2003, Guidetti et al. 2011). For our purposes, we consider this to be an effect that is `intrinsic' to the radio source, which we discuss in Section \ref{sec-screendiscussion-intrinsic}.

 \subsubsection{Intervening galaxies}\label{sec-screendiscussion-intervenegal}

The disordered magnetic field components of normal galaxies can act as foreground Faraday screens (e.g. Schulman \& Fomalont 1992, Kronberg et al. 1992, Bernet et al. 2008 \& 2012, Farnes et al. 2014b), potentially generating complexity in background sources. Typical values of $\sigma_{\phi,gal}$ inferred from sight lines through these objects are on the order of 10s -- 100s rad m$^{-2}$; this is sufficient to produce the complexity we observe, but only if the expected number of intervenors along the LOS is also sufficient. Using the prevalence of strong Mg II absorption systems as a proxy for the expected number of Faraday active intervening galaxies along the LOS (Bernet et al. 2010), the probability of having $>0$ intervenors out to redshift $z$ is described by Poisson statistics: 
 
 \begin{eqnarray}
P_{int}=1-e^{-\nu}
\end{eqnarray}

where

\begin{eqnarray}
\nu=\int_0^{z_{src}} \frac{dN}{dz}dz
\end{eqnarray}

$dN/dz$ is the expected number of intervenors along the LOS. For strong Mg II systems, this is given by Nestor et al. (2005):

\begin{multline}
 \frac{dN}{dz} = 1.001(1+z)^{0.226} \times \\
\bigg[\text{exp}\bigg(\frac{-0.3}{0.442(1+z)^{0.634}}\bigg) - \text{exp}\bigg(\frac{-6.0}{0.442(1+z)^{0.634}}\bigg)\bigg]
\end{multline}

We have 21 polarized sources with redshifts. Calculating $P_{int}$ towards each of these sources and summing the results gives the expected number of sources having intervenors along the LOS. The result is that while 8 of the polarized objects with redshifts are classified as complex, only $\sim2.5$ intervenors are expected --- insufficient to explain our results. We conclude that this mechanism is probably not an important contributor to complexity in our sample. 

\subsubsection{Cosmic filamentary structure}

Recent studies have argued that Faraday depths through cosmic filaments should be of order 1 rad m$^{-2}$, which, when integrated over redshift, can generate $\sigma_{\phi,filament}$ that saturates at 7--8 rad m$^{-2}$ (Akahori \& Ryu 2011). This is of the same order as the values of $\sigma_\phi$ required to fit the depolarization behaviour of our sources in Section \ref{sec-observables-depolrepol}. However, Akahori \& Ryu (2011) \& Akahori et al. (2014) show that only at $z \gg1$ does this variance build up sufficiently at the $\sim$ few $\times10$ arcseconds scales required to depolarize our sources given their angular diameter. Since the bulk of our sources appear to be at significantly lower redshift than this (Table \ref{tab:rs}), we discount cosmic filamentary structure as a cause of complexity in our sample. 

\subsubsection{Intrinsic to source}\label{sec-screendiscussion-intrinsic}

We consider an intrinsic Faraday screen to be a magnetized plasma that either interacts directly with the synchrotron emitting electrons, or is part of the host galaxy ISM. VLBI observations routinely observe intrinsic screens (e.g. G\'{o}mez et al. 2011, Zavala \& Taylor 2004), and such structure has likely been observed using non-VLBI spectropolarimetry by authors such as Johnston et al. (2010), O'Sullivan et al. (2012, 2013) and Farnes et al. (2014a).

The available evidence suggests that the complexity in 033653-361606 (Fig. \ref{fig:033653-361606}) is likely to be internally generated. Its inverted spectrum ($\alpha=+0.26$) implies it is core-dominated, while its repolarization must be generated either by multiple Faraday-thin components or a Faraday thick component. Whilst the source appears to show faint resolved components surrounding a bright unresolved core in our source finding image, the resolved components contribute $< 0.5\%$ of the total flux at 1.7 GHz and thus cannot generate the observed degree of fractional polarization. We are unaware of other structures along the LOS that could add additional polarized components, such as radio relics from the clusters in the region (e.g. Ferrari et al. 2008). Our best fit model of $p(\lambda^2)$ for this source (Section \ref{sec-observables-depolrepol}; Table \ref{tab:depolfits}) requires two polarized components with a difference in $\phi$ of $\sim64$ rad m$^{-2}$. Since differences of this magnitude are routinely identified in high resolution observations of AGN (e.g. Algaba 2013), we suggest that the complexity in this object is generated by interference of polarized components located in the compact inner regions of the AGN itself. Similar arguments have been made before by Law et al. (2011) and O'Sullivan et al. (2012), while Farnes et al. (2014a) have argued that sources of this type might be commonplace.

Our data are poorly suited to studying the contribution of internal depolarization to the Faraday complexity of compact sources. This will require object samples of greater size plus extensive supporting observations at other wavelengths (e.g. O'Sullivan et al. 2015), which must be addressed in future work.
 
\subsection{Summary of Section \ref{sec-discussion}} \label{sec-discussion-summary}

We have discussed the prevalence of Faraday complex sources and the physical origin of the observed complexity. In Section \ref{sec-fcvsfluxdens}, motivated by our result that the majority of our complex detections were bright sources (Section \ref{sec-observables-bright}), we described an experiment designed to determine whether S/N effects could explain this result. We concluded that this was the case, and that it implied that complex sources may be substantially more common than the 12\% of polarized sources detected as complex in our sample. In addition, we found that several of our faint complex sources must radiate a substantially higher proportion of their emission as complex components (i.e. Faraday depths removed in $\phi$ space from $\phi_{peak}$), otherwise S/N constraints would have caused the sources to appear Faraday simple.

We interpreted our results from Section \ref{sec-results} to suggest that the predominant origin of complexity in our sample is depolarization by large scale, spatially structured foreground screens. Examining specific scenarios, we found that pervasive turbulent structure in the Galactic ISM was insufficient to generate complexity in our sample. However, we found a large, statistically significant tendency for complex sources to lie in regions possessing H I column depths of $1.4\times10^{20}$ -- $1.65\times10^{20} ~\text{cm}^{-2}$. Noting that H I column densities in this range appear to trace interfaces in the ISM between neutral and ionized hydrogen structures, we interpreted this result to mean that H I densities in this range are characteristic of / act as a proxy for complexity inducing regions in our field. We speculated that the complexity inducing screens might be associated with turbulent interfaces between phases of the ISM, smooth RM gradients generated by these same interfaces, or warm partially ionized phases of the ISM. Noting that the Fornax galaxy cluster (and at least one background cluster) lies in our field, we estimated the Faraday dispersion generated by the Fornax ICM, concluding that it probably contributes to the complexity of at least four of our sources. Finally, we examined whether cosmic filamentary structure, intervening galaxies and internal /  intrinsic mechanisms might generate complexity in our sources. We were able to effectively rule out the first two possibilities, while our data are poorly suited to an in-depth analysis of the latter. We were however able to identify one repolarizing source as a probable candidate for internally-generated complexity.
 
\section{Summary \& Conclusions}\label{sec-conclusion}

We have presented a spectropolarimetric analysis of 563 radio sources between 1.3 and 2.0 GHz. We used RM synthesis and the second moment of the {\sc rmclean} component distribution to identify Faraday complexity in these objects. We have also considered whether a range of different radio and multiwavelength source properties correlate with Faraday complexity / simplicity, and the likely nature of the Faraday screens responsible for inducing Faraday complexity. Our principle findings are:

1) Of the 160 polarized sources in our sample, we detect Faraday complexity in 19 (12\%), with the additional polarized emission components contributing an average of 45\% of the total polarized flux. We have shown that S/N effects limit our ability to detect these sources in up to 80\% of our sample, so the true prevalence of complex sources is likely to be significantly higher. 

2) Faraday complexity between 1.3 and 2.0 GHz at $\sim$arcminute resolutions is predominantly associated with bright, partially resolved radio sources possessing steep spectral index. It appears to be associated with an enhancement in $\phi$ in a minority of cases. Burn or Tribble depolarization models generally describe $p(\lambda^2)$ well with Faraday dispersions in the range 0--50 rad m$^{-2}$, as do double Faraday-thin component models in several cases. We find no evidence to suggest that Faraday complexity is strongly tied to redshift, specific source morphologies, or that complex sources are more or less likely than simple sources to have counterparts in specific wavelength bands. However, Faraday complex objects show substantial diversity: One object possesses an inverted spectral index and shows strong repolarization. Oscillatory depolarization (as a function of $\lambda^2$) is found in some sources. A number have multiple resolved peaks in their FDS, and several show enhancements in $\phi$ well beyond that which is easy to explain by average Galactic or intrinsic  contributions. This diversity implies that Faraday complexity is a general phenomenon, possibly generated in diverse cosmic environments.

3) We claim that Galactic Faraday screens may generate a significant amount of the Faraday complexity we observe in our sample (at frequencies of 1.3--2.0 GHz and $\sim$arcminute resolutions). We base this claim on an observed difference (occurring with 99.8\% confidence) between the distributions of Galactic H I column density at the positions of complex sources vs. randomly sampled locations in the field. While we caution that these results should be taken as indicative only at this stage, we proposed that in our field, the H I column density range in which the complex sources are predominantly found mark out regions in which turbulent structures, RM gradients or a partially ionized ISM phase exist at interfaces between neutral and ionized hydrogen structures. These structures then act as a depolarizing Faraday screen for background sources. Galaxy cluster environments also likely contribute to the complexity of several sources in our sample. We demonstrated that the $\phi$ dispersion of the cluster ICM should be sufficient to induce complexity in embedded and background sources if the inner turbulence scale of the ICM is small enough. The inverted spectrum and strong repolarization of one of our complex sources (033653-361606) suggests that its complexity is caused by interference of polarized components in the compact inner regions of this AGN.

4) We have demonstrated that RM synthesis \& {\sc rmclean}, in combination with the novel method of examining the second moment of the {\sc rmclean} component distribution, is a reliable and powerful method for detecting Faraday complexity in survey-type observations.

In future work we intend to robustly test our claim that the Galactic ISM is responsible for generating Faraday complexity in extragalactic radio sources, as well as conducting a search for Faraday complex polarization structure originating in the immediate environments of AGN by observing with broader wavelength coverage at higher frequencies.

\section{Acknowledgments}\label{sec-acknowledgements}

The authors would like to thank the anonymous referee for their careful reading of the manuscript and for the constructive feedback they provided. B.~M.~G. and C.~S.~A.  acknowledge the support of the Australian Research Council (ARC) through grant FL100100114. T.~M.~O.~F. acknowledges support from an ARC Super Science Fellowship. The Dunlap Institute is funded through an endowment established by the David Dunlap family and the University of Toronto. The Australia Telescope Compact Array is part of the Australia Telescope National Facility which is funded by the Commonwealth of Australia for operation as a National Facility managed by CSIRO.


\section{REFERENCES}
Abergel, A., et al. 2014, A\&A, 571, A11\\
Akahori, T., Ryu, D., 2011, ApJ, 738, 134\\
Akahori, T., Gaensler, B. M., Ryu, D., 2014, ApJ, 790, 123\\
Akaike, H., IEEE Transactions on Automatic Control 19 (6) (1974) 716Ð723\\
Algaba, J. C., 2013, MNRAS, 429, 3551\\
Bahcall, J. N., et al. 1991, ApJ, 377, L5\\
Banfield, J. K., George, S. J., Taylor, A. R., et al. 2011, ApJ, 733,69\\
Banfield, J. K., Schnitzeler, D. H. F. M., George, S. J., et al. 2014, MNRAS, 444, 700\\
Beck, R., Gaensler B. M., 2004, New Astron. Rev., 48, 1289\\
Bernet, M. L., Miniati, F., \& Lilly, S. J. 2010, ApJ, 711, 380\\
Bernet, M. L., Miniati, F., \& Lilly, S. J. 2012, ApJ, 761, 144\\
Bernet, M. L., Miniati, F., Lilly, S. J., Kronberg, P. P., \& Dessauges-Zavadsky, M. 2008, Nature, 454, 302\\
Bland-Hawthorn, J., Ekers, R. D., van Breugel, W., Koekemoer, A., Taylor, K., 1995, ApJ, 447, L77\\
Brentjens, M. A., de Bruyn A. G., 2005, A\&A, 441, 1217\\
Briggs, D. S. 1995, BAAS, Vol. 27,1444\\
Brown, S., 2011, Internal POSSUM Report \#9: Assess Complexity of RM Synthesis Spectrum\\
Burn, B. J., 1966, MNRAS, 133, 67\\
Dolag, K., Schindler, S., Govoni, F., Feretti, L., 2001, A\&A, 378, 777\\
Faraday, M. 1844, Experimental Researches in Electricity (London: Taylor), reprinted 1942 (New York: Dover)\\
Farnes, J. S.; Gaensler, B. M.; Carretti, E., 2014a, ApJS, 212, 15\\
Farnes, J. S. , O'Sullivan, S. P., Corrigan, M. E., Gaensler, B. M, 2014b, ApJ, 795, 63\\ 
Farnsworth, D., Rudnick, L., Brown, S., 2011, AJ, 141, 191\\
Feain, I. J. et al., 2009, ApJ, 707, 114\\
Felten, J. E. 1996, in Clusters, Lensing, and the Future of the Universe, ed. V. Trimble, \& A. Reisenegger, ASP Conf. Ser., 88, 271\\
Ferrari C., Govoni F., Schindler S., Bykov A. M., Rephaeli Y., 2008, Space Sci. Rev., 134, 9\\
Feretti, L., Giovannini G., Govoni F., Murgia M., 2012, A\&ARv, 20, 54\\
Finkbeiner, D. P.,  2003, ApJS, 146, 407\\
Foreman-Mackey D., Hogg D. W., Lang D., Goodman J., 2013, PASP,125, 306\\
Foster, T., Kothes, R., Brown, J. C., 2013, ApJ, 773, L11\\
Garrington S. T. Leahy J. P., Conway R. G., Laing R. A., 1988, Nature, 331, 147\\
Gaensler, B. M. 2009, in IAU Symp. Cosmic Magnetic Fields: From Planets, to Stars and Galaxies, ed. K. G. Strassmeier, 259, 645\\ 
Gaensler, B. M., Landecker, T. L., Taylor, A. R., \& POSSUM Collaboration, 2010, BAAS, 42, 515.\\
Gaustad, J. E., McCullough, P. R., Rosing, W., Van Buren, D., 2001, PASP, 113, 1326\\
Goldstein S. J., Jr, Reed J. A., 1984, ApJ, 283, 540\\
G\'{o}mez, J. L., Roca-Sogorb M., Agudo I., Marscher A. P., Jorstad S. G., 2011, ApJ, 733, 11\\
Goodlet, J. A., Kaiser, C. R., Best, P. N., Dennett-Thorpe, J., 2004, MNRAS, 347, 508\\
Goodlet, J. A., Kaiser, C. R., 2005, MNRAS, 359, 1456\\
Gritschneder, M., Naab, T., Walch, S., Bukert, A., Heitsch, F., 2009, ApJL, 694, L26\\
Guidetti D., Laing R. A., Bridle A. H., Parma P., Gregorini L., 2011, MNRAS, 413, 2525\\
Hales, C. A., Gaensler, B. M., Norris, R. P., E. Middelberg, 2012, MNRAS, 424, 2160\\
Hambly, N. C et al. 2001, MNRAS, 326, 1279\\
Hammond, A. M., Robishaw, T., \& Gaensler, B. M. 2012, arXiv:1209.1438\\
Harvey-Smith, L., Madsen, G. J., Gaensler, B. M., 2011, ApJ, 736, 83\\
Haverkorn, M., Gaensler, B. M., McClure-Griffiths, N. M., Dickey, J. M., Green, A. J., 2004, ApJ, 609, 776\\
Haverkorn, M., et al., 2006, ApJ, 637, L33\\
Haverkorn, M., 2014, in ÒMagnetic Fields in Diffuse MediaÓ,  eds E. Gouveia dal Pino and A. Lazarian, Springer, arXiv:1406.0283\\
Heald, G., Braun R., Edmonds R., 2009, A\&A, 503, 409\\
Heiles, C., 1998, ApJ, 498, 689\\
Heiles, C., 2001, in Tetons 4: Galactic Structure, Stars and the Interstellar Medium, ed. by C.E. Woodward, M.D.
Bicay, J.M. Shull. ASP Conference Series, vol. 231 (Astronomical Society of the Pacific, San Francisco,
2001), p. 294. ISBN: 1-58381-064-1\\
Heiles, C., Haverkorn, M., 2012, Space Sci. Rev., 166, 293\\
Hennebelle, P., Audit, E., 2007, A\&A, 465, 431\\
Hilker, M., Infante, L., Vieira, G., Kissler-Patig, M., \& Richtler, T. 1999, A\&AS, 134, 75\\
Hovatta, T., Lister, M. L., Aller, M. F., et al. 2012, AJ, 144, 105\\
Johnston, H. M., Broderick, J. W., Cotter, G., Morganti, R., Hunstead, R.W., 2010, MNRAS, 407, 721\\
Kronberg, P. P., Perry, J. J., Zukowski, E. L. H., 1992, ApJ, 387, 528\\
Laing R. A., 1988, Nature, 331, 149\\
Lallement, R., Welsh, B. Y., Vergely, J. L., Crifo, F., Sfeir, D., 2003, A\&A, 411, 447\\
Law, C. J. et al., 2011, ApJ, 728, 57\\
Leahy, J. P., 1987, MNRAS, 226, 433\\
Macquart, J. P., Ekers, R. D., Feain, I. J., Johnston-Hollitt, M, 2012, ApJ, 750, 139\\
Mauch, T., et al., 2003, MNRAS, 342, 1117\\
McClure-Griffiths, N. M., et al. 2009, ApJS, 181, 398\\
Mesa, D., Baccigalupi, C., De Zotti, G., et al. 2002, A\&A, 396,463\\
Mooley, K. P., Frail, D. A., Ofek, E. O., Miller, N. A., Kulkarni, S. R., Horesh, A., 2013, ApJ, 768, 165\\
Morris, S. L., Weyman, R. J., Savage, B. D., Gilliland, R. L., 1991, ApJ, 377, L21\\
Murgia, M. et al., 2004, A\&A, 424, 429\\
Nestor, D. B., Turnshek, D. A., \& Rao, S. M. 2005, ApJ, 628, 637\\
Ntormousi, E., Burkert, A., Fierlinger, K., Heitsch, F., 2011, ApJ, 173, 13\\
Ofek, Eran O., Frail, Dale A., 2011, ApJ, 737, 45\
Offringa, A. R. et al., 2010, MNRAS, 405, 155\\
Oppermann, N., Junklewitz, H., Greiner, M., et al. 2015, A\&A, 575, A118\\
O'Sullivan, S. P. et al., 2012, MNRAS, 421, 3300\\
O'Sullivan, S. P. et al., 2013, ApJ, 764, 162\\
O'Sullivan, S. P. et al., 2015, arXiv:1504:06679\\
Paolillo, M., Fabbiano, G., Peres, G., Kim, D. W., 2002, ApJ, 565, 883\\
Parejko, J. K., Constantin, A., Vogeley, M. S., Hoyle, F., 2008, ApJ, 135, 10\\
Reynolds, R. J., Tufte, S. L., Haffner, L. M., Jaehnig, K., Percival,  J. W., 1998,  PASA, 15, 14\\
Ripley, B.D., 1976, Journal of Applied Probability, 13, 255\\
Rudnick, L. \& Blundell, K., 2003, ApJ, 588, 143\\
Sault R. J., Teuben P. J., Wright M. C. H., 1995, in R. A. Shaw, H. E. Payne, \& J. J. E. Hayes ed., Astronomical Data Analysis Software and Systems IV Vol. 77 of ASP Conference Series. p. 433\\
Scharf, C. A., Zurek, D. R., Bureau, M., 2005, ApJ, 633, 154\\
Schnitzeler, D. H. F. M., 2010, MNRAS, 409, L99\\
Schnitzeler, D. H. F. M., et al., 2011, Australia Telescope Technical memo \# AT/39.9/129\\
Schulman, E., Fomalont, E. B., 1992, AJ, 103, 1138S\\
Simmons, J. F. L., \& Stewart, B. G. 1985, A\&A, 142, 100\\
Slysh V. I., 1965, AZh, 42, 689\\
Snowden, S. L., et al. 1995, ApJ, 454, 643\\
Sokoloff, D. D., Bykov A. A., Shukurov A., Berkhuijsen E. M., Beck R., Poezd A. D., 1998, MNRAS, 299, 189\\
Stil, J. M., Keller, B. W., George, S. J., Taylor, A. R., 2014, ApJ, 787, 99\\
Subrahmanyan, R., Ekers, R. D., Saripalli, L., \& Sadler, E. M. 2010, MNRAS, 402, 2792\\
Sun, X. H. et al. 2015, AJ, 149, 60\\
Taylor, A. R., Stil, J. M., Sunstrum, C., 2009, ApJ, 702, 1230\\
Tribble, P. C., 1991, MNRAS, 250, 726\\
Tucci,M., Mart\'{õ}nez-Gonz\'{a}lez, E., Toffolatti, L., Gonz\'{a}lez-Nuevo, J., \& De Zotti, G. 2004, MNRAS, 349, 1267\\
Vacca, V., et al., 2012, A\&A, 540, A38\\
Voges, W. et al., 1999, A\&A 349, 389\\
Voges, W., et al. 2000, IAU Circ., 7432, 3\\
Whiting, M. \& Humphreys, B., 2012, PASA, 29, 371\\
Wilson, W. E. et al., 2011, MNRAS, 416, 832\\
Wright E. L. et al., 2010, AJ, 140, 1868\\
Zavala R. T., Taylor G. B., 2004, ApJ, 612, 749\\

\appendix \label{sec-append}

\section{Arguments for the reliability of our complex detections}\label{sec-discussion-soundness}

False detections of Faraday complexity could arise for a number of reasons, including calibration inaccuracy / errors, polarization leakage, imaging artefacts or poor implementation of analysis techniques such as RM synthesis / {\sc rmclean}. In this section, we argue that each of these is either unlikely in our data, or has been appropriately considered in our analysis.

\subsection{Calibration}\label{sec-discussion-soundness-cal}

 For the following reasons, we consider calibration inaccuracies / limitations to be an unlikely cause of complexity in our data: 1) Complex sources are distributed throughout the mosaic field (Figure \ref{fig:Faradaypropsvsmosposish}): 5 of the 7 separately-calibrated mosaic days contain complex detections at the $10\sigma$ GES level, and all but one of the sub-mosaics (delimited by the purple dashed lines) contain at least one complex source. 2) The observed complexity are unchanged when the complex sources are re-imaged using data from the May and June epochs individually. Practically identical errors must then have been made in the 14 separate sub-mosaics to produce this behaviour. 3) Calibration errors should affect \emph{all} sources to a similar extent. Given the amplitude of the complex components observed in our complex sources, there are numerous unpolarized and Faraday simple sources observed with sufficient S/N to observe these additional components, yet this is not seen. 4) Complex sources found together within the same sub-mosaic show differences in the manifestation of their complex behaviour (see for example the differences between sources 034205-370322 (Fig. \ref{fig:034205-370322}) and 033653-361606 (Fig. \ref{fig:033653-361606}) in section \ref{sec-indiv}), which is difficult to explain in terms of calibration errors.

\subsection{Polarization leakage}\label{sec-discussion-soundness-leak}

At the time of writing, calibration of ATCA CABB data for off-axis polarization leakage is not possible. Instead, we estimate upper limits on the off-axis polarization leakage using a statistical analysis of our sample sources, as follows: 1) We selected the 100 brightest sources in the field, identified every pointing in which each such source was observed, then calculated the radial and azimuthal angular position of the source relative to the antenna feeds in each. 2) For each $uv$ cut in the visibility data for that pointing, we phase-shifted the source to the phase centre and calculated the average Stokes $Q$ \& $U$ visibility amplitudes. Since the polarized source density is less than one per pointing on average, the foregoing procedure provides a reasonable estimate of the linearly polarized intensities. 3) We extracted the integrated Stokes $I$ flux of the sources in the image domain, since the total intensity source density precludes a $uv$ plane analysis. We then divided the values of Stokes $Q$ \& $U$ by I, thereby obtaining an estimate of Stokes $q$ \& $u$ as the source rotates through the polarized beam during the observations. 4) We repeated this procedure every 32 MHz through the 16cm CABB band, resulting in a large number of independent probes of the polarized beam as a function of frequency and beam position. The only signal common to these data points should be polarization leakage.
 
\begin{figure}[h!]
\includegraphics[scale=1]{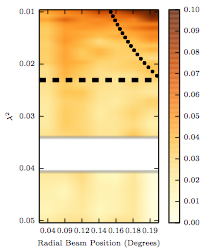}
\caption{Estimated frequency-dependant polarization leakage for individual $uv$ measurements in the ATCA beam, plotted as a function of $\lambda^2$ and radial beam position. The leakage estimates were derived as described in the main body of text. The dotted line is the primary beam HWHM at wavelength $\lambda$, while the heavy dashed line is the 2 GHz upper frequency cutoff that we employ for the bulk of analysis in this paper. The gap in the data for $0.034 < \lambda^2 < 0.04$ is due to RFI flagging. It can be seen that the estimated leakage approaches $\sim10\%$ of Stokes $I$ at a distance of 0.2 degrees from the beam centre at $\lambda^2=0.01$ (3 GHz), but decreases as we observe closer to the beam centre and at lower frequencies. Despite this overall trend, isolated islands of increased leakage appear above approximately 2 GHz (heavy dashed line in figure), even relatively close to the beam centre.}
\label{fig:leakage_uvdata}
\end{figure}

 We illustrate these results in Figure \ref{fig:leakage_uvdata}, where the median absolute deviation (MAD) of Stokes $q$ \& $u$ is plotted against $\lambda^2$ and radial distance from the nearest pointing centre. It is evident that the MAD (and thus the polarization leakage) increases sharply in severity at high frequencies and large radial separation from the beam centre. However, there remain isolated regions of enhanced leakage throughout the beam above $\sim2$ GHz. We thus adopt 2.0 GHz as the upper frequency limit on data considered in this paper.
 
\begin{figure}[h!]
\includegraphics[scale=0.5]{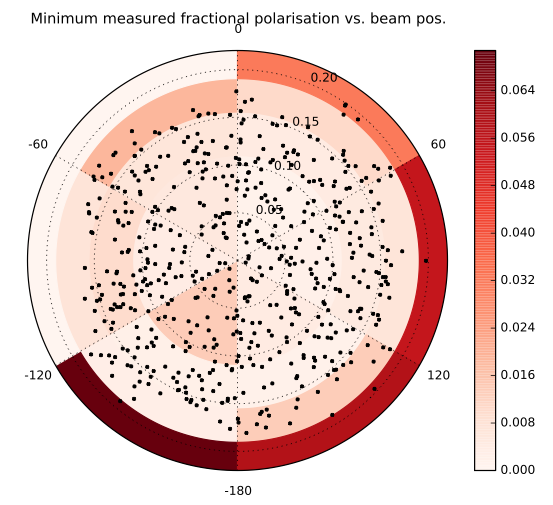}
\caption{A direct measurement from our data giving upper limits on leakage in the imaged data as a function of mosaic pointing centre-relative source position. For each sector of the beam shown in the figure, the maximum of the {\sc clean}ed FDS was extracted to yield the fractional polarization of the radio source. The minimum such value obtained through analysis of all sources in the sector was taken to be an upper limit on leakage, since leakage cannot give rise to a higher amplitude in the FDS than either real polarized signal or noise alone. We find that in almost every sector, the upper limit on leakage is typically less than 0.7\% of Stokes $I$ where sources bright enough to probe this deep exist.}
\label{fig:leakage_FDSdata}
\end{figure}
  
Having chosen to discard data above 2.0 GHz, we estimated upper limits on the leakage that would sum coherently when RM synthesis is applied to the imaged data. We applied RM synthesis to each source in our sample, then binned the sources based on their location in the primary beam, both in terms of angular distance from the phase centre and azimuthal angle from due North. In each of the 24 resulting bins, we identified the two lowest values of max($|FDS_{cln}|$) for $|\phi| < 1000$ rad m$^{-2}$, then adopted their average as an estimate of the upper limit on leakage within the sector. The results are shown in figure \ref{fig:leakage_FDSdata}, where we plot the positions of sources used for this calculation in the beam, representing the upper leakage estimate in each beam sector as a colorscale overlay. Taking the median as a function of radial angular separation from the pointing centre, leakage is limited to $< \sim 0.3\%$ of Stokes $I$ per RMSF beam for sources $<0.155^\circ$ from pointing centres where 14 of our 19 complex sources are found. At angular separations greater than this, we have very few sources with which to perform the estimate, and can only derive a weak upper limit of 1\% of Stokes $I$ in the same way. This demonstrates that only a small amount of leakage adds coherently across the band, and that any frequency dependent leakage that is present must get averaged down in the FDS during the application of RM synthesis.
 
Could leakage of this magnitude give rise to the complexity we observe? Table \ref{tab:allsourcedata} contains the fractional polarized flux found in `off-peak' or `complex' {\sc rmclean} components for all complex sources --- that is, the set of all {\sc rmclean} components after excluding the one with the highest amplitude. On average, these components contribute almost half of the polarized flux of the radio source. After convolving these off-peak {\sc rmclean} components with the FDS restoring beam, leakage cannot be responsible for the observed complexity if the amplitude of the resulting FDS is greater than the leakage limits derived in Section \ref{sec-impol}. We find this the case for all but four of our complex sources: The contribution to the fractional polarized signal by complex {\sc rmclean} components exceeds the upper limits on leakage by multiples of up to $\sim20$ (mean $= 5.6$, median $= 3.7$). Thus, the amplitude of the complex polarized emission components in our complex source sample generally exceeds the local upper estimates on leakage in the beam.

\begin{figure}[h]
\includegraphics[width=0.475\textwidth]{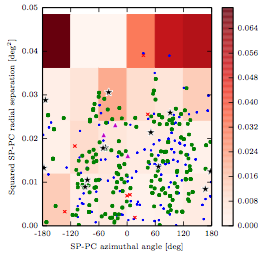}
\caption{A plot of $d^2$ versus $\theta$, where $d$ is the distance of the source from the pointing centre and $\theta$ is its position angle (East of North) with respect to the pointing centre. All complex detections are plotted, as are simple and unpolarized sources that 1) have angular separations less than $0.1^\circ$ to the complex detections, and 2) have fractional polarization {\sc rmclean} cutoffs set to less than the complex emission amplitude of complex sources. Plotting symbols are according to the same convention used throughout this paper. Coloured regions indicate the beam position leakage limits derived in section \ref{sec-impol}. If the Faraday complexity of the complex sources is caused by leakage, then all sources here are bright enough that they should also be detected as complex. As can be seen, many unpolarized and simple sources populate the plot, including many sitting at beam positions almost identical to the complex detections. This indicates that the complexity seen in these sources cannot be due to instrumental polarization leakage.}
\label{fig:leakmapnonpolar}
\end{figure}

As further evidence that leakage is not the cause of our observed complexity, we note that if a source's complexity is caused by leakage, then all other sufficiently bright sources in similar beam positions will also appear complex. Specifically, if we take such as source and, under {\sc rmclean}, restore its FDS using only the `off-peak' or `complex' {\sc rmclean} components, then the magnitude of the leakage is max($|$FDS$|$). All nearby sources which are bright enough for the {\sc rmclean} threshold to be set below this value should also then be detected as complex due to leakage. In Figure \ref{fig:leakmapnonpolar}, we plot the radial and azimuthal position of all sources for which this  {\sc rmclean} threshold criteria is met that are less than $0.1^\circ$ from a complex source. Data points follow the same marker conventions used throughout this paper. Clearly, a large number of sources do not show complexity on this plot, again indicating that leakage is not responsible for the observed complexity.

To conclude our argument, we make the following observations: While polarization leakage is expected to become more severe off-axis, our assigned Faraday complexity categories show no such trends. We compared the radial and azimuthal distribution functions of our complex detections vs. both our unpolarized and simple sources using 2-sample K-S tests. The resulting D and p values for comparison against the simple sources are 0.19 and 0.50 respectively, and the corresponding values are 0.13 and 0.88 for the unpolarized sources. There is thus no statistically significant differences between the radial distribution of complex and non-complex sources. The azimuthal dependences also show no significant differences. Furthermore, we observe sources which depolarize smoothly to $p=0$, sources that depolarize smoothly to some non-zero limiting value of $p$, sources which re-polarize, and sources that appear to show oscillatory depolarization. It is unlikely that polarization leakage could produce depolarization behaviour so varied in the same set of observations.

\subsection{Imaging artefacts}\label{sec-discussion-soundness-imart}

Many types of artefacts can affect aperture synthesis images. While it is impossible to guarantee that any data set is fully free of such artefacts, we have undertaken numerous consistency checks and experiments that show these are unlikely to contribute to the complexity that we observe. We provide the following basic tests / arguments to support this:

\begin{itemize}
\item All of our final mosaics were first visually examined for imaging artefacts or RFI. Images showing any problems were discarded or re-imaged.

\item We checked that the mean pixel values of the Stokes $Q$, $U$ and $V$ {\sc clean} residual maps was consistent with zero, and the distribution of pixel values in signal-free regions was well-fit by a Gaussian.

\item We cross-checked Stokes q, u \& v spectra from adjacent pointings and found them to be consistent with each other within errors. This indicates that the primary beam correction is accurate and the measured polarimetric structure of the sources is beam position-independent.

\item We confirmed that the $\alpha$ distribution of the sample was independent of beam position, and was consistent with spectral index distributions derived from similar surveys in the literature. Again, this shows that the primary beam model is sound, as well as our data reduction, calibration and imaging in general.

\item We observed that the Stokes $I$ spectra of sources in our sample were generally well-fit by a power law model, as expected. We checked that no significant `jumps' were apparent in Stokes $I$, $Q$, $U$ or $V$ spectra between separately calibrated adjacent frequency bins. Inspection of the Stokes $I$ data $-$ Stokes $I$ Model data point residuals shows some evidence of non-random `wiggles', but generally only at a low level - around $\sim 2-5$ mJy. Such wiggles could not be the cause of the complexity we observe in most of our objects, since 1) the frequency scale over which these wiggles occur would place power in the FDS well beyond the range at which the Faraday complex emission components is generally observed ($|\phi|>1000$ rad m$^{-2}$), and 2) the wiggle amplitude is such that the percentage change they induce in Stokes $q$ \& $u$ following the Stokes $I$ model division is generally negligible compared to the values of Stokes $q$ \& $u$ for the polarized sources.

\item We have compared the total intensity flux densities of the complex sources and found no difference within the errors between epochs. Furthermore on empirical grounds, we expect less than 1\% of sources to show variability at 1.4 GHz over $\sim$month-long time frames --- e.g. Mooley et al. (2013) and Ofek \& Frail (2011). 

\end{itemize}

\subsection{{\sc rmclean} depth}\label{sec-discussion-soundness-cldepth}

Deciding whether features in the FDS can be considered statistically significant detections or not is particularly important for our work, since our method of identifying complexity relies directly on the {\sc rmclean} components derived from the FDS features. It is therefore crucial that the lowest {\sc rmclean} cutoff threshold we adopt is set above the amplitude to which noise peaks are statistically likely to rise.
 
\begin{figure}[h]
\includegraphics[width=0.475\textwidth]{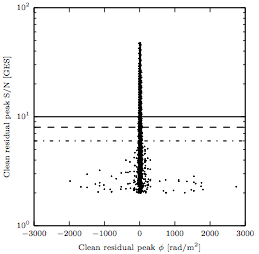}
\caption{The GES signal / noise of {\sc rmclean} components after having been multiplied by the inverse of the {\sc rmclean} loop gain, plotted against their Faraday depth. The dot-dash, dashed and solid lines represent the $6\sigma$, $8\sigma$ and $10\sigma${\sc rmclean} cutoffs (respectively) employed in the work we present in this paper. The {\sc rmclean} components show nearly constant variance down to just above the $5\sigma$ GES level, below which {\sc rmclean} components are found at increasingly varied Faraday depths. While some of these points may represent genuine detections, it is probable that the majority represent noise peaks above the {\sc rmclean} threshold. It is clear that the cutoffs that we employ in our analysis occur well above this level, particularly for the $8\sigma$ and $10\sigma$ GES levels.  }
\label{fig:cleancompGESvsFD}
\end{figure}
 
 The reliability of RM synthesis detections as a function of  S/N in the FDS has been assessed before. Brentjens \& deBruyn (2005) show that detection reliability falters at $6\sigma$ peak-to-noise, while Macquart et al. (2012) observe the same effect at $5$--$6\sigma$ peak-to-noise. Converting these numbers to GES values that are independent of experimental setup yields $4.75\sigma$ GES and $5\sigma$ GES respectively. Accordingly, in any RM synthesis experiment employing {\sc rmclean} deconvolution, the {\sc rmclean} cutoff should be set above these GES values plus some margin to ensure that noise peaks in the FDS are not being {\sc clean}ed. 
 
 As explained in section \ref{sec-compcat}, we adopt three different cutoffs in our analysis: $6\sigma$ GES, $8\sigma$ GES \& $10\sigma$ GES. These are above the GES-equivalent limits that Brentjens \& deBruyn (2005) or Macquart et al. (2012) recommend. We demonstrate that our data are consistent with their findings in the following way. We took the {\sc rmclean} components from Faraday simple sources, multiplied by the inverse of the {\sc rmclean} loop gain and calculated the GES of the resulting amplitude / noise ratio, then plotted this against the Faraday depth at which the {\sc rmclean} components were found. We deliberately set the {\sc rmclean} depth to $2\sigma$ GES --- well below the threshold where detections can be considered reliable. We plot the results in Figure \ref{fig:cleancompGESvsFD}, emulating Figure 4 in Macquart et al. (2012). It is evident that the Faraday depths are tightly clustered with more or less constant scatter down to $\sim 5\sigma$ GES, Between $5\sigma$ GES and $\sim 3.5\sigma$ GES, the scatter increases but is not uniformly distributed along the x-axis. It is not clear whether these points represent genuine low-level peak detections or noise peaks. Below $3\sigma$ GES, the points are almost randomly distributed along the $\phi$ axis, and we can therefore state with certainty that noise peaks are being {\sc clean}ed. It is clear that $6\sigma$, $8\sigma$ \& $10\sigma$ GES cutoffs are all above the noise regime.

\end{document}